\documentclass[Conference,a4paper]{IEEEtran}
\IEEEoverridecommandlockouts
\usepackage{color}
\usepackage{graphicx}
\usepackage{epstopdf}
\usepackage{amsmath}
\usepackage{amssymb}
\usepackage{algorithm}
\usepackage{algorithmic}
\usepackage{amsmath}
\usepackage{multirow}
\usepackage{booktabs}
\usepackage{array}
\usepackage{amsthm}
\usepackage{stfloats}
\usepackage{caption}
\usepackage{subfigure}
\usepackage{bm}
\usepackage{booktabs}
\usepackage{setspace}
\usepackage{diagbox}
\usepackage{enumerate}
\usepackage{ulem}
\usepackage{url}

\usepackage{hyperref}
\allowdisplaybreaks[4]

{\bgroup
 \addtolength\abovedisplayshortskip{#1}
 \addtolength\abovedisplayskip{#1}
 \addtolength\belowdisplayshortskip{#1}
 \addtolength\belowdisplayskip{#1}
 }
{\egroup\ignorespacesafterend}

\newtheorem{lemma}{Lemma}

\newcommand{\be}{\begin{equation}}
\newcommand{\ee}{\end{equation}}
\newcommand{\bea}{\begin{eqnarray}}
\newcommand{\eea}{\end{eqnarray}}
\newcommand{\ba}{\begin{array}}
\newcommand{\ea}{\end{array}}

\newcommand{\non}{\nonumber}

\title{Movable Antenna Enhanced  Networked Integrated Sensing and Communication System}
\author{\IEEEauthorblockN{Yuan Guo, Wen Chen, Qingqing Wu, Yang Liu, Qiong Wu, Kunlun Wang, Jun Li, and Lexi Xu
\thanks{
The work of Wen Chen is supported by NSFC 62531015 and Shanghai Kewei 24DP1500500.
The work of Qingqing Wu is supported by NSFC 62371289 and NSFC 62331022. 
The work of Yang Liu is supported in part by the open research fund of National Mobile Communications Research Laboratory, Southeast University (No. 2025D06).
The work of Qiong Wu is supported in part by Basic Research Program of Jiangsu under Grant BK20252084.
Wen Chen is the corresponding author.}
\thanks{
Y. Guo, W. Chen and
Q. Wu are with Department of Electronic Engineering, Shanghai Jiao Tong University, Shanghai, China, 
email:
yuanguo26@sjtu.edu.cn,
wenchen@sjtu.edu.cn,
qingqingwu@sjtu.edu.cn.
Y. Liu  is with the School of Information
and Communication Engineering, Dalian University of Technology, Dalian, China, 
email:
yangliu\_613@dlut.edu.cn.
Q. Wu is with the School of Internet of Things Engineering, 
Jiangnan University, Wuxi, China, 
emali: qiongwu@jiangnan.edu.cn.
K. Wang is with 
the School of Communication and Electronic Engineering, 
East China Normal University, Shanghai, China,
email: klwang@cee.ecnu.edu.cn.
J. Li is with the School of Information Science and Engineering, Southeast University, Nanjing, China, 
email: jleesr80@gmail.com.
L. Xu is with the Research Institute,
China United Network Communications Corporation, Beijing,
China, email: davidlexi@hotmail.com.
}
}
}

\begin{document}
\maketitle
\pagestyle{empty}
\thispagestyle{empty}

\begin{abstract}
 Integrated sensing and communication (ISAC) is a key technology for future 6G networks. 
 Most existing studies focus on monostatic and/or bistatic setups with limited coverage and capabilities. 
 Networked ISAC systems with distributed base stations (BSs) can overcome these limitations. 
 Moreover, movable antenna (MA) architectures offer improved ISAC performance over fixed-position antennas (FPAs) by enabling adaptable antenna movement.
In this paper,
we utilize the MA to promote communication capability
with guaranteed sensing performance
via jointly designing beamforming, power allocation,
receiving filters
and position configuration of transmit/receive MA towards maximizing the sum rate for both downlink (DL) and uplink (UL) users.
The optimization problem is  highly difficult
due to the unique channel model derived from the position coefficient of the MA.
To resolve this challenge,
via leveraging the cutting-the-edge majorization-minimization (MM) method,
we develop an efficient solution that optimizes all variables via 
convex optimization techniques.
Extensive simulation results verify 
the effectiveness of our proposed algorithms
and
demonstrate
the substantial performance promotion by deploying the MA framework in the networked ISAC system.

\end{abstract}


\maketitle
\section{Introduction}
Recently,
with the rapid increase in devices requiring precision sensing and efficient communication,
integrated sensing and communication (ISAC) \cite{ref_ISAC_F_Liu}$-$\cite{ref_ISAC_J. A. Zhang}
technologies have attracted great interest from both industry and academia.
In this context,
ISAC is envisioned as a promising solution,
which aims to deploy both communication and sensing functionalities on one unified hardware platform,
allowing the sharing of frequency spectrum, hardware resources, and signal-processing units.
Many recent advancements in the joint design of radar sensing and communication
are extensively documented in \cite{ref_ISAC_F_Liu}$-$\cite{ref_ISAC_J. A. Zhang}
and their associated references.

However,
the conventional fixed-position antenna (FPA) at the base station (BS) and/or mobile user in the ISAC system
generally fail to fully exploit the degrees of freedom (DoFs) in the continuous spatial domain,
leading to performance loss in both sensing and communication tasks.
A novel movable antenna (MA) technology \cite{ref_MA_z}$-$\cite{ref_channel model}
is expected to  overcome the limitation.
The MA is considered as
a viable approach for enhancing wireless network performance,
linked to a radio frequency (RF) chain through a flexible cable.
It can reconstruct channel conditions by flexibly adjusting its position in a given spatial area
with the aid of a driver component or by other means.
The various aspects of wireless systems by applying the MA framework  have been extensively studied recently,
see \cite{ref_ot work 1}$-$\cite{ref_Movable antennas Magazine} and their references.

\subsection{Related Works}

 In the mono-static ISAC system, 
the full-duplex (FD) operation presents a significant challenge due to self-interference \cite{ref_ISAC_F_Liu}, 
which cannot be overlooked.
The work \cite{ref_FD_mono_ISAC_1} aimed to 
deal with the challenges associated with the FD operation in mono-static  ISAC systems operating at  millimeter-Wave (mm-Wave) frequencies,
and proposed the
beamforming solutions designed to maximize the beamformer power in the sensing direction 
while simultaneously constraining the beamformer power in the communication directions.
The authors of \cite{ref_FD_mono_ISAC_2} and \cite{ref_FD_mono_ISAC_3} 
 adopted the Orthogonal Frequency Division Multiplexing (OFDM) waveform to 
estimate the Direction of Arrival (DoA), range, and relative velocity of the radar targets while maximizing the downlink (DL) communication rate
in the  mono-static FD ISAC network operating at  the mm-Wave frequencies.
The paper \cite{ref_FD_mono_ISAC_4} investigated a  mono-static FD ISAC system,
in which the BS simultaneously conducts target detection 
while communicating with multiple DL and uplink (UL) users, 
sharing the same time-frequency resources.
The authors of \cite{ref_FD_mono_ISAC_5} utilized the FD in the ISAC transceiver to solve the echo-miss problem.

The networked ISAC system
with multiple transmitters and/or receivers
can significantly improve both the communication and radar sensing performance \cite{ref_ISAC_F_Liu}.
In \cite{ref_Networked_ISAC_1},
the work investigated a networked ISAC system
that considers two types of users: 
those capable of canceling interference from dedicated sensing signals and those that are not.
The paper \cite{ref_Networked_ISAC_2}  presented 
a cell-free ISAC MIMO system, 
deriving sensing 
signal-to-noise (SNR)
 and developing joint beamforming and power allocation for optimal communication and sensing performance.
The authors of \cite{ref_Networked_ISAC_3}   simultaneously considered  
the active and passive sensing tasks in a downlink distributed ISAC system, 
adopting different fusion strategies based on the backhaul capacity and enhancing sensing performance 
while meeting minimum signal-to-interference-plus-noise (SINR) requirements.
The work \cite{ref_Networked_ISAC_4} demonstrated 
a multi-static ISAC design using network-assisted full-duplex (NAFD) cell-free networks,
 proposing a deep Q-network for optimizing access point (AP) duplex modes.

Due to the advantages of MA technology, 
a substantial body of literature has explored deploying MA in the ISAC systems
and utilizing joint beamformer and position design to enhance both sensing and communication performance,
e.g.,
\cite{ref_related_work_1}$-$\cite{ref_related_work_9_1}.
For instance,
the authors of \cite{ref_related_work_1} and \cite{ref_related_work_9_4}
investigated the deployment of the MA in the ISAC system
and
showed
it could significantly boost
sum-rate of  the DL users
while guaranteeing  the sensing performance constraint.
The work \cite{ref_related_work_2}
designed an MA-aided bi-static ISAC system's flexible waveform to enhance
communication rate and sensing mutual information (MI).
The paper \cite{ref_related_work_3}
demonstrated that
the implementation of the MA
could significantly inflate the data transmission rate with the transmit beamforming gain in
the unmanned aerial vehicle (UAV)-enabled  low-altitude platform (LAP) system.
The authors of \cite{ref_related_work_4}
considered the sensing SNR
maximization while constraining the minimum SINR per user.
The works \cite{ref_related_work_6} and \cite{ref_related_work_9_5} showed that MA could effectively reduce 
the Cram$ \acute{\text{e}}$r-Rao bound (CRB) for target parameter estimation,
which is generally a good  performance metric of sensing.
The recent work \cite{ref_related_work_7}  was the first one to deploy RIS
to enhance communication and sensing performance in the MA-aided ISAC system.
The authors of \cite{ref_related_work_8}
aimed to improve the energy efficiency in the MA-assisted ISAC system accounting for
 dynamic radar cross-section (RCS) coefficient. 
 The papers \cite{ref_related_work_9}$-$\cite{ref_related_work_9_3}
adopted the emerging MA technology in a mono-static FD ISAC system for efficiently suppressing self-interference.
The work \cite{ref_related_work_9_1} utilizes the novel MA architecture in the ISAC system for supporting low-space vehicles.

\subsection{Motivations and Contributions}

 As shown previously,
although several studies \cite{ref_FD_mono_ISAC_1}$-$\cite{ref_Networked_ISAC_4} 
have investigated joint beamforming optimization in the ISAC systems, 
most have relied on traditional FPAs, 
limiting their adaptability and performance. 
Since the MA technology can effectively utilize spatial DoFs in practical networks 
 \cite{ref_Lemma1_1}$-$\cite{ref_channel model},
extensive literature \cite{ref_related_work_1}$-$\cite{ref_related_work_9_1}
has explored the joint beamforming and MA's position optimization for MA-aided ISAC systems
but primarily focused on monostatic and/or bistatic configurations.
These settings have limited service coverage and inadequate performance in both sensing and communication within complex environments.
However,
the networked ISAC  is
a promising solution for enhancing coverage, spectral efficiency, and reliability
 \cite{ref_ISAC_F_Liu}$-$\cite{ref_ISAC_J. A. Zhang}.
Moreover,
these works
\cite{ref_related_work_1}$-$\cite{ref_related_work_8},
\cite{ref_related_work_9_1},
primarily concentrate on DL communication, 
overlooking the crucial aspect of UL communication
which is essential for comprehensive ISAC functionality. 
This oversight can restrict the practical application of existing systems in dynamic environments 
where bidirectional communication is vital.
Therefore,
we are motivated to study an MA-aided networked ISAC system,
that incorporates multi-static configurations and accommodates both DL and UL users,
aiming to enhance its communication and sensing capabilities.
Specifically,  
the contributions of this paper are elaborated as follows:
\begin{itemize}
\item
This paper considers the joint beamforming and antennas' position design
in a networked ISAC system enabled by the MA structure
to promote simultaneous DL and UL communication and target sensing functionalities,
which has not yet been thoroughly investigated.
This novel framework can
i) effectively enlarge the service region for both communication and sensing;
ii) supply more spatial DoFs by leveraging the MA technology to reconstruct the wireless channel condition;
iii) provide comprehensive bidirectional communication (i.e., DL and UL communication) transmission.

\item
We investigate the maximization of the sum-rate for both DL and UL users
while guaranteeing target detection quality via jointly optimizing
the BSs' beamforming,
UL users' power allocation,
receive filter
and MAs' positions.
Besides, 
this paper considers a very generic signal propagation model,
which fully takes into account both downlink and uplink communication and reflected radar probing signals.
As will be seen, this consideration in modeling significantly complicates the beamforming and antennas' position design task.
To the best of our knowledge,
this problem has not been considered in the existing literature, e.g.,
\cite{ref_related_work_1}$-$\cite{ref_related_work_9_1}.

\item
The joint beamforming and MA position's configuration design to improve sum-rate of both DL and UL users is
a highly non-convex problem
due to the position coefficients of MA.
Via utilizing cutting-the-edge optimization frameworks including the
weighted minimum mean squared error (WMMSE) \cite{ref_WMMSE} and majorization-minimization (MM) \cite{ref_MM},
we have developed a closed solution for updating MA's position of the DL users.
Furthermore,
we successfully develop an effective algorithm that updates the position coefficients of MA
by  convex optimization techniques.

\item
Last but not least,
extensive numerical results are provided to verify the effectiveness of our proposed solution,
which demonstrates as: 
1) our proposed algorithm can converge within several iterations; 
2) the proposed algorithm can achieve much better than the benchmark cases (especially its
counterpart with the FPA-based array at both BSs and users) in terms of sum-rate of both DL
and UL users under various settings.
Experiment results
validate that the deployment of MA can significantly benefit communication performance
in the networked  ISAC system.

\end{itemize}

\section{System Model and Problem Formulation}

\begin{figure}[ht]
    \centering
    \subfigure[{}]{
        \includegraphics[width=0.32\textwidth]{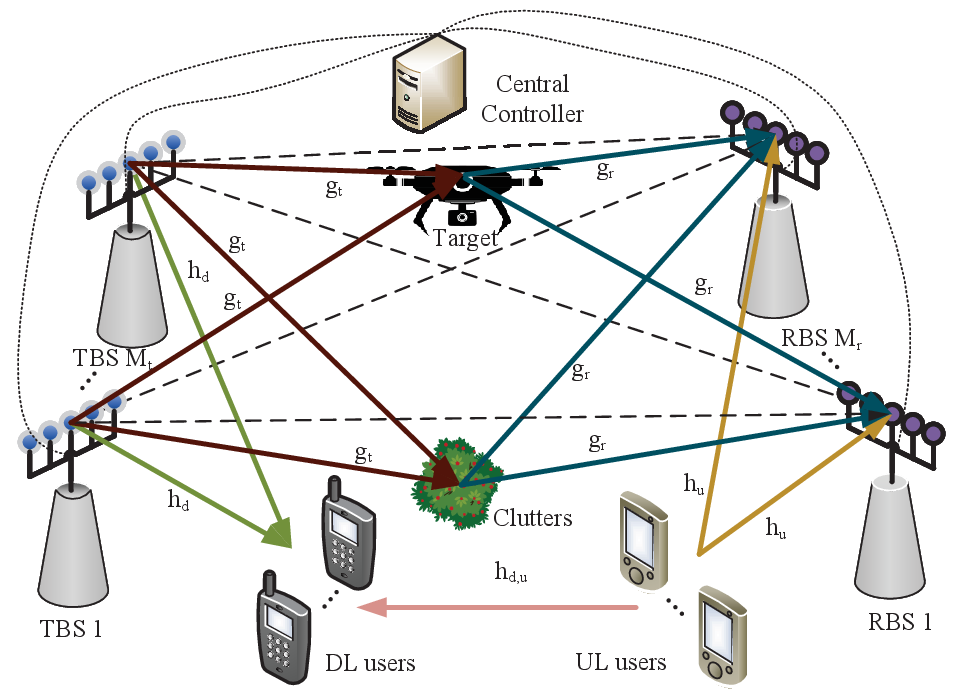}
    }
    \hfill
    \subfigure[{}]{
        \includegraphics[width=0.32\textwidth]{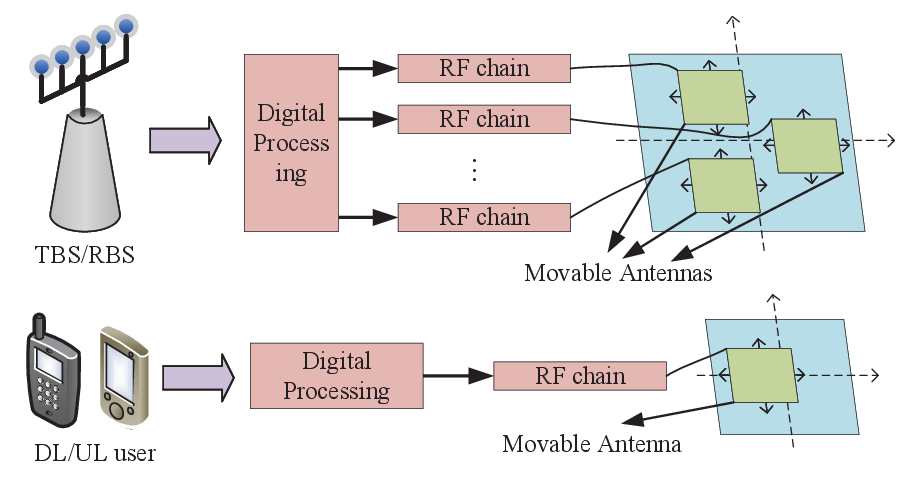}
    }
    \caption{A  MA-aided networked  ISAC  system.
     (a): System model 
    (b): Illustrations of the MAs at the BS and users, respectively.
    }
    \label{fig.1}
\end{figure}

As shown in Fig. \ref{fig.1}\footnote{ The dotted line represents the link between TBS and RBS. },
we consider an MA-enhanced  networked ISAC system
that comprises $M_t$ transmit BSs (TBSs),
$M_r$ receive BSs (RBSs),
$K_d$ single MA DL mobile users,
$K_u$ single MA UL mobile users,
one point-like sensing target,
and $K_t$ clutter sources.
Moreover,
TBSs and RBSs are respectively equipped with $N_t$ and $N_r$ MAs.
Let
$\mathcal{M}_t \triangleq \{1,\cdots,M_t\} $,
$\mathcal{M}_r \triangleq \{1,\cdots,M_r\} $,
$\mathcal{N}_t \triangleq \{1,\cdots,N_t\} $,
$\mathcal{N}_r \triangleq \{1,\cdots,N_r\} $,
$\mathcal{K}_d \triangleq \{1,\cdots,K_d\} $,
$\mathcal{K}_u \triangleq \{1,\cdots,K_u\} $
and
$\mathcal{K}_t \triangleq \{1,\cdots,K_t\} $
denoted the sets of
TBSs,
RBSs,
TBSs' MAs,
RBSs' MAs,
DL users,
UL users
and
clutter sources,
respectively.

Moreover,
the MAs are connected to RF chains by flexible cables 
\footnote{
The response times of motor-based mechanisms is the order of milliseconds to seconds \cite{ref_MA_z}.},
and can move within local region in real time \cite{ref_Lemma1_1}$-$\cite{ref_channel model}.
Let 
$\mathbf{t}^{0}_{m,n} = [x^{0}_{m,n},y^{0}_{m,n} ]^T \in \mathcal{C}$
denote the  position of the $n$-th MA of $m$-th  TBS.
Similarly,
the  $n$-th MA' position  of $m$-th  RBS is given by 
$\mathbf{t}^{1}_{m,n} = [x^{1}_{m,n},y^{1}_{m,n} ]^T \in \mathcal{C}$.
The positions of MA at the  $k$-th  DL and $l$-th UL users
are respectively written as
$\mathbf{r}_{0,k} = [x^d_{k},y^d_{k} ]^T \in \mathcal{C}$
and
$\mathbf{r}_{1,l} = [x^u_{l},y^u_{l} ]^T \in \mathcal{C}$,
where
$\mathcal{C}$  denotes the given two-dimensional (2D) moving
square region with a size of  $A \times A$ for both the BSs' and  users' MAs,
while the corresponding reference point's position is $[0,0]^T$.

In this network,
target detection and communication are performed in a time-division manner \cite{ref_ISAC_F_Liu}.
TBSs,
with the aid of MA,
service DL users and send probing waveform to detect the target,
and RBSs,
with the assistance of MA,
receive the information from UL users and the echo signal from the sensing target.
We assumed that the BSs are connected to the central controller via fronthaul links
and that cooperative sensing of the target through data-level fusion
\cite{ref_Networked_ISAC_1}$-$\cite{ref_Networked_ISAC_2}.
All TBSs and RBSs can achieve perfect synchronization and joint signal processing facilitated by the central controller,
and the signals transmitted by the TBSs can be made available at all RBSs.
Moreover,
the link between BSs (i.e., intra-BS channel) is assumed to be quasi-static.
Therefore,
the received signal from the direct link between the TBS and the RBS 
can be perfectly eliminated using prior knowledge obtained through the fronthaul links, 
i.e.,  passive optical network (PON), 
coax, 
cable \cite{ref_CRAN}.
Since the users are located in a complex environment with multiple reflected and scattered paths (containing the target/clutters-users links),
the channels of the users are assumed to be multiple paths channel model \cite{ref_channel model},
that can be estimated via least square method \cite{ref_MA_z}.

\subsection{Channel Model}

Since the MA's  moving area is usually much smaller than the signal propagation distance,
the channels between the BSs/users and the users are assumed to follow the far-field wireless channel model 
\cite{ref_Lemma1_1}$-$\cite{ref_channel model}.
Therefore,
changing the position of MA has no effect on the angle of departure (AoD), the angle of arrival (AoA),
and the amplitude of the complex path coefficient,
while only the phase of the complex path coefficient varies with changes in MA's position.

\underline{\textit{a) Communication Channel:}}

\textit{1) TBS-DL User Channel:}
Let $L^{t,0}_{m,k}$ and $L^{r,0}_{m,k}$
denote the total number of transmit and receive channel paths at
the $m$-th  TBS from the $k$-th DL user, respectively.
The  elevation and azimuth AoDs of the $i$-th transmit path between TBS $m$ and DL user $k$  are respectively denoted as
$\theta^{t,0}_{m,k,i}$
and
$\phi^{t,0}_{m,k,i}$,
while the corresponding elevation and azimuth  AoAs of the $i$-th receive path are represented as
$\theta^{r,0}_{m,k,i}$
and
$\phi^{r,0}_{m,k,i}$, respectively.
In the $i$-th transmit path for $m$-th TBS-$k$-th DL user link,
the propagation distance difference between the $n$-th MA and its reference point
can be given as
\begin{align}
&\rho^{t,0}_{m,k,n,i}(\mathbf{t}^{0}_{m,n}) = ({\mathbf{t}_{m,n}^{{0}}})^T\mathbf{a}^{t,0}_{m,k,i} \\
&= x^{0}_{m,n}\cos(\theta^{t,0}_{m,k,i})\sin(\phi^{t,0}_{m,k,i})  +   y^{0}_{m,n}\sin(\theta^{t,0}_{m,k,i}),\nonumber
\end{align}
and
its phase difference can be obtained as $2\pi\rho^{t,0}_{m,k,n,i}(\mathbf{t}^{0}_{m,n})/\lambda$,
where $\lambda$ denotes the carrier wavelength.

Thus,
the field-response vector (FRV) of the transmit channel paths
between the $k$-th DL user and the  $m$-th TBS' $n$-th MA is given by
\vspace{-0.2cm}
\begin{small}
\begin{align}
\mathbf{h}^{0}_{m,k,n}\!\! =\!\! [e^{j\frac{2\pi}{\lambda}\rho^{t,0}_{m,k,n,1}}, e^{j\frac{2\pi}{\lambda}\rho^{t,0}_{m,k,n,2}},\cdots,
e^{j\frac{2\pi}{\lambda}\rho^{t,0}_{m,k,n,L^{t,0}_{m,k}}}]^T,
\end{align}
\end{small}

\noindent
and then
the corresponding field response matrix (FRM) of all $N_t$  MAs
 can be written as
 \vspace{-0.2cm}
\begin{small}
\begin{align}
&\mathbf{H}^{0}_{m,k}(\mathbf{t}^0_m) = [\mathbf{h}^{0}_{m,k,1},\mathbf{h}^{0}_{m,k,2}, \cdots, \mathbf{h}^{0}_{m,k,N_t} ] \in \mathbb{C}^{L^{t,0}_{m,k} \times N_t },
\end{align}
\end{small}

 \vspace{-0.2cm}
\noindent
where $\mathbf{t}^0_m \triangleq \{\mathbf{t}^0_{m,n}\}$.
The FRV of the receive channel paths
between DL user $k$ and  TBS $m$
is defined as
 \vspace{-0.2cm}
\begin{small}
\begin{align}
&\mathbf{h}^{0,0}_{m,k}(\mathbf{r}_{0,k}) = [ e^{j\frac{2\pi}{\lambda} \rho^{r,0}_{m,k,1} },
e^{j\frac{2\pi}{\lambda} \rho^{r,0}_{m,k,2} }, \cdots,
e^{j\frac{2\pi}{\lambda} \rho^{r,0}_{m,k, L^{r,0}_{m,k}} }
 ]^T,
\end{align}
\end{small}

\vspace{-0.2cm}
\noindent
where $\rho^{r,0}_{m,k,i}(\mathbf{r}_{0,k}) \!\triangleq\! \mathbf{r}_{0,k}^T \mathbf{a}^{r,0}_{m,k,i}
\!=\!  x^d_{k}\cos(\theta^{r,0}_{m,k,i})\sin(\phi^{r,0}_{m,k,i})  +   y^d_{k}\sin(\theta^{r,0}_{m,k,i})$.
Therefore,
the channel vector between
the $m$-th TBS and the $k$-th DL user
is given by
 \vspace{-0.2cm}
 \begin{small}
\begin{align}
&\mathbf{h}_{d,m,k}^H(\mathbf{t}^0_{m},\mathbf{r}_{0,k})
 = \mathbf{h}^{0,0}_{m,k}(\mathbf{r}_{0,k})^H\mathbf{\Sigma}^{0}_{m,k}\mathbf{H}^{0}_{m,k}(\mathbf{t}^0_m) \in
 \mathbb{C}^{1 \times N_t},
\end{align}
\end{small}

 \vspace{-0.2cm}
\noindent
where
the  matrix
$\mathbf{\Sigma}^{0}_{m,k}$
is denoted as the response of all
transmit and receive paths from  TBS  $m$ to  DL user $k$.

\textit{2) RBS-UL User Channel:}
Let $L^{t,1}_{m,k}$ and $L^{r,1}_{m,k}$
denote the total number of transmit and receive channel paths at
  RBS $m$ from  UL user $k$, respectively.
The  elevation and azimuth AoDs of the $i$-th transmit path between RBS $m$ and UL user $k$  are respectively denoted as
$\theta^{t,1}_{m,k,i}$
and
$\phi^{t,1}_{m,k,i}$,
while the corresponding elevation and azimuth  AoAs of the $i$-th receive path are represented as
$\theta^{r,1}_{m,k,i}$
and
$\phi^{r,1}_{m,k,i}$, respectively.
In the $i$-th transmit path for $m$-th RBS-$k$-th UL user link,
the phase difference of the $n$-th MA and its reference point 
is given by
\begin{align}
&2\pi\rho^{t,1}_{m,k,n,i}(\mathbf{t}^{1}_{m,n})/\lambda 
= 2\pi({\mathbf{t}_{m,n}^{1}})^T\mathbf{a}^{t,1}_{m,k,i}/\lambda \\
&= 2\pi\big(x^{1}_{m,n}\cos(\theta^{t,1}_{m,k,i})\sin(\phi^{t,1}_{m,k,i})  +   y^{1}_{m,n}\sin(\theta^{t,1}_{m,k,i})\big)/\lambda.\nonumber
\end{align}

Thus,
the FRV of the transmit channel paths
between the $k$-th UL user and the  $m$-th RBS' $n$-th MA is given by
\begin{small}
\begin{align}
\mathbf{h}^{1}_{m,k,n}\!\! =\!\! [e^{j\frac{2\pi}{\lambda}\rho^{t,1}_{m,k,n,1}}, e^{j\frac{2\pi}{\lambda}\rho^{t,1}_{m,k,n,2}},\cdots,
e^{j\frac{2\pi}{\lambda}\rho^{t,1}_{m,k,n,L^{t,1}_{m,k}}}]^T,
\end{align}
\end{small}

\noindent
and then
the corresponding FRM of all $N_r$  MAs
is written as
\begin{small}
\begin{align}
&\mathbf{H}^{1}_{m,k}(\mathbf{t}^0_m) = [\mathbf{h}^{1}_{m,k,1},\mathbf{h}^{1}_{m,k,2}, \cdots, \mathbf{h}^{1}_{m,k,N_r} ] \in \mathbb{C}^{L^{t,1}_{m,k} \times N_r },
\end{align}
\end{small}

\noindent
where $\mathbf{t}^1_m \triangleq \{\mathbf{t}^1_{m,n}\}$.
The FRV of the receive channel paths
 between UL user $k$ and RBS $m$ is given as
\begin{small}
\begin{align}
& \mathbf{h}^{0,1}_{m,k}(\mathbf{r}_{1,k}) = [ e^{j\frac{2\pi}{\lambda} \rho^{r,1}_{m,k,1} },
e^{j\frac{2\pi}{\lambda} \rho^{r,1}_{m,k,2} }, \cdots,
e^{j\frac{2\pi}{\lambda} \rho^{r,1}_{m,k, L^{r,1}_{m,k}} }
 ]^T,
\end{align}
\end{small}

\noindent
where 
$\rho^{r,1}_{m,k,i}(\mathbf{r}_{1,k}) \!\!\triangleq\!\! \mathbf{r}_{1,k}^T \mathbf{a}^{r,1}_{m,k,i}
\!=\!  x^u_{k}\cos(\theta^{r,1}_{m,k,i})\sin(\phi^{r,1}_{m,k,i})+y^u_{k}\sin(\theta^{r,1}_{m,k,i}) 
$.
Therefore,
the channel vector 
between
the $m$-th RBS and the $k$-th UL user
is represented as
 \begin{small}
\begin{align}
& \mathbf{h}_{u,m,k}^H(\mathbf{t}^1_{m},\mathbf{r}_{1,k})
\! =\! \mathbf{h}^{0,1}_{m,k}(\mathbf{r}_{1,k})^H\mathbf{\Sigma}^{1}_{m,k}\mathbf{H}^{1}_{m,k}(\mathbf{t}^1_m) \in
 \mathbb{C}^{1 \times N_r},
\end{align}
\end{small}

\noindent
with
the  matrix
 $\mathbf{\Sigma}^{1}_{m,k}$
 denoting as the response of all
transmit and receive paths 
from the $m$-th RBS to the $k$-th UL user.

\textit{3) DL User-UL User Channel:}
$L^{t,2}_{k,l}$ and $L^{r,2}_{k,l}$ are respectively denoted as
the total number of transmit and receive channel paths between DL user $k$  and UL user $l$.
Denote the elevation and azimuth AoDs of the $i$-th transmit path between UL user $l$  and DL user $k$
as $\theta^{t,2}_{l,k,i}$
and
$\phi^{t,2}_{l,k,i}$, respectively,
while the corresponding  elevation and azimuth AoAs for the 
$i$-th receive path are represented as
 $\theta^{r,2}_{l,k,i}$
and
$\phi^{r,2}_{l,k,i}$, respectively.

The FRVs of the transmit and receive channel paths
between DL user $k$ and  UL user $l$ are respectively
expressed as
\begin{small}
\begin{align}
\mathbf{h}^{2}_{k,l}(\mathbf{r}_{1,l}) = [ e^{j\frac{2\pi}{\lambda} \rho^{t,2}_{k,l,1} },
e^{j\frac{2\pi}{\lambda} \rho^{t,2}_{k,l,2} }, \cdots,
e^{j\frac{2\pi}{\lambda} \rho^{t,2}_{k,l, L^{t,2}_{k,l}} }
 ]^T,\\
 \mathbf{h}^{3}_{k,l}(\mathbf{r}_{0,k}) = [ e^{j\frac{2\pi}{\lambda} \rho^{r,2}_{k,l,1} },
e^{j\frac{2\pi}{\lambda} \rho^{r,2}_{k,l,2} }, \cdots,
e^{j\frac{2\pi}{\lambda} \rho^{r,2}_{k,l, L^{r,2}_{k,l}} }
 ]^T,
\end{align}
\end{small}

\noindent
where $\rho^{t,2}_{k,l,i}(\mathbf{r}_{1,l}) \triangleq \mathbf{r}_{1,l}^T \mathbf{a}^{t,2}_{k,l,i}
=  x^u_{l}\cos(\theta^{t,2}_{k,l,i})\sin(\phi^{t,2}_{k,l,i})  +   y^u_{l}\sin(\theta^{t,2}_{k,l,i})$
and
$\rho^{r,2}_{k,l,i}(\mathbf{r}_{0,k}) \triangleq \mathbf{r}_{0,k}^T \mathbf{a}^{r,2}_{k,l,i}
=  y^d_{k}\sin(\theta^{r,2}_{k,l,i}) + x^d_{k}\cos(\theta^{r,2}_{k,l,i})\sin(\phi^{r,2}_{k,l,i})  $.
And,
the channel  between the $k$-th DL user and the $l$-th UL user can be written as
\begin{align}
{h}_{du,k,l}^{*}(\mathbf{r}_{0,k},\mathbf{r}_{1,l})
 = (\mathbf{h}^{3}_{k,l}(\mathbf{r}_{0,k}))^{H}\mathbf{\Sigma}^{2}_{k,l}\mathbf{h}^{2}_{k,l}(\mathbf{r}_{1,l}),
\end{align}
where
the matrix
$\mathbf{\Sigma}^{2}_{k,l}$
is denoted as the response of all
transmit and receive paths 
from the $l$-th UL user to  the $k$-th DL user.

\underline{\textit{b) Radar Channel:}}
The steering vector of the $m$-th TBS towards the sensing target and/or clutters is represented by
\begin{align}
&\mathbf{g}^t_{m,j} = [ e^{j\frac{2\pi}{\lambda} (\mathbf{t}^{0}_{m,1})^T \mathbf{a}^4_{m,j} },
e^{j\frac{2\pi}{\lambda} (\mathbf{t}^{0}_{m,2})^T \mathbf{a}^4_{m,j} },\\
&\cdots,
e^{j\frac{2\pi}{\lambda} (\mathbf{t}^{0}_{m,N_t})^T \mathbf{a}^4_{m,j} }
],
\forall j \in \mathcal{J} \triangleq \{0\} \cup \mathcal{K}_t, \nonumber
\end{align}
where index $0$ represents the sensing target,
$\mathbf{a}^4_{m,j} \triangleq [\cos(\theta^{t,4}_{m,j})\sin(\phi^{t,4}_{m,j}); \sin(\theta^{t,4}_{m,j})]^T$,
and
$\theta^{t,4}_{m,j}$
and
$\phi^{t,4}_{m,j}$
denote
the
elevation and azimuth AoDs  between TBS $m$ and  target/clutters, respectively.
Similarly,
the steering vector of RBS $p$  towards the sensing target/clutters  can also be written as
\begin{align}
\mathbf{g}^r_{p,j} = [ e^{j\frac{2\pi}{\lambda} (\mathbf{t}^{1}_{p,1})^T \mathbf{a}^5_{p,j} },
e^{j\frac{2\pi}{\lambda} (\mathbf{t}^{1}_{p,2})^T \mathbf{a}^5_{p,j} },\\
\cdots,
e^{j\frac{2\pi}{\lambda} (\mathbf{t}^{1}_{p,N_r})^T \mathbf{a}^5_{p,j} }
],
\forall j \in \mathcal{J}, \nonumber
\end{align}
where
$\mathbf{a}^5_{p,j} \triangleq [\cos(\theta^{r,4}_{p,j})\sin(\phi^{r,4}_{p,j}); \sin(\theta^{r,4}_{p,j})]^T$,
and
$\theta^{r,4}_{p,j}$
and
$\phi^{r,4}_{p,j}$
denote respectively
the
elevation and azimuth AoAs  between RBS $p$ and target/clutters.
Therefore, the $m$-th TBS-target/clutters and the $p$-th RBS-target/clutters  channels are respectively given as
\begin{align}
\mathbf{g}_{t,m,j} = \beta^{t}_{m,j}\mathbf{g}^t_{m,j},\
\mathbf{g}_{r,p,j} = \beta^{r}_{p,j}\mathbf{g}^r_{p,j},
\end{align}
where $\{\beta^{t}_{m,j}\}$ and $\{\beta^{r}_{p,j}\}$ are complex fading
coefficients and are assumed to be known \cite{ref_ISAC_F_Liu}.

\subsection{Signal Model}

Denote
$\mathbf{s} \triangleq [s_1,\cdots, s_k, \cdots,  s_{K_d}]^T  \in \mathbb{C}^{K_d \times 1} $ and $\mathbf{s}_r \in \mathbb{C}^{N_t \times 1}$ as the communication symbol
transmitted to DL users and radar probing signal
from TBS, respectively.
For simplicity, it is assumed that
$\mathbb{E}\{\mathbf{s}\mathbf{s}^{H}\}=\mathbf{I}$,
$\mathbb{E}\{\mathbf{s}_r\mathbf{s}_r^{H}\}=\mathbf{I}$
and
$\mathbb{E}\{\mathbf{s}\mathbf{s}_r^{H}\}=\mathbf{0}$.
The transmit signal of the $m$-th TBS can be written as 
\footnote{
The joint communication and sensing signal
model effectively exploits the available DoF \cite{ref_DoF}.}

\begin{align}
\mathbf{x}_m = {\sum}_{k=1}^{K_d}\mathbf{w}_{m,k}s_k + \mathbf{W}^r_m\mathbf{s}_r,
\forall m \in \mathcal{M}_t,
\end{align}
where  $\mathbf{w}_{m,k} \in \mathbb{C}^{N_t \times 1}$ and $\mathbf{W}^r_m \in \mathbb{C}^{N_t \times N_t}$
represent the transmit beamformer of the $m$-th TBS
for
the $k$-th DL user's
communication signal and the probing
signal, respectively.

The $l$-th UL user's uplink signal can be written as
\begin{align}
x^u_l = \sqrt{q_l}s_l^u,\
\forall l \in \mathcal{K}_u,
\end{align}
where $s_l^u$ and $q_l$ denote  the information
symbol and transmission power of UL user $l$,
respectively,
and $s_l^u$ are mutually uncorrelated and
each has zero mean and unit variance.

The received signal at DL user $k$ can be written as
\begin{align}
&y_{D,k} = \underbrace{{\sum}_{m=1}^{M_t}\mathbf{h}_{d,m,k}^H\mathbf{x}_m}
\limits_{\textrm{DL users signals}}
 + \underbrace{{\sum}_{l=1}^{K_u} {h}_{du,k,l}x^u_l}
 \limits_{\textrm{UL users signals}}
  + n_{d,k}\\
&= {\sum}_{i=1}^{K_d}\hat{\mathbf{h}}_{d,k}^H\hat{\mathbf{w}}_{i}s_i +
\hat{\mathbf{h}}_{d,k}^H \hat{\mathbf{W}}^r\mathbf{s}^r
+ {\sum}_{l=1}^{K_u} {h}_{du,k,l}x^u_l + n_{d,k},\nonumber
\end{align}
where
$\hat{\mathbf{h}}_{d,k} \triangleq [\mathbf{h}_{d,1,k}^T,\mathbf{h}_{d,2,k}^T,\cdots,\mathbf{h}_{d,M_t,k}^T]^T \in
\mathbb{C}^{M_tN_t \times 1} $,
$ \hat{\mathbf{w}}_{k} \triangleq [\mathbf{w}_{1,k}^T,\mathbf{w}_{2,k}^T,\cdots,\mathbf{w}_{M_t,k}^T]^T\in
\mathbb{C}^{M_tN_t \times 1} $,
$ \bar{\mathbf{W}}_r \triangleq [(\mathbf{W}^r_1)^T,(\mathbf{W}^r_2)^T, \cdots,(\mathbf{W}^r_{M_t})^T]^T
\in \mathbb{C}^{M_tN_t \times N_t} $,
$\hat{\mathbf{W}}_r \triangleq \mathrm{blkdiag}(\bar{\mathbf{W}}_r) $
and $n_{d,k}\!\! \sim\!\! \mathcal{C\!N}(0,\sigma_{d,k}^2)$ is denoted as 
the complex additive white Gaussian noise (AWGN) at DL user $k$.

\subsection{Problem Formulation}

\underline{\textit{a) Communication Rate:}}
The received signal at the $p$-th RBS can be expressed as 
\begin{align}
\mathbf{y}_{r,p} &= \underbrace{{\sum}_{j=0}^{J}\alpha_{t,j}\mathbf{g}_{r,p,j}( {\sum}_{m=1}^{M_t} \mathbf{g}_{t,m,j}^H\mathbf{x}_m  )}
\limits_{\textrm{sensing echoes}}
\\
&+ \underbrace{{\sum}_{l=1}^{K_u} \mathbf{h}_{u,p,l}x^u_l}
\limits_{\textrm{UL users signals}}
 + \mathbf{n}_{r,p}, \nonumber
\end{align}
where
$\mathbf{n}_{r,p} \sim \mathcal{CN}(0,\sigma_{r}^2\mathbf{I})$ is the AWGN at the $p$-th  RBS,
$\alpha_{t,j}$
represents  the target RCS between TBS and RBS
and $\alpha_{t,j} \sim \mathcal{CN}(0,\sigma_{t,j}^2) $ \cite{ref_RCS}.

To obtain the information of UL users and target separately,
RBS $p$ will adopt $K_u + 1$ linear filters $\mathbf{u}_{p,i} \in \mathbb{C}^{M_r \times 1}$,
$\forall i \in \mathcal{J}_u \triangleq \{0\} \cup \mathcal{K}_u$,
to  post-process the received signal,
where $\mathbf{u}_{p,0}$ corresponds to the radar sensing filter.
Therefore,
the output information of the $i$-th filter at the $p$-th RBS can be written as
\begin{align}
\hat{y}_{r,p,i} =
\mathbf{u}_{p,i}^H\mathbf{y}_{r,p},
\forall i \in \mathcal{J}_u,
\end{align}
and then
 the information from  all $M_r$ RBS collected at the central controller is given by
 \begin{small}
\begin{align}
&\hat{y}^u_{r,i} =
{\sum}_{p=1}^{M_r}\hat{y}_{r,p,i}
={\sum}_{l=1}^{K_u}\hat{\mathbf{u}}_{i}^H\hat{\mathbf{h}}_{u,l}x^u_l\\
&+ {\sum}_{j=0}^{K_t}\alpha_{j}\hat{\mathbf{u}}_{i}^H\hat{\mathbf{g}}_{r,j}\hat{\mathbf{g}}_{t,j}^H({\sum}_{k=1}^{K_d} \hat{\mathbf{w}}_ks_k + \hat{\mathbf{W}}_r\hat{\mathbf{s}}^r ) + \hat{\mathbf{u}}_{i}^H\mathbf{n}_r
, \nonumber
\end{align}
\end{small}

\vspace{-0.3cm}

\noindent
where
\begin{small}
\begin{align}
&\hat{\mathbf{u}}_{i} \triangleq [\mathbf{u}_{1,i}^T,\mathbf{u}_{2,i}^T,\cdots,\mathbf{u}_{M_r,i}^T]^T \in \mathbb{C}^{M_rN_r \times 1},\\
&\hat{\mathbf{h}}_{u,l} \triangleq [\mathbf{h}_{u,1,l}^T,\mathbf{h}_{u,2,l}^T,\cdots,\mathbf{h}_{u,M_r,l}^T]^T\in \mathbb{C}^{M_rN_r \times 1},\nonumber\\
&\hat{\mathbf{g}}_{t,j} \triangleq [\mathbf{g}_{t,1,j}^T,\mathbf{g}_{t,2,j}^T,\cdots,\mathbf{g}_{t,M_t,j}^T]^T\in \mathbb{C}^{M_tN_t \times 1},\nonumber\\
&\hat{\mathbf{g}}_{r,j} \triangleq [\mathbf{g}_{r,1,j}^T,\mathbf{g}_{r,2,j}^T,\cdots,\mathbf{g}_{r,M_r,j}^T]^T\in \mathbb{C}^{M_rN_r \times 1},\nonumber\\
&\mathbf{n}_r \triangleq [\mathbf{n}_{r,1}^T,\mathbf{n}_{r,2}^T,\cdots,\mathbf{n}_{r,M_r}^T]^T\in \mathbb{C}^{M_rN_r \times 1}.\nonumber
\end{align}
\end{small}

The achievable rate of the $k$-th DL user and the $l$-th UL user can be  obtained as
\begin{align}
&\textrm{R}_{d,k} = \textrm{log}(1+\textrm{SINR}_{d,k}), \forall k \in \mathcal{K}_d,\\
&\textrm{R}_{u,l} = \textrm{log}(1+\textrm{SINR}_{u,l}), \forall l \in \mathcal{K}_u,
\end{align}
respectively,
where
\begin{small}
\begin{align}
&\textrm{SINR}_{d,k} = \\
&\frac{\vert\hat{\mathbf{h}}_{d,k}^H\hat{\mathbf{w}}_{k}\vert^2}
{  {\sum}_{i\neq k}^{K_d}\vert\hat{\mathbf{h}}_{d,k}^H\hat{\mathbf{w}}_{i}\vert^2
 + \Vert\hat{\mathbf{h}}_{d,k}^H \hat{\mathbf{W}}^r\Vert^2_2
 + {\sum}_{l=1}^{K_u} q_l\vert{h}_{du,k,l}\vert^{2} + \sigma_{d,k}^2
   }, \nonumber
\end{align}
\end{small}

\vspace{-0.1cm}
\noindent
and
the SINR formulation of UL user is written in (\ref{UL_SINR}),
as shown on the top of this page.

\begin{figure*}
\begin{small}
\begin{align}
\textrm{SINR}_{u,l}
=  \frac{ q_l \vert \hat{\mathbf{u}}_{l}^H\hat{\mathbf{h}}_{u,l} \vert^2}
{
\sum_{i\neq l}^{K_u}q_i\vert\hat{\mathbf{u}}_{l}^H\hat{\mathbf{h}}_{u,i}\vert^2
+ \sum_{j=0}^{K_t}\sum_{k=1}^{K_d}\sigma_{t,j}^2\vert\hat{\mathbf{u}}_{l}^H \hat{\mathbf{g}}_{r,j}\hat{\mathbf{g}}_{t,j}^H\hat{\mathbf{w}}_k \vert^2
+ \sum_{j=0}^{K_t}\sigma_{t,j}^2\Vert\hat{\mathbf{u}}_{l}^H \hat{\mathbf{g}}_{r,j}\hat{\mathbf{g}}_{t,j}^H\hat{\mathbf{W}}_r \Vert^2_2
+ \sigma_{r}^2\Vert\hat{\mathbf{u}}_{l}^H\Vert^2_2
} \label{UL_SINR}
\end{align}
\end{small}
\boldsymbol{\hrule}
\end{figure*}

\underline{\textit{b) Sensing SINR:}}
In the target detection problem, 
since the detection probability for a given false alarm probability monotonically increases with the sensing SINR, 
we select the sensing SINR as the metric \cite{ref_RadarSINR}$-$\cite{ref_RadarSINR_1}.
The radar SINR \cite{ref_RadarSINR}$-$\cite{ref_RadarSINR_1} for sensing can be derived as (\ref{Radar_SINR}),
as shown on the top of this page.

\begin{figure*}
\begin{small}
\begin{align}
\textrm{SINR}_{t} =
\frac{\sum_{k=1}^{K_d}\sigma_{t,0}^2\vert\hat{\mathbf{u}}_{0}^H \hat{\mathbf{g}}_{r,0}\hat{\mathbf{g}}_{t,0}^H\hat{\mathbf{w}}_k \vert^2
+
\sigma_{t,0}^2\Vert\hat{\mathbf{u}}_{0}^H \hat{\mathbf{g}}_{r,0}\hat{\mathbf{g}}_{t,0}^H\hat{\mathbf{W}}_r \Vert^2_F
}
{
\sum_{i=1}^{K_u}q_i\vert\hat{\mathbf{u}}_{0}^H\hat{\mathbf{h}}_{u,i}\vert^2
+
\sum_{j=1}^{K_t}\sum_{k=1}^{K_d}\sigma_{t,j}^2\vert\hat{\mathbf{u}}_{0}^H \hat{\mathbf{g}}_{r,j}\hat{\mathbf{g}}_{t,j}^H\hat{\mathbf{w}}_k \vert^2
+ \sum_{j=1}^{K_t}\sigma_{t,j}^2\Vert\hat{\mathbf{u}}_{0}^H \hat{\mathbf{g}}_{r,j}\hat{\mathbf{g}}_{t,j}^H\hat{\mathbf{W}}_r \Vert^2_2
+ \sigma_{r}^2\Vert\hat{\mathbf{u}}_{0}^H\Vert^2_2
}
\label{Radar_SINR}
\end{align}
\end{small}
\boldsymbol{\hrule}
\end{figure*}

\underline{\textit{c) Objective Function and Constraints:}}
Our goal is to maximize the sum-rate of all DL and UL users
via jointly optimizing the transmit beamformer $\{ \mathbf{w}_{m,k}, \mathbf{W}^r_m \}$,
the linear filters $\{\mathbf{u}_{p,i}\}$,
the DL users' uplink transmit power $\{q_l\}$,
the positions of all  MAs $\{\mathbf{t}^0_{m,n}, 
\mathbf{t}^1_{p,j}, \mathbf{r}_{0,k}, \mathbf{r}_{1,l}\}$.
The weighted sum-rate optimization problem  can be formulated as
\begin{subequations}
\begin{align}
\textrm{(P0)}:&
\mathop{\textrm{max}}
\limits_{\{\mathbf{w}_{m,k}, \mathbf{W}^r_m\},
\{\mathbf{u}_{p,i}\},
\{q_l\},
\atop
\{\mathbf{t}^0_{m,n}, 
\mathbf{t}^1_{p,j},
\mathbf{r}_{0,k}, 
\mathbf{r}_{1,l}\}
}\ \!\!\!
{\sum}_{k=1}^{K_d} \mu_{D,k} \textrm{R}_{d,k}\!\! +\!\!  {\sum}_{l=1}^{K_u} \mu_{U,l} \textrm{R}_{u,l}\\
\textrm{s.t.}\ &\textrm{SINR}_{t} \geq \Gamma_r,\\
&{\sum}_{k=1}^{K_d} \Vert\mathbf{w}_{m,k} \Vert_2^2
 + \Vert\mathbf{W}^r_{m} \Vert_F^2 \leq {P}^{BS}_{m}, \forall m \in \mathcal{M}_t,\\
& 0 \leq q_l  \leq P_{u,l}, \forall l \in \mathcal{K}_u,\\
& \Vert \mathbf{t}^0_{m,n} \!-\! \mathbf{t}^0_{m,i} \Vert_2\! \geq\! D_{t}, \forall n, i \!\in\! \mathcal{N}_t,
n \!\neq\! i,
\forall m \!\in\! \mathcal{M}_t,\\
& \Vert \mathbf{t}^1_{p,j}\! -\! \mathbf{t}^1_{p,s} \Vert_2 \!\geq\! D_{t}, \forall j, s \in \mathcal{N}_r,
j \neq s,
\forall p \in \mathcal{M}_r,\\
&\mathbf{t}^0_{m,n} \in \mathcal{C},\
\mathbf{t}^1_{p,j} \in \mathcal{C},\
\mathbf{r}_{0,k} \in \mathcal{C},\
\mathbf{r}_{1,l} \in \mathcal{C},
\end{align}
\end{subequations}
where
$\mu_{D,k} \in (0,1)$
and
$\mu_{U,l} \in (0,1)$,
and ${\sum}_{k=1}^{K_d} \mu_{D,k} + {\sum}_{l=1}^{K_u} \mu_{U,l} = 1$
\footnote{Since the weight factors are treated as constants in the optimization process,
we will omit them in the following.},
$\Gamma_r$ denotes the minimum target sensing SINR threshold,
$P^{BS}_{m}$ and $P_{u,l}$ represent the maximum transmission power of the $m$-th TBS and the $l$-th UL user,
respectively,
and
$D_{t}$ is the minimum distance between MAs of BSs to avoid the coupling effect.
Obviously,
due to the non-convex objective function and constraints,
the optimization problem (P0) is highly challenging.

\section{Algorithm Design}
\subsection{Problem Reformulation}
To make the problem (P0) more tractable,
we first employ
the WMMSE method \cite{ref_WMMSE}  to equivalently  transform
the objective functions
$\textrm{R}_{d,k}$
and
$ \textrm{R}_{u,l}$ into
(\ref{MSE_Rd})
and
(\ref{MSE_Ru}),
respectively,
at the top of the next page,
where $\{\omega_{d,k}\}$,
$\{\beta_{d,k}\}$,
$\{\omega_{u,l}\}$
and
$\{\beta_{u,l}\}$
are the introduced auxiliary variables.
Therefore,
the original problem (P0) can be reexpressed as
\begin{subequations}
\begin{align}
\textrm{(P1)}:&\mathop{\textrm{max}}
\limits_{\{\mathbf{w}_{m,k}, \mathbf{W}^r_m\},
\{\mathbf{u}_{p,i}\},
\{q_l\},
\atop
\{\mathbf{t}^0_{m,n}, \mathbf{t}^1_{p,j},
\mathbf{r}_{0,k}, \mathbf{r}_{1,l}\}
}\
\!\!{\sum}_{k=1}^{K_d} \tilde{\mathrm{R}}_{d,k}\! +\!  {\sum}_{l=1}^{K_u} \tilde{\mathrm{R}}_{u,l}\\
\textrm{s.t.}\ 
&\textrm{SINR}_{t} \geq \Gamma_r, \label{P1_c_1} \\
&{\sum}_{k=1}^{K_d} \Vert\mathbf{w}_{m,k} \Vert_2^2
 + \Vert\mathbf{W}^r_{m} \Vert_F^2 \leq \textrm{P}^{BS}_{m}, \forall m \in \mathcal{M}_t,\label{P1_c_2}\\
& 0 \leq q_l  \leq P_{u,l}, \forall l \in \mathcal{K}_u,\label{P1_c_3}\\
& \Vert \mathbf{t}^0_{m,n} \!-\! \mathbf{t}^0_{m,i} \Vert_2\! \geq\! D_{t}, \forall n, i \!\in\! \mathcal{N}_t,
n \!\neq\! i,
\forall m \!\in\! \mathcal{M}_t,\label{P1_c_4}\\
& \Vert \mathbf{t}^1_{p,j}\! -\! \mathbf{t}^1_{p,s} \Vert_2 \!\geq\! D_{t}, \forall j, s \in \mathcal{N}_r,
j \neq s,
\forall p \in \mathcal{M}_r,\label{P1_c_5}\\
&\mathbf{t}^0_{m,n} \in \mathcal{C},\
\mathbf{t}^1_{p,j} \in \mathcal{C},\
\mathbf{r}_{0,k} \in \mathcal{C},\
\mathbf{r}_{1,l} \in \mathcal{C},\label{P1_c_6}
\end{align}
\end{subequations}

Next,
we will adopt block coordinate ascent
(BCA) method \cite{ref_BCA} to efficiently solve problem (P1) by
updating different blocks of variables.

\begin{figure*}
\begin{small}
\begin{align}
\tilde{\mathrm{R}}_{d,k}
\triangleq
\textrm{log}(\omega_{d,k})
- \omega_{d,k}\big(  1 - 2\textrm{Re}\{ \beta_{d,k}^{\ast}\hat{\mathbf{h}}_{d,k}^H\hat{\mathbf{w}}_{k}  \}
+ \vert\beta_{d,k}\vert^2
(
{\sum}_{i = 1}^{K_d}\vert\hat{\mathbf{h}}_{d,k}^H\hat{\mathbf{w}}_{i}\vert^2
+ \Vert\hat{\mathbf{h}}_{d,k}^H \hat{\mathbf{W}}^r\Vert^2_2
+ {\sum}_{l=1}^{K_u} q_l\vert{h}_{du,k,l}\vert^{2} + \sigma_{d,k}^2
)
 \big) +1
\label{MSE_Rd}
\end{align}
\end{small}
\boldsymbol{\hrule}
\end{figure*}

\begin{figure*}
\begin{small}
\begin{align}
&\tilde{\mathrm{R}}_{u,l}
\triangleq
\textrm{log}(\omega_{u,l})
-
\omega_{u,l}
\bigg(  1 - 2\textrm{Re}\{ \beta_{u,l}^{\ast}\sqrt{q_l}\hat{\mathbf{u}}_{l}^H\hat{\mathbf{h}}_{u,l}  \} \label{MSE_Ru}\\
&
+
\vert\beta_{u,l}\vert^2
\big(
{\sum}_{i=1}^{K_u}q_i\vert\hat{\mathbf{u}}_{l}^H\hat{\mathbf{h}}_{u,i}\vert^2
+
{\sum}_{j=0}^{K_t}\sigma_{t,j}^2
({\sum}_{k=1}^{K_d}\vert\hat{\mathbf{u}}_{l}^H \hat{\mathbf{g}}_{r,j}\hat{\mathbf{g}}_{t,j}^H\hat{\mathbf{w}}_k \vert^2
+
\Vert\hat{\mathbf{u}}_{l}^H \hat{\mathbf{g}}_{r,j}\hat{\mathbf{g}}_{t,j}^H\hat{\mathbf{W}}_r \Vert^2_2
)
+
\sigma_{r}^2\Vert\hat{\mathbf{u}}_{l}^H\Vert^2_2
\big)
\bigg)
+1
\nonumber
\end{align}
\end{small}
\boldsymbol{\hrule}
\end{figure*}

\subsection{Optimizing auxiliary variables}

With other variables being fixed,
 the analytical  solutions of the auxiliary variables,
 i.e.,
$\{\omega_{d,k}\}$,
$\{\beta_{d,k}\}$,
$\{\omega_{u,l}\}$
and
$\{\beta_{u,l}\}$,
can be obtained from the derivation of WMMSE transformation proven in \cite{ref_WMMSE},
 as shown in (\ref{beta_dk})$-$(\ref{omega_ul}).

\begin{figure*}
\begin{small}
\begin{align}
&
\beta^{\star}_{d,k} =
{
\hat{\mathbf{h}}_{d,k}^H\hat{\mathbf{w}}_{k}
}
/({
{\sum}_{i = 1}^{K_d}\vert\hat{\mathbf{h}}_{d,k}^H\hat{\mathbf{w}}_{i}\vert^2
+
\Vert\hat{\mathbf{h}}_{d,k}^H \hat{\mathbf{W}}^r\Vert^2_2
+
{\sum}_{l=1}^{K_u} q_l\vert{h}_{du,k,l}\vert^{2} + \sigma_{d,k}^2
}),\label{beta_dk}\\
&
\omega^{\star}_{d,k}=
\big(
\vert 1 - \beta^{\ast}_{d,k} \hat{\mathbf{h}}_{d,k}^H\hat{\mathbf{w}}_{k} \vert^2
+
\vert \beta^{\ast}_{d,k} \vert^2
(
{\sum}_{i \neq k}^{K_d}\vert\hat{\mathbf{h}}_{d,k}^H\hat{\mathbf{w}}_{i}\vert^2
+
\Vert\hat{\mathbf{h}}_{d,k}^H \hat{\mathbf{W}}^r\Vert^2_2
+
{\sum}_{l=1}^{K_u} q_l\vert{h}_{du,k,l}\vert^{2} + \sigma_{d,k}^2
)
\big)^{-1},\label{omega_dk}\\
&\beta^{\star}_{u,l} =
{
\sqrt{q_l}\hat{\mathbf{u}}_{l}^H\hat{\mathbf{h}}_{u,l}
}
/({
{\sum}_{i=1}^{K_u}q_i\vert\hat{\mathbf{u}}_{l}^H\hat{\mathbf{h}}_{u,i}\vert^2
+
{\sum}_{j=0}^{K_t}\sigma_{t,j}^2
({\sum}_{k=1}^{K_d}\vert\hat{\mathbf{u}}_{l}^H \hat{\mathbf{g}}_{r,j}\hat{\mathbf{g}}_{t,j}^H\hat{\mathbf{w}}_k \vert^2
+
\Vert\hat{\mathbf{u}}_{l}^H \hat{\mathbf{g}}_{r,j}\hat{\mathbf{g}}_{t,j}^H\hat{\mathbf{W}}_r \Vert^2_2
)
+
\sigma_{r}^2\Vert\hat{\mathbf{u}}_{l}^H\Vert^2_2
}),\label{beta_ul}\\
&
\omega^{\star}_{u,l}
=
\bigg(
\vert
1 -  \beta^{\ast}_{u,l}\sqrt{q_l}\hat{\mathbf{u}}_{l}^H\hat{\mathbf{h}}_{u,l}
\vert^2
+
\vert\beta^{\ast}_{u,l}\vert^2
\big(
{\sum}_{j=0}^{K_t}\sigma_{t,j}^2
({\sum}_{k=1}^{K_d}\vert\hat{\mathbf{u}}_{l}^H \hat{\mathbf{g}}_{r,j}\hat{\mathbf{g}}_{t,j}^H\hat{\mathbf{w}}_k \vert^2
+
\Vert\hat{\mathbf{u}}_{l}^H \hat{\mathbf{g}}_{r,j}\hat{\mathbf{g}}_{t,j}^H\hat{\mathbf{W}}_r \Vert^2_2
)
+
\sigma_{r}^2\Vert\hat{\mathbf{u}}_{l}^H\Vert^2_2
\big)
\bigg)^{-1}\label{omega_ul}
\end{align}
\end{small}
\boldsymbol{\hrule}
\end{figure*}

\subsection{Updating The TBS Beamformer}
In this subsection,
we investigate the optimization of all $M_t$
TBS beamformer $\{\mathbf{w}_{m,k}, \mathbf{W}^r_m\}$
with  other variables being given.
Firstly,
the functions $\tilde{\mathrm{R}}_{d,k}$ and
$\tilde{\mathrm{R}}_{u,l}$,
which are respectively defined in (\ref{MSE_Rd}) and (\ref{MSE_Ru}),
are rewritten as
\begin{small}
\begin{align}
&-\!\sum_{k=1}^{K_d}\! \tilde{\mathrm{R}}_{\!d\!,k}
\!\!= \!\!\! \sum_{k=1}^{K_d}\! (\!\hat{\mathbf{w}}^H_{k}\!\!\mathbf{A}_1\!\hat{\mathbf{w}}_{k}
\!\!- \!\! 2\textrm{Re}\{\! \mathbf{a}^H_{1,k}\hat{\mathbf{w}}_{k}\! \}\!)
\!\!+\!\!\! \sum_{i=1}^{M_t}\! (\!\bar{\mathbf{w}}^r_{i}\!)^{\!H\!}\!\mathbf{A}_{\!2\!,i}\!\bar{\mathbf{w}}^r_{i}
\!\! -\!\! c_{\!1\!,\!1},\label{TBS_Rd_1}
  \\
&-\sum_{l=1}^{K_u} \tilde{\mathrm{R}}_{u,l}
\!=\! \sum_{k=1}^{K_d}\! \hat{\mathbf{w}}^H_{k}\mathbf{A}_3\hat{\mathbf{w}}_{k}
\!+\! \sum_{i=1}^{M_t}\! (\bar{\mathbf{w}}^r_{i})^H\mathbf{A}_{4,i}\bar{\mathbf{w}}^r_{i}
\!-\! c_{1,2}, \label{TBS_Ru_1}
\end{align}
\end{small}

\noindent
where $c_{1,1}$ and $c_{1,2}$ are constant terms
and  the newly introduced parameters in (\ref{TBS_Rd_1}) and (\ref{TBS_Ru_1}) being defined as follows
\begin{small}
\begin{align}
&\mathbf{a}_{1,k} \triangleq \omega_{d,k}\beta_{d,k}\hat{\mathbf{h}}_{d,k},
\bar{\mathbf{w}}^r_{i} \triangleq \textrm{vec}({\mathbf{W}^r_i}),\\
&\mathbf{A}_1 \triangleq {\sum}_{k=1}^{K_d}\omega_{d,k}\vert\beta_{d,k}\vert^2\hat{\mathbf{h}}_{d,k}\hat{\mathbf{h}}_{d,k}^H,
\mathbf{A}_{2,i} \triangleq \mathbf{I}\otimes {\mathbf{h}}_{d,i,k}{\mathbf{h}}_{d,i,k}^H,\nonumber\\
&\mathbf{A}_{3} \triangleq
{\sum}_{l=1}^{K_u}
{\sum}_{j=0}^{K_t}
\beta_{u,l}\vert\beta_{u,l}\vert^2
\sigma_{t,j}^2
\hat{\mathbf{g}}_{t,j}\hat{\mathbf{g}}_{r,j}^H \hat{\mathbf{u}}_{l}
 \hat{\mathbf{u}}_{l}^H\hat{\mathbf{g}}_{r,j}\hat{\mathbf{g}}_{t,j}^H,\nonumber\\
& \mathbf{A}_{4,i} \!\triangleq \!\!
 {\sum}_{l=1}^{K_u}
  {\sum}_{j=0}^{K_t}
 \beta_{u,l}\vert\beta_{u,l}\vert^2
\big(\sigma_{t,j}^2 \mathbf{I}\! \otimes\!
 (  {\mathbf{g}}_{t,i,j} \hat{\mathbf{g}}_{r,j}^H \hat{\mathbf{u}}_{l}
 \hat{\mathbf{u}}_{l}^H \hat{\mathbf{g}}_{r,j} {\mathbf{g}}_{t,i,j}^H  ) \big ).\nonumber
\end{align}
\end{small}

Similarly,
the constraint (\ref{P1_c_1}) can be rewritten as
\begin{small}
\begin{align}
&\textrm{SINR}_{t} \geq \Gamma_r
\Longleftrightarrow
{\sum}_{k=1}^{K_d}\hat{\mathbf{w}}_k^H \mathbf{A}_{11}\hat{\mathbf{w}}_k
+
{\sum}_{i=1}^{M_t} (\bar{\mathbf{w}}^r_{i})^H\mathbf{A}_{12,i}\bar{\mathbf{w}}^r_{i}
\nonumber\\
&+ c_{1,3}- {\sum}_{k=1}^{K_d}\hat{\mathbf{w}}_k^H \mathbf{A}_{9}\hat{\mathbf{w}}_k
-{\sum}_{i=1}^{M_t} (\bar{\mathbf{w}}^r_{i})^H\mathbf{A}_{10,i}\bar{\mathbf{w}}^r_{i}
\leq 0,\label{TBS_SINR_1}
\end{align}
\end{small}

\noindent
where $c_{1,3}$ is constant  and the above newly
introduced coefficients are defined as
\begin{small}
\begin{align}
&\mathbf{A}_{9} \triangleq \sigma^2_{t,0} \hat{\mathbf{g}}_{t,0}\hat{\mathbf{g}}_{r,0}^H\hat{\mathbf{u}}_{0}
  \hat{\mathbf{u}}_{0}^H \hat{\mathbf{g}}_{r,0}\hat{\mathbf{g}}_{t,0}^H/\Gamma_r,\\
&\mathbf{A}_{10,i} \triangleq \mathbf{I} \otimes \sigma^2_{t,0}   {\mathbf{g}}_{t,i,0}\hat{\mathbf{g}}_{r,0}^H\hat{\mathbf{u}}_{0}
  \hat{\mathbf{u}}_{0}^H \hat{\mathbf{g}}_{r,0}{\mathbf{g}}_{t,i,0}^H/\Gamma_r,\nonumber
 \\
&\mathbf{A}_{11} \triangleq {\sum}_{j=1}^{K_t}\sigma^2_{t,j} \hat{\mathbf{g}}_{t,j}\hat{\mathbf{g}}_{r,j}^H\hat{\mathbf{u}}_{0}
  \hat{\mathbf{u}}_{0}^H \hat{\mathbf{g}}_{r,j}\hat{\mathbf{g}}_{t,j}^H,\nonumber\\
&  \mathbf{A}_{12,i} \triangleq {\sum}_{j=1}^{K_t}\mathbf{I} \otimes \sigma^2_{t,j}   {\mathbf{g}}_{t,i,j}\hat{\mathbf{g}}_{r,j}^H\hat{\mathbf{u}}_{0}
  \hat{\mathbf{u}}_{0}^H \hat{\mathbf{g}}_{r,j}{\mathbf{g}}_{t,i,j}^H.\nonumber
\end{align}
\end{small}

Based on the above equivalent  transformation,
the  problem of updating $\{\hat{\mathbf{w}}_{k}, \bar{\mathbf{w}}^r_i\}$
can be formulated as
\begin{subequations}
\begin{align}
\textrm{(P2)}:&\mathop{\textrm{min}}
\limits_{\{\hat{\mathbf{w}}_{k}, \bar{\mathbf{w}}_{r,i}\}
}\
 {\sum}_{k=1}^{K_d} (\hat{\mathbf{w}}^H_{k}\mathbf{A}_5\hat{\mathbf{w}}_{k}
-  2\textrm{Re}\{ \mathbf{a}^H_{1,k}\hat{\mathbf{w}}_{k} \})\\
&
+ {\sum}_{i=1}^{M_t} (\bar{\mathbf{w}}^r_{i})^H\mathbf{A}_{6,i}\bar{\mathbf{w}}^r_{i}
- \hat{c}_{1,2}\nonumber\\
\textrm{s.t.}\ &{\sum}_{k=1}^{K_d}\hat{\mathbf{w}}_k^H \mathbf{A}_{11}\hat{\mathbf{w}}_k
\!+\!\! {\sum}_{i=1}^{M_t} (\bar{\mathbf{w}}^r_{i})^H\!\mathbf{A}_{12,i}\bar{\mathbf{w}}^r_{i}
\!+\! c_{1,3} \label{P2_c_1} \\
- &{\sum}_{k=1}^{K_d}\hat{\mathbf{w}}_k^H \mathbf{A}_{9}\hat{\mathbf{w}}_k
-{\sum}_{i=1}^{M_t} (\bar{\mathbf{w}}^r_{i})^H\mathbf{A}_{10,i}\bar{\mathbf{w}}^r_{i}
\leq 0, \nonumber\\
&{\sum}_{k=1}^{K_d} \Vert\hat{\mathbf{w}}_{k}\mathbf{R}_i \Vert_2^2
 + \Vert\bar{\mathbf{w}}^r_{i} \Vert_2^2 \leq \textrm{P}^{BS}_{i}, \forall i \in \mathcal{M}_t,\label{P2_c_2}
\end{align}
\end{subequations}
where
$\mathbf{A}_5 \triangleq \mathbf{A}_1 + \mathbf{A}_3$,
$\mathbf{A}_{6,i} \triangleq \mathbf{A}_{2,i} + \mathbf{A}_{4,i}$,
$\hat{c}_{1,2} \triangleq c_{1,2} + c_{1,1}$,
$\mathbf{R}_i \triangleq \textrm{blkdiag}( [\mathbf{0},\cdots,\mathbf{0}_{i-1,i-1}, \mathbf{I}_{i,i}, \mathbf{0}_{i+1,i+1},\cdots,\mathbf{0}  ]  )  \in \mathbb{C}^{M_tN_t\times M_tN_t} $,
$\mathbf{0} \in \mathbb{C}^{N_t\times N_t}$,
$\mathbf{I} \in \mathbb{C}^{N_t\times N_t}$.

It is obvious that the non-convex constraint (\ref{P2_c_1}) makes the
problem (P2) intractable.
Inspired by the MM framework \cite{ref_MM},
we respectively linearize the quadratic terms
$\hat{\mathbf{w}}_k^H \mathbf{A}_{9}\hat{\mathbf{w}}_k$
and
$\bar{\mathbf{w}}^H_{r,i}\mathbf{A}_{10,i}\bar{\mathbf{w}}_{r,i}$
 as follows
 \begin{small}
\begin{align}
&\hat{\mathbf{w}}_k^H \mathbf{A}_{9}\hat{\mathbf{w}}_k
\geq
2\textrm{Re}\{ \hat{\mathbf{w}}_{k,0}^H \mathbf{A}_{9}(\hat{\mathbf{w}}_k - \hat{\mathbf{w}}_{k,0})  \}
+ \hat{\mathbf{w}}_{k,0}^H \mathbf{A}_{9}\hat{\mathbf{w}}_{k,0}\nonumber\\
&= 2\textrm{Re}\{ \hat{\mathbf{w}}_{k,0}^H \mathbf{A}_{9}\hat{\mathbf{w}}_k   \}
- (\hat{\mathbf{w}}_{k,0}^H \mathbf{A}_{9}\hat{\mathbf{w}}_{k,0})^{\ast},\label{P2_c_2_1}\\
&(\bar{\mathbf{w}}^r_{i})^H\mathbf{A}_{10,i}\bar{\mathbf{w}}^r_{i}
\!\!\geq \!\!
2\textrm{Re}\{\! (\bar{\mathbf{w}}^r_{i,0})^H\!\mathbf{A}_{10,i}(\! \bar{\mathbf{w}}^r_{i}\!\! -\!\! \bar{\mathbf{w}}^r_{i,0}  \!)  \!\}
\!\!+\!\! (\bar{\mathbf{w}}^r_{i,0})^H\!\mathbf{A}_{10,i}\bar{\mathbf{w}}^r_{i,0}\nonumber\\
&= 2\textrm{Re}\{ (\bar{\mathbf{w}}^r_{i,0})^H\mathbf{A}_{10,i} \bar{\mathbf{w}}^r_{i}    \}
- ((\bar{\mathbf{w}}^r_{i,0})^H\mathbf{A}_{10,i}\bar{\mathbf{w}}^r_{i,0})^{\ast},\label{P2_c_2_2}
\end{align}
\end{small}

\vspace{-0.2cm}
\noindent
where
$\hat{\mathbf{w}}_{k,0}$
and
$\bar{\mathbf{w}}^r_{i,0}$ are feasible solutions obtained
in the last iteration.
Therefore,
by replacing the non-convex terms of constraint (\ref{P2_c_1})
by
(\ref{P2_c_2_1})
and
(\ref{P2_c_2_2}),
the problem (P2) can be rewritten as
\begin{subequations}
\begin{align}
\textrm{(P3)}:&\mathop{\textrm{min}}
\limits_{\{\hat{\mathbf{w}}_{k}, \bar{\mathbf{w}}_{r,i}\}
}\
 {\sum}_{k=1}^{K_d} (\hat{\mathbf{w}}^H_{k}\mathbf{A}_5\hat{\mathbf{w}}_{k}
-  2\textrm{Re}\{ \mathbf{a}^H_{1,k}\hat{\mathbf{w}}_{k} \})\\
&
+ {\sum}_{i=1}^{M_t} (\bar{\mathbf{w}}^r_{i})^H\mathbf{A}_{6,i}\bar{\mathbf{w}}^r_{i}
- \hat{c}_{1,2}\nonumber\\
\textrm{s.t.}\ &{\sum}_{k=1}^{K_d}\hat{\mathbf{w}}_k^H \mathbf{A}_{11}\hat{\mathbf{w}}_k
\!+\!\! {\sum}_{i=1}^{M_t} (\bar{\mathbf{w}}^r_{i})^H\!\mathbf{A}_{12,i}\bar{\mathbf{w}}^r_{i}
\!+\! \hat{c}_{1,3}\label{P2_c_1}\\
- &\!{\sum}_{k=1}^{K_d}2\textrm{Re}\{ \mathbf{a}_{2,k}^H \hat{\mathbf{w}}_k   \}
\!-\!{\sum}_{i=1}^{M_t} 2\textrm{Re}\{ \mathbf{a}_{3,i}^H \bar{\mathbf{w}}^r_{i}    \}
\!\leq\! 0, \nonumber\\
&{\sum}_{k=1}^{K_d} \Vert\hat{\mathbf{w}}_{k}\mathbf{R}_i \Vert_2^2
 + \Vert\bar{\mathbf{w}}^r_{i} \Vert_2^2 \leq \textrm{P}^{BS}_{i}, \forall i \in \mathcal{M}_t,\label{P2_c_2}
\end{align}
\end{subequations}
where
$
\mathbf{a}_{2,k} \triangleq\mathbf{A}_{9}^H \hat{\mathbf{w}}_{k,0}
$,
$
\mathbf{a}_{3,i}\triangleq \mathbf{A}_{10,i}^H \bar{\mathbf{w}}^r_{i,0}
$,
$\hat{c}_{1,3} \triangleq
\! - \!
{\sum}_{k=1}^{K_d}\!  (\hat{\mathbf{w}}_{k,0}^H \mathbf{A}_{9}\hat{\mathbf{w}}_{k,0})^{\ast}
\!-\! {\sum}_{i=1}^{M_t}\!((\bar{\mathbf{w}}^r_{i,0})^H\mathbf{A}_{10,i}\bar{\mathbf{w}}^r_{i,0})^{\ast}
+{c}_{1,3}
$.
The problem (P3)
is a typical second order cone program (SOCP) and can be efficiently
solved by standard numerical solvers, such as CVX \cite{ref_CVX}.

\subsection{Optimizing The UL User Transmission Power}

In this subsection,
we present the method for optimizing the UL users' transmission power allocation.
With other variables being fixed,
the optimization problem of updating $ \{q_l\} $ is formulated as
\begin{subequations}
\begin{align}
\textrm{(P4)}:&\mathop{\textrm{min}}
\limits_{
\{q_l\}
}\
{\sum}_{l=1}^{K_u}({b}_{1,l}q_l
-  b_{2,l}\sqrt{q_l})
+ c_{2,1}
\\
\textrm{s.t.}\ &
{\sum}_{l=1}^{K_u} b_{3,l}q_l - c_{2,2} \leq 0,\\
& 0 \leq  q_l \leq P_{u,l}, \forall l \in \mathcal{K}_u,
\end{align}
\end{subequations}
where $c_{2,1}$ and $c_{2,2}$ are constant terms,
and the new parameters in problem (P4) are defined as
\begin{align}
& b_{2,l} \triangleq
2\textrm{Re}\{ \omega_{u,l}\beta_{u,l}^{\ast}\hat{\mathbf{u}}_{i}^H\hat{\mathbf{h}}_{u,l} \},
b_{3,l} \triangleq
\vert \hat{\mathbf{u}}_{0}^H\hat{\mathbf{h}}_{u,l} \vert,\\
&b_{1,l}\! \triangleq \!\!
{\sum}_{i=1}^{K_u} \omega_{u,l}\vert \beta_{u,l} \vert^2\vert \hat{\mathbf{u}}_{i}^H\hat{\mathbf{h}}_{u,l} \vert^2
\!\!+\!\!\!
{\sum}_{k=1}^{K_d}  \omega_{d,k}\vert\beta_{d,k}\vert^2\vert{h}_{du,k,l}\vert^{2}
.\nonumber
\end{align}

The problem (P4) is an SOCP and can be directly solved by CVX.

\subsection{Optimizing The UL User Receiver Filter}
With the given other variables,
the UL user receive filters $\{\hat{\mathbf{u}}_l\}$ optimization reduces to solving the following problem
\begin{subequations}
\begin{align}
\textrm{(P5)}:&\mathop{\textrm{min}}
\limits_{
\{\hat{\mathbf{u}}_l\}
}\
{\sum}_{l=1}^{K_u}\{ \hat{\mathbf{u}}_l^H \mathbf{D}_l \hat{\mathbf{u}}_l - 2\textrm{Re}\{ \hat{\mathbf{u}}_l^H\mathbf{d}_{1,l} \} \}
+ c_{3,1}\label{P5_1}
\end{align}
\end{subequations}
where the above newly introduced coefficients are given as
\vspace{-0.2cm}
\begin{small}
\begin{align}
&\mathbf{d}_{1,l} \triangleq
\omega_{u,l}\beta_{u,l}^{\ast}\sqrt{q_l}\hat{\mathbf{h}}_{u,l},
\mathbf{D}_{1} \triangleq
\omega_{u,l}\vert\beta_{u,l}\vert^2
\big(
{\sum}_{i=1}^{K_u}q_i\hat{\mathbf{h}}_{u,i}\hat{\mathbf{h}}_{u,i}^H\\
&\!+\!\!\!\!
{\sum}_{j\!=\!0}^{K_t}\!\!\sigma_{t,j}^2\!
(\!{\sum}_{k\!=\!1}^{K_d} \!\hat{\mathbf{g}}_{\!r,j}\!\hat{\mathbf{g}}_{\!t,j}^H\!\hat{\mathbf{w}}_k\!
\hat{\mathbf{w}}_k^H\!\hat{\mathbf{g}}_{\!t,j}\hat{\mathbf{g}}_{\!r,j}^H
\!\!+\!\!
 \hat{\mathbf{g}}_{\!r,j}\!\hat{\mathbf{g}}_{\!t,j}^H\!\hat{\mathbf{W}}_{\!r}\!
 \hat{\mathbf{W}}_{\!r}^H\!\hat{\mathbf{g}}_{\!t,j}\hat{\mathbf{g}}_{\!r,j}^H
\!\!+\!\!
\sigma_{r}^2\mathbf{I}\!
\big).\nonumber
\end{align}
\end{small}

Since the variables $\{\hat{\mathbf{u}}_l\}$
are not intricately coupled in (\ref{P5_1}),
the problem (P5)
can be decomposed
into $K_u$ independent subproblems,
which each subproblem can be written as
\vspace{-0.2cm}
\begin{subequations}
\begin{align}
\textrm{(P6$_l$)}:&\mathop{\textrm{min}}
\limits_{
\hat{\mathbf{u}}_l
}\
 \hat{\mathbf{u}}_l^H \mathbf{D}_l \hat{\mathbf{u}}_l - 2\textrm{Re}\{ \hat{\mathbf{u}}_l^H\mathbf{d}_{1,l} \}
\end{align}
\end{subequations}

The subproblem (P6$_l$)
is a typical unconstrained
convex quadratic problem.
And then,
by setting problem (P6$_l$)'s derivative to zero,
the optimal solution can be
obtained as
\vspace{-0.2cm}
\begin{align}
\hat{\mathbf{u}}_l^{\star} = \mathbf{D}_l^{-1}\mathbf{d}_{1,l}, \forall l \in \mathcal{K}_u. \label{U_l}
\end{align}

\subsection{Optimizing The Sensing  Filter}
Given fixed other variables,
the problem with respect to (w.r.t.) $\hat{\mathbf{u}}_0$
is  reduced to a typical feasibility check problem,
i.e.,
Phase-I problem \cite{ref_Convex Optimization},
which is given as
\vspace{-0.2cm}
\begin{subequations}
\begin{align}
\textrm{(P7)}:&\mathop{\textrm{Find}}
\limits_{
\hat{\mathbf{u}}_0
}\
 \hat{\mathbf{u}}_0\\
& \textrm{s.t.}\ \frac{\hat{\mathbf{u}}_0^H \mathbf{E}_1  \hat{\mathbf{u}}_0}
 {\hat{\mathbf{u}}_0^H \mathbf{E}_2  \hat{\mathbf{u}}_0} \geq \Gamma, \label{P7_1}
\end{align}
\end{subequations}
where the   newly defined  coefficients are given as
\begin{small}
\begin{align}
&\mathbf{E}_1 \!\!\triangleq\!\! \sigma_{t,0}^2( \!
{\sum}_{k=1}^{K_d}\!\hat{\mathbf{g}}_{r,0}\hat{\mathbf{g}}_{t,0}^H\!\hat{\mathbf{w}}_k  \!
\hat{\mathbf{w}}_k^H\!\hat{\mathbf{g}}_{t,0}\hat{\mathbf{g}}_{r,0}^H
\!\!+\!\!\hat{\mathbf{g}}_{r,0}\hat{\mathbf{g}}_{t,0}^H\hat{\mathbf{W}}_r\!
\hat{\mathbf{W}}_r^H\hat{\mathbf{g}}_{t,0}\hat{\mathbf{g}}_{r,0}^H
),\nonumber
\\
&\mathbf{E}_2 \!\triangleq\!\!
{\sum}_{i=1}^{K_u}q_i\hat{\mathbf{h}}_{u,i}\hat{\mathbf{h}}_{u,i}^H
\!\!+\!\! {\sum}_{j=1}^{K_t}\!\sigma_{t,j}^2\!(\!
{\sum}_{k=1}^{K_d}\hat{\mathbf{g}}_{r,j}\hat{\mathbf{g}}_{t,j}^H\hat{\mathbf{w}}_k
\hat{\mathbf{w}}_k^H\hat{\mathbf{g}}_{t,j}\hat{\mathbf{g}}_{r,j}^H\non\\
&+
\hat{\mathbf{g}}_{r,j}\hat{\mathbf{g}}_{t,j}^H\hat{\mathbf{W}}_r\!
\hat{\mathbf{W}}_r^H\hat{\mathbf{g}}_{t,j}\hat{\mathbf{g}}_{r,j}^H
)
+ \sigma_{r}^2\mathbf{I}.
\end{align}
\end{small}

Since the variable $\hat{\mathbf{u}}_0$  only appears in the constraint (\ref{P7_1}),
we can turn to consider  a sensing SINR  maximization problem,
which is formulated as
\vspace{-0.2cm}
\begin{subequations}
\begin{align}
\textrm{(P8)}:&\mathop{\textrm{max}}
\limits_{
\hat{\mathbf{u}}_0
}\
\frac{\hat{\mathbf{u}}_0^H \mathbf{E}_1  \hat{\mathbf{u}}_0}
 {\hat{\mathbf{u}}_0^H \mathbf{E}_2  \hat{\mathbf{u}}_0}
\end{align}
\end{subequations}

The above problem (P8) is a typical Rayleigh quotient maximization.
By applying the Rayleigh-Ritz theorem,
the optimal $\hat{\mathbf{u}}_0$ can be obtained by
aligning with the eigenvector associated with the largest eigenvalue of
the matrix $\mathbf{E}_2^{-1}\mathbf{E}_1$ \cite{ref_Rayleigh}.

\subsection{Optimizing The TBS MAs' Position}
In this subsection,
when other variables are given,
we  discuss  updating the position of $m$-th TBS' $n$-th MA,
i.e., $\mathbf{t}^0_{m,n}$
\footnote{To enhance readability, we have omitted the subscripts, i.e., $m$ and $n$, from the variables and/or coefficients. 
Specifically, we rewrite:
$\mathbf{t}^0_{m,n}$ as $\mathbf{t}^0$,
$\mathbf{h}^{0}_{m,k,n}$ as $\mathbf{h}^{0}_{k}$, and
$\mathbf{g}_{t,m,j}[n]$ as ${g}_{t,j}$.}.
Firstly,
the function $\tilde{\mathrm{R}}_{d,k}$,
 which is defined in (\ref{MSE_Rd}),
can be rewritten as
\vspace{-0.2cm}
\begin{small}
\begin{align}
&\!-\!\tilde{\mathrm{R}}_{d,k}
\!\!=\!\!
(\mathbf{h}^{0}_{k})^H
\mathbf{{F}}^{1}_{k}
\mathbf{h}^{0}_{k}
\!\!+\!\!
\textrm{Re}\{ (\mathbf{f}^1_{k})^H\mathbf{h}^{0}_{k} \}
\!\!- \!\!
\textrm{Re}\{ (\mathbf{f}^2_{k})^H\mathbf{h}^{0}_{k} \}
\!\!+\!\! c_{4,1,k},
\end{align}
\end{small}

\vspace{-0.2cm}
\noindent
where $c_{4,1,k}$ is a constant
and the above new coefficients are given as follows
\vspace{-0.2cm}
\begin{small}
\begin{align}
&\mathbf{\bar{F}}^{1}_{m,n,k,i} \triangleq
( \mathbf{w}_{m,i}\mathbf{w}_{n,i}^H )^T
\otimes
( (\mathbf{\Sigma}^0_{n,k})^H \mathbf{h}^{0,0}_{n,k} (\mathbf{h}^{0,0}_{m,k})^H \mathbf{\Sigma}^0_{n,k} ),\\
&\tilde{\mathbf{h}}^{0}_{m,k} \triangleq \textrm{vec}({\mathbf{H}^{0}_{m,k}}),
\mathbf{f}_{m,k,j}^{15} \triangleq {\sum}_{j\neq m }^{M_t}(\mathbf{\bar{F}}^{1}_{m,j,k,i})^H \tilde{\mathbf{h}}^{0}_{j,k},\nonumber\\
&\mathbf{F}^{2}_{m,k} \triangleq (\bar{\mathbf{w}}^r_{m}(\bar{\mathbf{w}}^r_{m})^H )^T
\otimes
( (\mathbf{\Sigma}^0_{m,k})^H \mathbf{h}^{0,0}_{m,k} (\mathbf{h}^{0,0}_{m,k})^H \mathbf{\Sigma}^0_{m,k} ),\nonumber\\
& \mathbf{f}^{16}_{m,k}\!\! \triangleq\!\! 2 \omega_{d,k} \beta_{d,k} ( \mathbf{w}_{m,k}^T \!\!\otimes\!\! ( \! (\mathbf{h}^{0,0}_{m,k})^H \!\mathbf{\Sigma}^0_{m,k}) \!)^H,
 \mathbf{F}^{4}_{m,k} \!\!\triangleq\! \omega_{d,k}\!\vert\!\beta_{d,k}\!\vert^2 \!\mathbf{F}^{2}_{m,k},\nonumber\\
&\mathbf{F}^{3}_{m,k} \!\!\triangleq\!\!\! {\sum}_{i=1}^{K_d}\omega_{d,k}\vert\!\beta_{d,k}\!\vert^2\mathbf{\bar{F}}^{1}_{m,m,k,i},
 \mathbf{f}^{17}_{m,k}\!\!\triangleq\!\!\!  {\sum}_{i=1}^{K_d}\!2\omega_{d,k}\vert\!\beta_{d,k}\!\vert^2\!\mathbf{f}_{m,k,i}^{15},\nonumber\\
& \mathbf{F}^{5}_{m,k}  \triangleq \mathbf{F}^{3}_{m,k} + \mathbf{F}^{4}_{m,k},
 \mathbf{f}^{2}_{k} \triangleq \mathbf{f}^{16}_{m,k}[ (n-1)L^{t,0}_{m,k}+1: nL^{t,0}_{m,k} ],\nonumber\\
& {\mathbf{F}}^{1}_{m,k,i,j}\! \triangleq \!\mathbf{F}^{5}_{m,k}[ (i-1)L^{t,0}_{m,k}\!+\!1\!:\! iL^{t,0}_{m,k}, (j-1)L^{t,0}_{m,k}\!+\!1\!: \! jL^{t,0}_{m,k} ],\nonumber\\
& {\mathbf{F}}^{1}_{k}\! \triangleq \!\mathbf{F}^{5}_{m,k}[ (n-1)L^{t,0}_{m,k}\!+\!1\!:\! nL^{t,0}_{m,k}, (n-1)L^{t,0}_{m,k}\!+\!1\!: \! nL^{t,0}_{m,k} ],\nonumber\\
&\mathbf{f}^{18}_{m,k,n} \triangleq  2( {\sum}_{i\neq n}^{N_t}  ({\mathbf{\bar{F}}}^{1}_{m,k,i,n})^H \mathbf{h}^{0}_{m,k,i} ),\nonumber\\
& \mathbf{f}^{1}_{k} \triangleq  \mathbf{f}^{18}_{m,k,n} +  \mathbf{f}^{17}_{m,k}[ (n-1)L^{t,0}_{m,k}+1: nL^{t,0}_{m,k} ].\nonumber
\end{align}
\end{small}

Furthermore,
the function $\tilde{\mathrm{R}}_{u,l}$ defined in (\ref{MSE_Ru})
is also rewritten as follows

\vspace{-0.2cm}
\begin{small}
\begin{align}
&{\sum}_{l=1}^{K_u}\tilde{\mathrm{R}}_{u,l}\!
=
 {\sum}_{j=0}^{K_t} -\textrm{Re}\{ (f^{3}_{j})^{\ast}  {g}_{t,j}  \} + c_{4,2},\label{t0_R_l_transformation}
\end{align}
\end{small}

\vspace{-0.2cm}
\noindent
where $c_{4,2}$ is a constant term
and the new parameters in (\ref{t0_R_l_transformation}) are defined as follows
\vspace{-0.2cm}
\begin{small}
\begin{align}
& \mathbf{F}^7_{l,j} \triangleq ( \hat{\mathbf{g}}_{r,j}^H\hat{\mathbf{u}}_{l}\hat{\mathbf{u}}_{l}^H\hat{\mathbf{g}}_{r,j}  )^T
\otimes
(\hat{\mathbf{W}}_r \hat{\mathbf{W}}_r^H),\\
& \mathbf{F}^6_{l,j,k} \triangleq ( \hat{\mathbf{g}}_{r,j}^H\hat{\mathbf{u}}_{l}\hat{\mathbf{u}}_{l}^H\hat{\mathbf{g}}_{r,j}  )^T
\otimes
(\hat{\mathbf{w}}_k \hat{\mathbf{w}}_k^H),\nonumber\\
& \mathbf{F}^8_{l,j} \triangleq {\sum}_{k=1}^{K_d} \mathbf{F}^6_{l,j,k} + \mathbf{F}^7_{l,j},
 \mathbf{F}^9_{j} \triangleq {\sum}_{l=1}^{K_u} \omega_{u,l}\vert\beta_{u,l}\vert^2\sigma_{t,j}^2\mathbf{F}^8_{l,j},\nonumber\\
& \bar{\mathbf{F}}^9_{j,m,n} \triangleq \mathbf{F}^9_{j}[ (m-1)N_t+1:mN_t , (n-1)N_t+1:nN_t ],\nonumber\\
& f^{20}_{m,j,n}\triangleq {\sum}_{i\neq n}^{N_t} (\bar{\mathbf{F}}^9_{j,m,m}[i,n])^{\ast} \mathbf{g}_{t,m,j}[i],\nonumber\\
&\mathbf{f}^{19}_{m,j} \triangleq {\sum}_{i\neq m}^{M_t}  2(\bar{\mathbf{F}}^9_{j,m,n})^H\mathbf{g}_{t,i,j},
f^{3}_{j} \triangleq f^{20}_{m,j,n} +  \mathbf{f}^{19}_{m,j}[n].\nonumber
\end{align}
\end{small}

Similarly,
the sensing SINR constraint (\ref{P1_c_1}) is rewritten as
\begin{small}
\begin{align}
&\textrm{SINR}_{t} \geq \Gamma_r
\Longleftrightarrow\\
&{\sum}_{i=1}^{K_t}\textrm{Re}\{ (f^{4}_{i})^{\ast} {g}_{t,i} \}
 - \textrm{Re}\{ (f^{5})^{\ast} \mathbf{g}_{t,0} \} + c_{4,3} \leq 0, \nonumber
\end{align}
\end{small}

\vspace{-0.2cm}
\noindent
where  $c_{4,3}$ is constant
and the newly introduced coefficients are defined as follows
\vspace{-0.2cm}
\begin{small}
\begin{align}
& \mathbf{F}^{10}_{k} \triangleq \sigma_{t,0}^2( \hat{\mathbf{g}}_{r,0}^H\hat{\mathbf{u}}_{0}\hat{\mathbf{u}}_{0}^H\hat{\mathbf{g}}_{r,0} )^T
\otimes ( \hat{\mathbf{w}}_k \hat{\mathbf{w}}_k^H ),\\
& \mathbf{F}^{11} \triangleq  \sigma_{t,0}^2( \hat{\mathbf{g}}_{r,0}^H\hat{\mathbf{u}}_{0}\hat{\mathbf{u}}_{0}^H\hat{\mathbf{g}}_{r,0} )^T
\otimes ( \hat{\mathbf{W}}_r \hat{\mathbf{W}}_r^H ),\nonumber\\
& \mathbf{F}^{13}_{k,j} \triangleq \sigma_{t,j}^2( \hat{\mathbf{g}}_{r,j}^H\hat{\mathbf{u}}_{0}\hat{\mathbf{u}}_{0}^H\hat{\mathbf{g}}_{r,j} )^T
\otimes ( \hat{\mathbf{w}}_k \hat{\mathbf{w}}_k^H ),\nonumber\\
& \mathbf{F}^{14}_{j} \triangleq \sigma_{t,j}^2( \hat{\mathbf{g}}_{r,j}^H\hat{\mathbf{u}}_{0}\hat{\mathbf{u}}_{0}^H\hat{\mathbf{g}}_{r,j} )^T
\otimes ( \hat{\mathbf{W}}_r \hat{\mathbf{W}}_r^H ),\nonumber\\
&\mathbf{F}^{12} \triangleq ({\sum}_{k=1}^{K_d} \mathbf{F}^{10}_{k} + \mathbf{F}^{11})/\Gamma,
 \mathbf{F}^{15}_{j} \triangleq {\sum}_{k=1}^{K_d} \mathbf{F}^{13}_{k,j} + \mathbf{F}^{14}_j,\nonumber\\
& \bar{\mathbf{F}}^{15}_{j,m,n} \triangleq \mathbf{F}^{15}_{j}[  (m-1)N_t+1:mN_t , (n-1)N_t+1:nN_t  ],\nonumber\\
&f^{22}_{j,m,n} \triangleq {\sum}_{z\neq n}^{N_t}  2(\bar{\mathbf{F}}^{15}_{j,m,m}[z,n])^{\ast} \mathbf{g}_{t,m,j}[z],\nonumber\\
& \bar{\mathbf{F}}^{12}_{m,n} \triangleq \mathbf{F}^{12}[  (m-1)N_t+1:mN_t , (n-1)N_t+1:nN_t ],\nonumber\\
&{f}^{24}_{m,n} \triangleq {\sum}_{i\neq n}^{N_t} 2 (\bar{\mathbf{F}}^{12}_{m,m}[i,n])^{\ast} \mathbf{g}_{t,m,0}[i],\nonumber\\
& \mathbf{f}^{21}_{j,m} \triangleq {\sum}_{i\neq m}^{M_t} 2(\bar{\mathbf{F}}^{15}_{j,i,m})^H\mathbf{g}_{t,i,j},
 \mathbf{f}^{23}_{m} \triangleq {\sum}_{n\neq m}^{M_t} 2 (\bar{\mathbf{F}}^{12}_{m,n})^H\mathbf{g}_{t,n,0},\nonumber\\
&  f^{4}_{j} \triangleq f^{22}_{j,m,n}  + \mathbf{f}^{21}_{j,m}[n],
  {f}^{5} \triangleq {f}^{24}_{m,n} + \mathbf{f}^{23}_{m}[n]\nonumber.
\end{align}
\end{small}

Based on the above equivalent transformation,
the subproblem w.r.t $\mathbf{t}^0$ is formulated as
\begin{small}
\begin{subequations}
\begin{align}
\textrm{(P9)}:&\mathop{\textrm{min}}
\limits_{
\mathbf{t}^0
}
{\sum}_{k=1}^{K_d}\big(
(\mathbf{h}^{0}_{k}(\mathbf{t}^0))^H
\mathbf{{F}}^{1}_{k}
\mathbf{h}^{0}_{k}(\mathbf{t}^0) \nonumber \\
&+
\textrm{Re}\{ (\mathbf{f}^{1}_{k})^H\mathbf{h}^{0}_{k}(\mathbf{t}^0) \}
- 
\textrm{Re}\{ (\mathbf{f}^{2}_{k})^H\mathbf{h}^{0}_{k}(\mathbf{t}^0) \}\big)
\nonumber\\
&+
{\sum}_{j=0}^{K_t} \textrm{Re}\{ ({f}^{3}_{j})^{\ast}  {g}_{t,j}(\mathbf{t}^0) \}+ {c}_{4,4} \label{P9_obj}\\
\textrm{s.t.}\
&{\sum}_{i=1}^{K_t}\textrm{Re}\{ (f^{4}_{i})^{\ast} {g}_{t,i}(\mathbf{t}^0) \}\label{P9_c_1}\\
& - \textrm{Re}\{ (f^{5})^{\ast} {g}_{t,0}(\mathbf{t}^0) \} + c_{4,3} \leq 0,\nonumber\\
&\Vert \mathbf{t}^0 - \mathbf{t}^0_{m,i} \Vert_2 \geq D_{t}, \forall n, i \in \mathcal{N}_t, n \neq i,\label{P9_c_2}\\
& \mathbf{t}^0 \in \mathcal{C},\label{P9_c_3}
\end{align}
\end{subequations}
\end{small}

\vspace{-0.2cm}
\noindent
where ${c}_{4,4}$ is a constant.
Obviously,
the current form of problem (P9) is difficult to obtain the optimal $\mathbf{t}^0$ due
to the non-convex objective function and constraints.
In the following,
we adopt the  MM methodology \cite{ref_MM} to resolve the above difficulty.
Firstly,
for any positive semidefinite matrix $\mathbf{F}$,
we have the following inequality
\vspace{-0.2cm}
\begin{small}
\begin{align}
&\mathbf{h}^H(\mathbf{t}^0)\mathbf{F}\mathbf{h}(\mathbf{t}^0)\label{MM}\\
&=(\mathbf{h}\!-\!\mathbf{h}_{0})^H\mathbf{F}(\mathbf{h}\!-\!\mathbf{h}_{0})
+2\mathrm{Re}\{\mathbf{h}_{0}^H\mathbf{F}(\mathbf{h}\!-\!\mathbf{h}_{0})\}
+\mathbf{h}_{0}^H\mathbf{F}\mathbf{h}_{0}\nonumber\\
&\leq \lambda_{max}(\mathbf{F})\Vert\mathbf{h}-\mathbf{h}_{0}\Vert^2_2+2\mathrm{Re}\{\mathbf{h}_{0}^H\mathbf{F}
(\mathbf{h}-\mathbf{h}_{0})\}+\mathbf{h}_{0}^H\mathbf{F}\mathbf{h}_{0}\nonumber\\
&=\lambda_{max}(\mathbf{F})\Vert\mathbf{h}\Vert^2_2
+2\mathrm{Re}\{(\mathbf{F}\mathbf{h}_{0}-\lambda_{max}(\mathbf{F})\mathbf{h}_{0})^H\mathbf{h}\}+c_{F},\nonumber
\end{align}
\end{small}

\noindent
with $\lambda_{max}(\mathbf{F})$ being the maximal eigenvalue of the matrix $\mathbf{F}$,
$c_{F}$ is  a constant,
$\mathbf{h}_0 \triangleq \mathbf{h}(\mathbf{t}^0_{0})$,
$ \mathbf{t}^0_{0} = [x^0_{0},y^0_{0} ]^T $  is the latest value
of
$ \mathbf{t}^0 $.

Next,
by leveraging the arguments presented in (\ref{MM}),
we can construct a tight upper-bound for  
$(\mathbf{h}^{0}_{k}(\mathbf{t}^0))^H
\mathbf{{F}}^{1}_{k}
\mathbf{h}^{0}_{k}(\mathbf{t}^0)$,
which can be formulated in
(\ref{T1_Rd_MM_1}),
where
$ \mathbf{h}^{0}_{k,0} \triangleq \mathbf{h}^{0}_{k}(\mathbf{t}^0_0) $,
$\Vert  \mathbf{h}^{0}_{k} \Vert^2_2 = \Vert  \mathbf{h}^{0}_{k,0}  \Vert^2_2 = L^{t,0}_{m,k}$,
$c_{4,5,k} \triangleq 2 \lambda_{max}(\mathbf{{F}}^{1}_{k})L^{t,0}_{m,k}
- ((\mathbf{h}^{0}_{k,0})^H  \mathbf{{F}}^{1}_{k}  \mathbf{h}^{0}_{k,0})^{\ast}
$,
$
\mathbf{f}^{6}_{k} \!\triangleq\! 2\big( (\mathbf{{F}}^{1}_{k})^H\mathbf{h}^{0}_{k,0}
\!-\! \lambda_{max}(\mathbf{{F}}^{1}_{k})\mathbf{h}^{0}_{k,0}  \big)
$
and
$\mathbf{f}^{7}_{k}\! \triangleq\! \mathbf{f}^{6}_{k} \!+\! \mathbf{f}^{1}_{k} $.

\begin{figure*}
\begin{align}
&(\mathbf{h}^{0}_{k}(\mathbf{t}^0))^H
\mathbf{{F}}^{1}_{k}
\mathbf{h}^{0}_{k}(\mathbf{t}^0)
+
\textrm{Re}\{ (\mathbf{f}^{1}_{k})^H\mathbf{h}^{0}_{k}(\mathbf{t}^0) \}
  \label{T1_Rd_MM_1} \\
&
\leq
\lambda_{max}(\mathbf{{F}}^{1}_{k})\Vert  \mathbf{h}^{0}_{k} \Vert^2_2
- 2\textrm{Re}\{\lambda_{max}(\mathbf{{F}}^{1}_{k}) (\mathbf{h}^{0}_{k,0})^H \mathbf{h}^{0}_{k}  \}
+\lambda_{max}(\mathbf{{F}}^{1}_{k})\Vert  \mathbf{h}^{0}_{k,0} \Vert^2_2 \nonumber  \\
&+2\textrm{Re }\{ (\mathbf{h}^{0}_{k,0})^H  \mathbf{{F}}^{1}_{k}  \mathbf{h}^{0}_{k}  \}
- 2\textrm{Re }\{ (\mathbf{h}^{0}_{k,0})^H  \mathbf{{F}}^{1}_{k} \mathbf{h}^{0}_{k,0}  \}
+(\mathbf{h}^{0}_{k,0})^H  \mathbf{{F}}^{1}_{k}  \mathbf{h}^{0}_{k,0}
+
\textrm{Re}\{ (\mathbf{f}^{1}_{k})^H\mathbf{h}^{0}_{k}(\mathbf{t}^0) \}
\nonumber\\
& = \textrm{Re}\{ (\mathbf{f}^{6}_{k})^H\mathbf{h}^{0}_{k} \} + c_{4,5,k}
+
\textrm{Re}\{ (\mathbf{f}^{1}_{k})^H\mathbf{h}^{0}_{k}(\mathbf{t}^0) \}
=\textrm{Re}\{ (\mathbf{f}^{7}_{k})^H\mathbf{h}^{0}_{k} \} + c_{4,5,k}\nonumber
\end{align}
\boldsymbol{\hrule}
\end{figure*}

Furthermore,
to convexity the challenging terms 
$\textrm{Re}\{ (\mathbf{f}^{7}_{k})^H\mathbf{h}^{0}_{k} \}$,
$\textrm{Re}\{ (\mathbf{f}^{2}_{k})^H\mathbf{h}^{0}_{k} \}$
and
$\textrm{Re}\{ ({f}^{3}_{j})^{\ast}  {g}_{t,j}(\mathbf{t}^0) \}$,
we introduce the following results that are proved in \cite{ref_Lemma1_1}.

\begin{lemma}
\label{Lemma_1}
Assume that $\mathbf{f} \in \mathbb{C}^{L\times 1}$,
$\mathbf{h}(\mathbf{t}) 
= [e^{j\!\frac{2\pi}{\lambda}\!(\cos(\!\theta_1\!) \sin(\!\phi_1\!)x\! +\! \sin(\theta_1)y )   },\cdots,
e^{j\!\frac{2\pi}{\lambda}\!(\cos(\!\theta_L\!) \sin(\!\phi_L\!)x\! +\! \sin(\theta_L)y )   }
]
 \in \mathbb{C}^{L\times 1}$
and $\mathbf{t} = [x,y]^T$.
The convex lower and upper bounds of
the function $h_t(\mathbf{t}) \triangleq \textrm{Re}\{ \mathbf{f}^H\mathbf{h}(\mathbf{t}) \}
= \sum_{l=1}^{L}  \vert\mathbf{f}[l]  \vert  \cos( \angle\mathbf{h}(\mathbf{t})[l]
- \angle\mathbf{f}[l]   )   $
( the variable $\mathbf{t}$ only exists in the phase part of $h_t(\mathbf{t})$  )
 can be written as
 \begin{align}
 &h_t(\mathbf{t})\! \geq\!
h_t(\mathbf{t}_0)\!\! +\! \nabla h_t^T(\mathbf{t}_0)(\mathbf{t}\!-\!\mathbf{t}_0)
\!-\! {\delta}/{2} (\mathbf{t}\!-\!\mathbf{t}_0)^T(\mathbf{t}\!-\!\mathbf{t}_0),\label{Lemma1_1}\\
& h_t(\mathbf{t})\! \leq \!
h_t(\mathbf{t}_0)\!\! +\! \nabla h_t^T(\mathbf{t}_0)(\mathbf{t}\!-\!\mathbf{t}_0)
\!+\! {\delta}/{2} (\mathbf{t}\!-\!\mathbf{t}_0)^T(\mathbf{t}\!-\!\mathbf{t}_0),\label{Lemma1_2}
 \end{align}
 respectively,
 where $\mathbf{t}_0$ is obtained from the last iteration,
 the symbols
$\vert \ \vert $
and
$\angle$
denote the  operations of obtaining  the amplitude and phase,
respectively,
 $\delta$ is a positive real number satisfying 
 $\delta = \frac{8\pi^2}{\lambda^2}  {\sum}_{l=1}^{L}\vert\mathbf{f}[l]\vert    $,
 $\nabla h_t(\mathbf{t})= \big[\frac{\partial h_t(\mathbf{t})}{\partial x}, \frac{\partial h_t(\mathbf{t})}{\partial y}\big]^T
  \in \mathbb{C}^{2\times 1}$  denotes
 the gradient vector  of $h_t(\mathbf{t})$ over $\mathbf{t}$.
\end{lemma}

Firstly,
by using the inequality (\ref{Lemma1_2}) in Lemma \ref{Lemma_1},
we can obtain a convex  upper-bound of
$\textrm{Re}\{ (\mathbf{f}^{7}_{k})^H\mathbf{h}^{0}_{k} \}$,
showing as 
\begin{small} 
\begin{align}
& \textrm{Re}\{ (\mathbf{f}^{7}_{k})^H\mathbf{h}^{0}_{k} \}
\triangleq h^{1}_{k}(\mathbf{t}^0)\label{Lemma1_Rd_1}\\
&= {\sum}_{l=1}^{L^{t,0}_{m,k}} \vert \mathbf{f}^{7}_{k}[l] \vert
\cos\big(  ({2\pi}/{\lambda}) (\mathbf{t}^0)^T  \mathbf{a}^{t,0}_{m,k,l} - \angle\mathbf{f}^{7}_{k}[l]  \big)
\nonumber  \\
&\leq
h^{1}_{k}(\mathbf{t}^0_{0})
+ \nabla^T h^{1}_{k}(\mathbf{t}^0_{0})(\mathbf{t}^0 - \mathbf{t}^0_{0})
+ 0.5{\delta_{1,k}} (\mathbf{t}^0 - \mathbf{t}^0_{0})^T(\mathbf{t}^0 - \mathbf{t}^0_{0})\nonumber\\
& =0.5 {\delta_{1,k}} \Vert\mathbf{t}^0\Vert^2_2 + (\mathbf{f}^{8}_{k})^T\mathbf{t}^0 + c_{4,6,k},\nonumber
\end{align}
\end{small}

\vspace{-0.3cm}
\noindent
where
$c_{4,6,k}$ is a constant term,
$\delta_{1,k} = \frac{8\pi^2}{\lambda^2} {\sum}_{l=1}^{L^{t,0}_{m,k}} \vert \mathbf{f}^{7}_{k}[l] \vert$,
$\mathbf{f}^{8}_{k} \triangleq \nabla h^{1}_{k}(\mathbf{t}^0_{0}) - \delta_{1,k}\mathbf{t}^0_{0}$,
and
the elements of the gradient vector
$\nabla h^{1}_{k}(\mathbf{t}^0_{0})
= \big[\frac{\partial h^{1}_{k} (\mathbf{t}^0_{0}) }{\partial x^0_{0}},
\frac{\partial h^{1}_{k}(\mathbf{t}^0_{0}) }{\partial y^0_{0} }\big]^T
$
are respectively given as
\footnote{In the following,
when applying the equations (\ref{Lemma1_1}) and/or (\ref{Lemma1_2}),
the derivation details  of the  gradient vector can refer to  (\ref{T1_Gra_1}) and (\ref{T1_Gra_2}). }
\begin{small}
\begin{align}
&\frac{\partial h^{1}_{k} (\mathbf{t}^0_{0}) }{\partial x^0_{0}}
=- {\sum}_{l=1}^{L^{t,0}_{m,k}}\frac{2\pi}{\lambda}\vert \mathbf{f}^{7}_{k}[l] \vert
\cos(\theta^{t,0}_{m,k,l})\label{T1_Gra_1}\\
&\sin(\phi^{t,0}_{m,k,l})
\sin\big(  \frac{2\pi}{\lambda} (\mathbf{t}^0_0)^T  \mathbf{a}^{t,0}_{m,k,l} - \angle\mathbf{f}^{7}_{k}[l]  \big),\nonumber
\\
&\frac{\partial h^{1}_{k}(\mathbf{t}^0_{0}) }{\partial y^0_{0} }
=
- {\sum}_{l=1}^{L^{t,0}_{m,k}}\frac{2\pi}{\lambda}\vert \mathbf{f}^{7}_{k}[l] \vert
\sin(\theta^{t,0}_{m,k,l})\label{T1_Gra_2}\\
&\sin\big(  \frac{2\pi}{\lambda} (\mathbf{t}^0_{0})^T  \mathbf{a}^{t,0}_{m,k,l} - \angle\mathbf{f}^{7}_{k}[l]  \big).\nonumber
\end{align}
\end{small}

Furthermore,
the non-convex term
$\textrm{Re}\{ (\mathbf{f}^{2}_{k})^H\mathbf{h}^{0}_{k} \}$
can be lower-bounded by the inequality (\ref{Lemma1_1}),
which is given as
\begin{small}
\begin{align}
& \textrm{Re}\{ (\mathbf{f}^{2}_{k})^H\mathbf{h}^{0}_{k} \}
\triangleq h^{2}_{k}(\mathbf{t}^0)\\
&= {\sum}_{l=1}^{L^{t,0}_{m,k}} \vert \mathbf{f}^{2}_{k}[l] \vert
\cos\big( ({2\pi}/{\lambda}) (\mathbf{t}^0)^T  \mathbf{a}^{t,0}_{m,k,l} - \angle\mathbf{f}^{2}_{k}[l]  \big)
\nonumber  \\
&\geq
h^{2}_{k}(\mathbf{t}^0_{0})
+ \nabla^T h^{2}_{k}(\mathbf{t}^0_{0})(\mathbf{t}^0 - \mathbf{t}^0_{0})
- 0.5{\bar{ \delta}_{1,k}} (\mathbf{t}^0 - \mathbf{t}^0_{0})^T(\mathbf{t}^0 - \mathbf{t}^0_{0})\nonumber\\
& = -0.5{\bar{\delta}_{1,k}} \Vert\mathbf{t}^0\Vert^2_2 + (\mathbf{{f}}^{9}_{k})^T\mathbf{t}^0 + {c}_{4,7,k},\nonumber
\end{align}
\end{small}

\vspace{-0.2cm}
\noindent
where
${c}_{4,7,k}$ is a constant term,
$\bar{\delta}_{1,k} = \frac{8\pi^2}{\lambda^2} {\sum}_{l=1}^{L^{t,0}_{m,k}} \vert \mathbf{f}^{2}_{k}[l] \vert$,
$\mathbf{{f}}^{9}_{k} \triangleq \nabla h^{2}_{k}(\mathbf{t}^0_{0}) + \bar{\delta}_{1,k}\mathbf{t}^0_{0}$,
$\nabla h^{2}_{k}(\mathbf{t}^0_{0})
$
is
the gradient vector.

Next,
the term 
$\textrm{Re}\{ ({f}^{3}_{j})^{\ast}  {g}_{t,j}(\mathbf{t}^0) \}$
be convexified by leveraging (\ref{Lemma1_2}) as follows
\begin{small} 
\begin{align}
& \textrm{Re}\{ ({f}^{10}_{j})^{\ast}  {g}_{t,j}(\mathbf{t}^0) \}
\triangleq g^{1}_{j}(\mathbf{t}^0)\\
&=  \vert {f}^{3}_{j} \vert\vert{g}_{t,j}\vert
\cos\big(  ({2\pi}/{\lambda}) (\mathbf{t}^0)^T  \mathbf{a}^{4}_{m,j} - \angle{f}^{3}_{j}  \big)
\nonumber  \\
&\leq
 g^{1}_{j}(\mathbf{t}^0_0)
+ \nabla^T g^{1}_{j}(\mathbf{t}^0_{0})(\mathbf{t}^0 - \mathbf{t}^0_{0})
+ 0.5{\delta_{2,j}} (\mathbf{t}^0 - \mathbf{t}^0_{0})^T(\mathbf{t}^0 - \mathbf{t}^0_{0})\nonumber\\
& = 0.5{\delta_{2,j}} \Vert\mathbf{t}^0\Vert^2_2 + (\mathbf{f}^{10}_{j})^T\mathbf{t}^0 + c_{4,8,j},\nonumber
\end{align}
\end{small}

\vspace{-0.2cm}
\noindent
with
$c_{4,8,j}$
being constant,
$\delta_{2,j} \triangleq \frac{8\pi^2}{\lambda^2}\vert{f}^{3}_{j}\vert \vert\mathbf{g}_{t,j}\vert$,
$\mathbf{f}^{10}_{j}\triangleq \nabla g^{1}_{j}(\mathbf{t}^0_{0}) - \delta_{2,j}\mathbf{t}^0_{0}$,
$\nabla g^{1}_{j}(\mathbf{t}^0_{0})$
is
the gradient vector over $\mathbf{t}^0_{0}$.

Based on the above transformation,
the upper-bound of the objective function (\ref{P9_obj}) is given as
\begin{small}
\begin{align}
&{\sum}_{k=1}^{K_d}\big(
(\mathbf{h}^{0}_{k}(\mathbf{t}^0))^H
\mathbf{{F}}_{1,k}
\mathbf{h}^{0}_{k}(\mathbf{t}^0) 
+
\textrm{Re}\{ (\mathbf{f}^1_{k})^H\mathbf{h}^{0}_{k}(\mathbf{t}^0) \}\\
&- 
\textrm{Re}\{ (\mathbf{f}^2_{k})^H\mathbf{h}^{0}_{k}(\mathbf{t}^0) \}\big)
+
{\sum}_{j=0}^{K_t} \textrm{Re}\{ ({f}^{3}_{j})^{\ast}  {g}_{t,j}(\mathbf{t}^0) \}+ \hat{c}_{4,2}\nonumber\\
&
\leq
{\sum}_{k=1}^{K_d}\big(0.5{\delta_{1,k}} \Vert\mathbf{t}^0\Vert^2_2 + (\mathbf{f}^{8}_{k})^T\mathbf{t}^0 + c_{4,6,k}\nonumber\\
&-
(-{0.5\bar{\delta}_{1,k}} \Vert\mathbf{t}^0\Vert^2_2 + (\mathbf{{f}}^{9}_{k})^T\mathbf{t}^0 + {c}_{4,7,k})\big)\nonumber\\
&+{\sum}_{j=0}^{K_t}({0.5\delta_{2,j}} \Vert\mathbf{t}^0\Vert^2_2 + (\mathbf{f}^{10}_{j})^T\mathbf{t}^0 + c_{4,8,j})
+ {c}_{4,4} \nonumber\\
&=
{\delta_{3}} \Vert\mathbf{t}^0\Vert^2_2 + (\mathbf{f}^{11})^T\mathbf{t}^0 + c_{4,9}\nonumber
\end{align}
\end{small}

\vspace{-0.2cm}
\noindent
where
$c_{4,9}$
is a constant term,
$\delta_{3} \triangleq \sum_{k=1}^{K_d}({\delta_{1,k}}/{2}+{\bar{\delta}_{1,k}}/{2}) 
+ \sum_{j=0}^{K_t} {\delta_{2,j}}/{2}  $
and
$\mathbf{f}^{11}\triangleq \sum_{k=1}^{K_d} (\mathbf{f}^{8}_{k} - \mathbf{{f}}^{9}_{k} )
+ \sum_{j=0}^{K_t} \mathbf{f}^{10}_{j} $.

In the following,
the term
$\textrm{Re}\{ (f^{13}_{i})^{\ast} {g}_{t,i}(\mathbf{t}^0) \}$ 
of the sensing SINR constraint (\ref{P9_c_1})
can be convexified via (\ref{Lemma1_2}) as follows
\begin{small}
\begin{align}
&\textrm{Re}\{ (f^{4}_{i})^{\ast} \mathbf{g}_{t,i}(\mathbf{t}^0) \}
\triangleq g^{2}_{i}(\mathbf{t}^0)\\
&= \vert f^{4}_{i}\vert \vert\mathbf{g}_{t,i}\vert 
\cos\big(  ({2\pi}/{\lambda}) (\mathbf{t}^0)^T  \mathbf{a}^{4}_{m,i} - \angle f^{4}_{i} \big)\nonumber\\
&\leq
 g^{2}_{i}(\mathbf{t}^0_0)
+ \nabla^T g^{2}_{i}(\mathbf{t}^0_{0})(\mathbf{t}^0 - \mathbf{t}^0_{0})
+ 0.5{\delta_{4,i}} (\mathbf{t}^0 - \mathbf{t}^0_{0})^T(\mathbf{t}^0 - \mathbf{t}^0_{0})\nonumber\\
&=
{0.5\delta_{4,i}} \Vert\mathbf{t}^0\Vert^2_2 + (\mathbf{f}^{12}_{i})^T\mathbf{t}^0 + c_{4,10,i},\nonumber
\end{align}
\end{small}

\vspace{-0.2cm}
\noindent
with
$c_{4,10,i}$
being constant,
$\delta_{4,i} \triangleq \frac{8\pi^2}{\lambda^2}\vert f^{4}_{i}\vert \vert\mathbf{g}_{t,i}\vert$,
$\mathbf{f}^{12}_{i}\triangleq \nabla g^{2}_{i}(\mathbf{t}^0_{0}) -  \delta_{4,i}\mathbf{t}^0_{0}$,
$\nabla  g^{2}_{i}(\mathbf{t}^0_{0})$
is
the gradient vector.

And then,
we proceed to cope with the non-convex term
$ \textrm{Re}\{ (f^{5})^{\ast} {g}_{t,0}(\mathbf{t}^0) \} $ 
of the sensing SINR constraint (\ref{P9_c_1}).
This time, we require a convex lower bound for
$ \textrm{Re}\{ (f^{5})^{\ast} {g}_{t,0}(\mathbf{t}^0) \} $ 
 by applying
(\ref{Lemma1_1}) in Lemma \ref{Lemma_1},
which is given as follows
\begin{small}
\begin{align}
&\textrm{Re}\{ (f^{5})^{\ast} {g}_{t,0}(\mathbf{t}^0) \}
\triangleq g^{3}(\mathbf{t}^0)\\
&=
\vert f^{5}  \vert \vert \mathbf{g}_{t,0}  \vert 
\cos\big(  ({2\pi}/{\lambda}) (\mathbf{t}^0)^T  \mathbf{a}^{4}_{m,0} - \angle f^{5} \big)
\nonumber\\
&\geq
 g^{3}(\mathbf{t}^0_0)
+ \nabla^T g^{3}(\mathbf{t}^0_{0})(\mathbf{t}^0 - \mathbf{t}^0_{0})
- {0.5\delta_{6}} (\mathbf{t}^0 - \mathbf{t}^0_{0})^T(\mathbf{t}^0 - \mathbf{t}^0_{0})\nonumber\\
&= - {0.5\delta_{6}} \Vert\mathbf{t}^0\Vert^2_2 + (\mathbf{f}^{13})^T\mathbf{t}^0 + c_{4,11},\nonumber
\end{align}
\end{small}

\vspace{-0.2cm}
\noindent
where
$c_{4,11}$ is constant,
$ \delta_{6} \triangleq \frac{8\pi^2}{\lambda^2} \vert f^{5}  \vert \vert \mathbf{g}_{t,0}  \vert$,
$  \mathbf{f}^{13} \triangleq \nabla g^{3}(\mathbf{t}^0_{0}) + \delta_{6} \mathbf{t}^0_{0} $,
$\nabla g^{3}(\mathbf{t}^0_{0})$ is the gradient vector.

Therefore,
a convex upper-bound of
the SINR constraint (\ref{P9_c_1}) can be written as 
\begin{small}
\begin{align}
&{\sum}_{i=1}^{K_t}\textrm{Re}\{ (f^{4}_{i})^{\ast} {g}_{t,i}(\mathbf{t}^0) \}
 - \textrm{Re}\{ (f^{5})^{\ast} {g}_{t,0}(\mathbf{t}^0) \} + c_{4,3}\\
& 
\leq
{\sum}_{i=1}^{K_t}
({\delta_{4,i}}/{2} \Vert\mathbf{t}^0\Vert^2_2 + (\mathbf{f}^{12}_{i})^T\mathbf{t}^0 + c_{4,10,i})\nonumber\\
&-
(- {\delta_{6}}/{2} \Vert\mathbf{t}^0\Vert^2_2 + (\mathbf{f}^{13})^T\mathbf{t}^0 + c_{4,11})
+ c_{4,3}\nonumber\\
&=\delta_{7} \Vert\mathbf{t}^0\Vert^2_2 + (\mathbf{f}^{14})^T\mathbf{t}^0 + c_{4,12},\nonumber
\end{align}
\end{small}

\vspace{-0.2cm}
\noindent
where
$c_{4,12}$ is constant term,
$\delta_{7} \triangleq \sum_{j=1}^{K_t} {\delta_{4,j}}/{2} +  {\delta_{6}}/{2}$
and
$\mathbf{f}^{14}\triangleq \sum_{j=1}^{K_t} \mathbf{f}^{12}_{j} - \mathbf{f}^{13}$.
To make (P9) tractable,
we still need to convexify the constraint (\ref{P9_c_2}).
Inspired by the MM framework \cite{ref_MM},
 the constraint (\ref{P9_c_2}) can be  linearized as follows
\begin{align}
\Vert \mathbf{t}^0 - \mathbf{t}^0_{m,i} \Vert_2
\geq \frac{(\mathbf{t}^0_{0} - \mathbf{t}^0_{m,i})^T}{\Vert \mathbf{t}^0_{0} - \mathbf{t}^0_{m,i} \Vert_2}
(\mathbf{t}^0 - \mathbf{t}^0_{m,i}). \label{T1_Dmin_1}
\end{align}
Based on the above MM transformation,
the update of $\mathbf{t}^0$
can be performed
by solving a convex surrogate of the original challenging problem (P9),
as follows
\begin{subequations}
\begin{align}
\textrm{(P10)}:&\mathop{\textrm{min}}
\limits_{
\mathbf{t}^0
}
\delta_{3} \Vert\mathbf{t}^0\Vert^2_2 + (\mathbf{f}^{11})^T\mathbf{t}^0_{m,n} + c_{4,9}\\
\textrm{s.t.}\
& \delta_{7} \Vert\mathbf{t}^0\Vert^2_2 + (\mathbf{f}^{14})^T\mathbf{t}^0 + c_{4,12}
\leq 0,\\
& \frac{(\mathbf{t}^0_{0} - \mathbf{t}^0_{m,i})^T}{\Vert \mathbf{t}^0_{0}\!\! - \!\mathbf{t}^0_{m,i} \Vert_2}
\!(\mathbf{t}^0 \!-\! \mathbf{t}^0_{m,i})  \geq D_{t}, \forall  i\in \mathcal{N}_t, n \neq i,\\
& \mathbf{t}^0 \in \mathcal{C},
\end{align}
\end{subequations}

Clearly,
problem (P10) is convex and can be numerically solved.

\subsection{Optimizing The RBS MAs' Position}

In this subsection,
with other variables being given,
we investigate the update of  the position of  the $m$-th RBS'  $n$-th MA,
i.e., $\mathbf{t}^1_{m,n}$
\footnote{
To improve readability, 
we have removed the subscripts, i.e., $m$ and $n$,
 from the variables and/or coefficients. 
Specifically, 
we represent:
$\mathbf{t}^1_{m,n}$ as $\mathbf{t}^1$,
$\mathbf{h}^{1}_{m,l,n}$ as $\mathbf{h}^{1}_{l}$, 
and
$\mathbf{g}_{r,m,j}[n]$ as ${g}_{r,j}$.
}.
Firstly,
the function $\tilde{\mathrm{R}}_{u,l}$ given in (\ref{MSE_Ru}) is rewritten as
 \vspace{-0.2cm}
\begin{small}
\begin{align}
&
-
{\sum}_{l=1}^{K_u}\tilde{\mathrm{R}}_{u,l}
= 
{\sum}_{l=1}^{K_u}
\big(
 (\mathbf{h}^{1}_{l})^H
 \mathbf{D}^{1}_{l}\mathbf{h}^{1}_{l} 
 +
 \textrm{Re}\{ (\mathbf{d}^{1}_{l})^H \mathbf{h}^{1}_{l}  \}\label{T2_Ru_1} \\
& -
 \textrm{Re}\{ (\mathbf{{d}}^{2}_{l})^H \mathbf{h}^{1}_{l}  \}
\big)
+
{\sum}_{j=0}^{K_t} \textrm{Re}\{ (d^{3}_{j})^{\ast}\mathbf{g}_{r,j} \}+c_{5,1},\nonumber
\end{align}
\end{small}

\vspace{-0.2cm}
\noindent
with $c_{5,1}$ being a constant
and
the parameters in  (\ref{T2_Ru_1})   defining  in following
\begin{small}
\begin{align}
& \mathbf{\bar{D}}^1_{m,n,l,i}
\!\!\triangleq \!\omega_{u,l}\vert\beta_{u,l}\vert^2
\!( \mathbf{u}_{m,l}\mathbf{u}_{n,l}^H )^T \!\!\!\otimes\!\! ( q_i ( \mathbf{\Sigma}^{1}_{n,i})^H \!  \mathbf{h}^{0,1}_{n,i}\!\mathbf{h}^{0,1}_{m,i}\mathbf{\Sigma}^{1}_{m,i} ),\\
& \hat{\mathbf{h}}^{1}_{m,l} \triangleq \textrm{vec} (\mathbf{H}^{1}_{m,l}(\mathbf{t}^1_m)),
 \mathbf{d}^{19}_{m,l,i}
\triangleq
2( {\sum}_{j \neq m}^{M_t} (\mathbf{\bar{D}}^1_{m,j,l,i})^H  \hat{\mathbf{h}}^{1}_{j,i}  ),\nonumber\\
& \mathbf{d}^{20}_{m,l}
\triangleq
2 \omega_{u,l}\beta_{u,l}\sqrt{q_l}
\mathbf{u}_{m,l}^H \otimes (  (\mathbf{h}^{0,1}_{m,l})^H \mathbf{\Sigma}^{1}_{m,l} ),\nonumber\\
 &\mathbf{\bar{D}}^2_{j,l}
\triangleq
{\sum}_{k=1}^{K_d}[ ( \hat{\mathbf{g}}_{t,j}^H\hat{\mathbf{w}}_k \hat{\mathbf{w}}_k^H \hat{\mathbf{g}}_{t,j}  )^T
\otimes \sigma^2_{t,j}(\hat{\mathbf{u}}_{l}\hat{\mathbf{u}}_{l}^H)  ],\nonumber\\
& \mathbf{D}^4_{j,l}  \triangleq \omega_{u,l}\vert\beta_{u,l}\vert^2( \mathbf{\bar{D}}^2_{j,l} + \mathbf{D}^3_{j,l}),\nonumber\\
& \mathbf{D}^3_{j,l}
\triangleq
( \hat{\mathbf{g}}_{t,j}^H\hat{\mathbf{W}}^r (\hat{\mathbf{W}}^r)^H \hat{\mathbf{g}}_{t,j}  )^T
\otimes \sigma^2_{t,j}(\hat{\mathbf{u}}_{l}\hat{\mathbf{u}}_{l}^H),\nonumber\\
&\bar{\mathbf{D}}^4_{j,l,m,n} \triangleq \mathbf{D}^4_{j,l}[ (m-1)N_r+1:mN_r, (n-1)N_r+1:nN_r   ],\nonumber\\
&\mathbf{d}^{21}_{j,l,m}
\triangleq
2({\sum}_{i\neq m}^{M_r} ( \bar{\mathbf{D}}^4_{j,l,i,m} )^H \mathbf{g}_{r,i,j} ),
 \mathbf{D}^5_{m,i} \triangleq  {\sum}_{l=1}^{K_u} \mathbf{\bar{D}}^1_{m,m,l,i},\nonumber\\
& \mathbf{d}^{22}_{m,i} \!\triangleq\! \! {\sum}_{l=1}^{K_u} \mathbf{d}^{19}_{m,l,i} ,
 \mathbf{D}^{6}_{j,m}\! \triangleq\!\! {\sum}_{l=1}^{K_u} \bar{\mathbf{D}}^4_{j,l,m,m},
 \mathbf{d}^{23}_{j,m}\! \triangleq\!\! {\sum}_{l=1}^{K_u}  \mathbf{d}^{21}_{j,l,m},\nonumber\\
& \mathbf{D}^1_{l} \triangleq \mathbf{D}^5_{m,l}  [ (n-1)L^{t,1}_{m,l}+1:nL^{t,1}_{m,l} , (j-1)L^{t,1}_{m,l}+1:jL^{t,1}_{m,l} ],\nonumber\\
& \mathbf{d}^{24}_{m,l,n}\triangleq 2({\sum}_{j \neq n}^{N_r}  (\mathbf{D}^7_{m,l,n,j})^H\mathbf{h}^{1}_{m,l,n}   ),\nonumber\\
& \mathbf{d}^{1}_{l}\!\triangleq\! \mathbf{d}^{24}_{m,l,n}\! \!+\! \mathbf{d}^{22}_{m,l}[(n\!-\!1)L^{t,1}_{m,l}\!+\!1\!:\!nL^{t,1}_{m,l}],\nonumber\\
& d^{3}_{j}\! \triangleq\! 2({\sum}_{i\neq n}^{N_r} (\mathbf{D}^{6}_{j,m}[i,n])^{\ast} {\mathbf{g}}_{r,m,j}[i]
)
\!+\!
\mathbf{d}^{23}_{j,m}[n],\nonumber\\
&\mathbf{{d}}^{2}_{l}\!\!\triangleq \! \mathbf{d}^{20}_{m,l}[(n\!-\!1)L^{t,1}_{m,l}+1\!:\!nL^{t,1}_{m,l}]
.\nonumber
\end{align}
\end{small}
And
the sensing SINR constraint (\ref{P1_c_1}) can be rewritten as
 \vspace{-0.2cm}
\begin{small}
\begin{align}
&\textrm{SINR}_{t} \geq \Gamma_r
\!\!\!\Longleftrightarrow\!\!\!
{\sum}_{l=1}^{K_u}
\big(
 (\mathbf{h}^{1}_{l})^H
 \mathbf{D}^{2}_{l}\mathbf{h}^{1}_{l} 
 \!+\!\!
 \textrm{Re}\{ (\mathbf{d}^{4}_{l})^H \mathbf{h}^{1}_{l}  \}
\big)
\label{T2_SINR_1} \\
&+ {\sum}_{j=1}^{K_t}
\big(
\textrm{Re}\{
(d^{5}_{j})^{\ast}\mathbf{g}_{r,j}
\}
\big)
 - \textrm{Re}\{
(d^{6})^{\ast}{g}_{r,0}
\}+ c_{5,2}
 \leq 0, \nonumber
\end{align}
\end{small}

\noindent
where  $c_{5,2}$ is a  constant term
and
the above newly introduced coefficients are given in 
\begin{small}
\begin{align}
& \mathbf{D}^{10} \triangleq
\big({\sum}_{k=1}^{K_d}[ ( \hat{\mathbf{g}}_{t,0}^H\hat{\mathbf{w}}_k \hat{\mathbf{w}}_k^H \hat{\mathbf{g}}_{t,0}  )^T
\otimes \sigma^2_{t,0}(\hat{\mathbf{u}}_{0}\hat{\mathbf{u}}_{0}^H)  ]\\
&+
( \hat{\mathbf{g}}_{t,0}^H\hat{\mathbf{W}}^r (\hat{\mathbf{W}}^r)^H \hat{\mathbf{g}}_{t,0}  )^T
\otimes \sigma^2_{t,0}(\hat{\mathbf{u}}_{0}\hat{\mathbf{u}}_{0}^H)\big)/\Gamma_t,\nonumber\\
& \bar{\mathbf{D}}^{10}_{n,m}
\triangleq
\mathbf{D}^{10} [ (n-1)N_r+1:nN_r  , (m-1)N_r+1:mN_r ],\nonumber\\
& \mathbf{D}^{11}_{l,m,n} \triangleq
( \mathbf{u}_{m,0}\mathbf{u}_{n,0}^H )^T \otimes ( q_l (\mathbf{\Sigma}^{1}_{n,l})^H   \mathbf{h}^{0,1}_{n,l}\mathbf{h}^{0,1}_{m,l}\Sigma^{1}_{m,l} ),\nonumber\\
& \mathbf{d}^{25}_{l,m}
\triangleq
2( {\sum}_{n \neq m}^{M_r}  (\mathbf{D}^{11}_{l,m,n})^H \hat{\mathbf{h}}^{1}_{n,l}),\nonumber\\
& \mathbf{D}^{14}_{j} \triangleq
{\sum}_{k=1}^{K_d}[ ( \hat{\mathbf{g}}_{t,j}^H\hat{\mathbf{w}}_k \hat{\mathbf{w}}_k^H \hat{\mathbf{g}}_{t,j}  )^T
\otimes \sigma^2_{t,j}(\hat{\mathbf{u}}_{0}\hat{\mathbf{u}}_{0}^H)  ]\nonumber\\
&+
( \hat{\mathbf{g}}_{t,j}^H\hat{\mathbf{W}}^r (\hat{\mathbf{W}}^r)^H \hat{\mathbf{g}}_{t,j}  )^T
\otimes \sigma^2_{t,j}(\hat{\mathbf{u}}_{0}\hat{\mathbf{u}}_{0}^H),\nonumber\\
& \bar{\mathbf{D}}^{14}_{j,i,z} \triangleq \mathbf{D}^{14}_{j}[ (i-1)N_r+1:iN_r  , (z-1)N_r+1:zN_r ],\nonumber\\
& \mathbf{d}^{26}_{j,m}
\triangleq
2( {\sum}_{n\neq m}^{M_r}  (  {\mathbf{g}}_{r,n,j}^H   \bar{\mathbf{D}}^{14}_{j,n,m})   ),
 \mathbf{d}^{27}_{m}
\triangleq
2( {\sum}_{n\neq m}^{M_r}(  {\mathbf{g}}_{r,n,0}^H   \bar{\mathbf{D}}^{10}_{n,m})   ),\nonumber\\
& {\mathbf{D}}^{2}_{l}
\triangleq
\mathbf{D}^{11}_{l,m,m}[ (n-1)L^{t,1}_{m,l}+1: nL^{t,1}_{m,l} , (i-1)L^{t,1}_{m,l}+1: iL^{t,1}_{m,l}],\nonumber\\
 &\mathbf{d}^{28}_{l,m,n}
\triangleq
2(  {\sum}_{j\neq n}^{N_r}   ({\mathbf{D}}^{15}_{l,m,j,n})^H \mathbf{h}^{1}_{m,l,j}   ),\nonumber\\
& \mathbf{d}^{4}_{l}
\!\triangleq \!
\mathbf{d}^{28}_{l,m,n}\! + \!\mathbf{d}^{25}_{l,m}[(n\!-\!1)L^{t,1}_{m,l}\!+\!1\!:\!nL^{t,1}_{m,l}],\nonumber\\
& d^{5}_{j}
\!\triangleq \!
2( {\sum}_{i \neq n}^{N_r} (\bar{\mathbf{D}}^{14}_{j,m,m}[i,n])^{\ast}{\mathbf{g}}_{r,m,j}[i]  )
\!+\!
 \mathbf{d}^{26}_{j,m}[n],\nonumber\\
&  d^{6}
 \!\triangleq \!
  2( {\sum}_{i \neq n}^{N_r} (\bar{\mathbf{D}}^{10}_{m,m}[i,n])^{\ast}{\mathbf{g}}_{r,m,0}[i]  )
 \! +\!
  \mathbf{d}^{27}_{m}[n].\nonumber
\end{align}
\end{small}

Therefore,
the variable $\mathbf{t}^1$
optimization reduces to solving the following problem
\vspace{-0.1cm}
\begin{small}
\begin{subequations}
\begin{align}
\textrm{(P11)}:&\mathop{\textrm{min}}
\limits_{
\mathbf{t}^1
}
{\sum}_{l=1}^{K_u}
\big(
 (\mathbf{h}^{1}_{l})^H
 \mathbf{D}^{1}_{l}\mathbf{h}^{1}_{l} 
 +
 \textrm{Re}\{ (\mathbf{d}^{1}_{l})^H \mathbf{h}^{1}_{l}  \}\label{P11_obj_1} \\
 &-
 \textrm{Re}\{ (\mathbf{{d}}^{2}_{l})^H \mathbf{h}^{1}_{l}  \}
\big)
+
{\sum}_{j=0}^{K_t} \textrm{Re}\{ (d^{3}_{j})^{\ast}\mathbf{g}_{r,j} \}+c_{5,1}\nonumber\\
\textrm{s.t.}\
&{\sum}_{l=1}^{K_u}
\big(
 (\mathbf{h}^{1}_{l})^H
 \mathbf{D}^{2}_{l}\mathbf{h}^{1}_{l} 
 +
 \textrm{Re}\{ (\mathbf{d}^{4}_{l})^H \mathbf{h}^{1}_{l}  \}
\big)\label{P11_c_1} \\
&\!+\!{\sum}_{j=1}^{K_t}\!
\big(\!
\textrm{Re}\{\!
(d^{5}_{j})^{\ast}\!{g}_{r,j}
\}\!
\big)
 - \textrm{Re}\{
(d^{6})^{\ast}\mathbf{g}_{r,0}
\}+ c_{5,2}
 \leq 0,\nonumber\\
&\Vert \mathbf{t}^1 - \mathbf{t}^1_{m,i} \Vert_2 \geq D_{t}, \forall n, i \in \mathcal{N}_t, n \neq i,\label{P11_c_2}\\
& \mathbf{t}^1\in \mathcal{C}.\label{P11_c_3}
\end{align}
\end{subequations}
\end{small}
In the following,
we will obtain the upper-bound of  the objective function (\ref{P11_obj_1}) via convexify it.
By leveraging the arguments presented  in (\ref{MM}),
 a tight upper-bound of  
$(\mathbf{h}^{1}_{l})^H
 \mathbf{D}^{1}_{l}\mathbf{h}^{1}_{l}
 +
 \textrm{Re}\{ (\mathbf{d}^{1}_{l})^H \mathbf{h}^{1}_{l}  \}$
 can be shown as follows
\begin{small}
\begin{align}
& (\mathbf{h}^{1}_{l})^H
 \mathbf{D}^{1}_{l}\mathbf{h}^{1}_{l} 
 +
 \textrm{Re}\{ (\mathbf{d}^{1}_{l})^H \mathbf{h}^{1}_{l}  \} \label{R1_Rd_MM_1} \\
&\leq
\lambda_{max}( \mathbf{D}^{1}_{l})(\Vert  \mathbf{h}^{1}_{l}  \Vert^2_2+\Vert  \mathbf{h}^{1}_{l,0}\Vert^2_2 )
- 2\textrm{Re}\{\lambda_{max}( \mathbf{D}^{1}_{l}) (\mathbf{h}^{1}_{l,0})^H \mathbf{h}^{1}_{l}  \}\nonumber\\
&\!+\!2\textrm{Re }\{ (\mathbf{h}^{1}_{l,0})^H \mathbf{D}^{1}_{l} (\mathbf{h}^{1}_{l}\!-\!\mathbf{h}^{1}_{l,0})  \}
+(\mathbf{h}^{1}_{l,0})^H  \mathbf{D}^{1}_{l}  \mathbf{h}^{1}_{l,0}
\!+\!
 \textrm{Re}\{ (\mathbf{d}^{1}_{l})^H \mathbf{h}^{1}_{l}  \}
\nonumber\\
& =\! \textrm{Re}\{ (\mathbf{d}^{7}_{l})^H\mathbf{h}^{1}_{l} \}\! +\! c_{5,3,l}
\!+\!
 \textrm{Re}\{ (\mathbf{d}^{1}_{l})^H\! \mathbf{h}^{1}_{l}  \}
\!=\!\textrm{Re}\{ (\mathbf{d}^{8}_{l})^H\! \mathbf{h}^{1}_{l}  \}\!\! +\!\! c_{5,3,l},\nonumber
\end{align}
\end{small}

\noindent
where
$\mathbf{t}^1_0$ is the feasible solution obtained in the last iteration,
$ \mathbf{h}^{1}_{l,0} \!\triangleq\! \mathbf{h}^{1}_{l}(\mathbf{t}^1_0) $,
$\Vert  \mathbf{h}^{1}_{l} \Vert^2_2 \!=\! \Vert  \mathbf{h}^{1}_{l,0}  \Vert^2_2 \!=\! L^{t,1}_{m,l}$,
$c_{5,3,l} \!\triangleq\! 2 \lambda_{max}(\mathbf{D}^{1}_{l})L^{t,1}_{m,l}
- ((\mathbf{h}^{1}_{l,0})^H  \mathbf{D}^{1}_{l}  \mathbf{h}^{1}_{l,0})^{\ast}
$,
$
\mathbf{d}^{7}_{l} \!\triangleq\! 2( (\mathbf{D}^{1}_{l})^H\mathbf{h}^{1}_{l,0}
\!-\! \lambda_{max}(\mathbf{D}^{1}_{l})\mathbf{h}^{1}_{l,0}  )
$
and
$\mathbf{d}^{8}_{l}\! \triangleq\! \mathbf{d}^{7}_{l} \!+\! \mathbf{d}^{1}_{l} $.
 
%
 Furthermore,
by using the inequality (\ref{Lemma1_2}) in Lemma \ref{Lemma_1},
we can obtain a convex  upper-bound of
$\textrm{Re}\{ (\mathbf{d}^{8}_{l})^H\mathbf{h}^{1}_{l} \}$
as follows
\vspace{-0.3cm}
\begin{small}
\begin{align}
& \textrm{Re}\{ (\mathbf{d}^{8}_{l})^H\mathbf{h}^{1}_{l} \}
\triangleq h^{3}_{l}(\mathbf{t}^1)\\
&= {\sum}_{i=1}^{L^{t,1}_{m,l}} \vert \mathbf{d}^{8}_{l}[i] \vert
\cos\big(  ({2\pi}/{\lambda}) (\mathbf{t}^1)^T  \mathbf{a}^{r,1}_{m,l,i} - \angle\mathbf{d}^{8}_{l}[i]  \big)
\nonumber \\
&\leq
h^{3}_{l}(\mathbf{t}^1_{0})
+ \nabla^T h^{3}_{l}(\mathbf{t}^1_{0})(\mathbf{t}^1 - \mathbf{t}^1_{0})
+ {0.
5\tau_{1,l}} (\mathbf{t}^1 - \mathbf{t}^1_{0})^T(\mathbf{t}^1 - \mathbf{t}^1_{0})\nonumber\\
& ={0.5\tau_{1,l}} \Vert\mathbf{t}^1\Vert^2_2 + (\mathbf{d}^{9}_{l})^T\mathbf{t}^1 + c_{5,4,l}\nonumber
\end{align}
\end{small}

\vspace{-0.2cm}
\noindent
where
$c_{5,4,l}$ is a constant term,
$\tau_{1,l} = \frac{8\pi^2}{\lambda^2} {\sum}_{i=1}^{L^{t,1}_{m,l}} \vert \mathbf{d}^{8}_{l}[i] \vert$,
$\mathbf{d}^{9}_{l} \triangleq \nabla h^{3}_{l}(\mathbf{t}^1_{0}) - \tau_{1,l} \mathbf{t}^1_{0}$,
$\nabla h^{3}_{l}(\mathbf{t}^1_{0})$
is the  gradient vector.
Furthermore,
by leveraging the inequality (\ref{Lemma1_1}),
the lower-bound of 
$\textrm{Re}\{ (\mathbf{d}^2_{l})^H \mathbf{h}^{1}_{l}  \}$
is given as
\vspace{-0.1cm}
\begin{small}
\begin{align}
& \textrm{Re}\{ (\mathbf{d}^{2}_{l})^H \mathbf{h}^{1}_{l}  \}
\triangleq h^{4}_{l}(\mathbf{t}^1)\\
&= {\sum}_{i=1}^{L^{t,1}_{m,l}} \vert \mathbf{d}^{2}_{l}[i] \vert
\cos\big( ({2\pi}/{\lambda}) (\mathbf{t}^1)^T  \mathbf{a}^{r,1}_{m,l,i} - \angle\mathbf{d}^{2}_{l}[i]  \big)
\nonumber \\
&\geq
h^{4}_{l}(\mathbf{t}^1_{0})
+ \nabla^T h^{4}_{l}(\mathbf{t}^1_{0})(\mathbf{t}^1 - \mathbf{t}^1_{0})
- {0.5\bar{\tau}_{1,l}} (\mathbf{t}^1 - \mathbf{t}^1_{0})^T(\mathbf{t}^1 - \mathbf{t}^1_{0})\nonumber\\
& =-{0.5\bar{\tau}_{1,l}} \Vert\mathbf{t}^1\Vert^2_2 + (\mathbf{d}^{10}_{l})^T\mathbf{t}^1 + c_{5,5,l},\nonumber
\end{align}
\end{small}

\vspace{-0.2cm}
\noindent
with
$c_{5,5,l}$ being constant term,
$\bar{\tau}_{1,l} = \frac{8\pi^2}{\lambda^2} {\sum}_{i=1}^{L^{t,1}_{m,l}} \vert \mathbf{d}^{2}_{l}[i] \vert$,
$\mathbf{d}^{10}_{l} \triangleq   \bar{\tau}_{1,l} \mathbf{t}^1_{0} + h^{4}_{l}(\mathbf{t}^1_{0})$,
$\nabla h^{4}_{l}(\mathbf{t}^1_{0})
$
is the  gradient vector.

Next,
the  term
$\textrm{Re}\{ (d^{3}_{j})^{\ast}{g}_{r,j} \}$
can
be convexified by leveraging (\ref{Lemma1_2}) as follows
\vspace{-0.1cm}
\begin{small}
\begin{align}
&\textrm{Re}\{ (d^{3}_{j})^{\ast}{g}_{r,j} \}\triangleq g^{4}_{j}(\mathbf{t}^1)\\
&= \vert d^{3}_{j}\vert \vert\mathbf{g}_{r,j}\vert 
\cos\big(  ({2\pi}/{\lambda}) (\mathbf{t}^1)^T  \mathbf{a}^{5}_{m,j} - \angle d^{3}_{j} \big)\nonumber\\
& \leq g^{4}_{j}(\mathbf{t}^1_{0})
\!+\! \nabla^T g^{4}_{j}(\mathbf{t}^1_{0})(\mathbf{t}^1 \!-\! \mathbf{t}^1_{0})
\!+\! {0.5{\tau}_{2,j}} (\mathbf{t}^1 \!-\! \mathbf{t}^1_{0})^T(\mathbf{t}^1 \!-\! \mathbf{t}^1_{0})\nonumber\\
&=
{0.5\tau_{2,j}} \Vert\mathbf{t}^1\Vert^2_2 
+ (\mathbf{d}^{11}_{j})^T\mathbf{t}^1 + c_{5,6,j},\nonumber
\end{align}
\end{small}

\vspace{-0.2cm}
\noindent
with
$c_{5,6,j}$
being constant,
$\tau_{2,j} \triangleq \frac{8\pi^2}{\lambda^2}\vert d^{3}_{j}\vert \vert\mathbf{g}_{r,j}\vert$,
$\mathbf{d}^{11}_{j}\triangleq \nabla  g^{4}_{j}(\mathbf{t}^1_{0}) - \tau_{2,j}\mathbf{t}^1_{0}$,
$\nabla g^{4}_{j}(\mathbf{t}^1_{0})$
is
the gradient vector.

Based on the above transformation,
the upper-bound of the objective function (\ref{P11_obj_1}) 
is given as
 \vspace{-0.2cm}
\begin{small}
\begin{align}
&{\sum}_{l=1}^{K_u}
\big(
 (\mathbf{h}^{1}_{l})^H
 \mathbf{D}^{1}_{l}\mathbf{h}^{1}_{l} 
 +
 \textrm{Re}\{ (\mathbf{d}^{1}_{l})^H \mathbf{h}^{1}_{l}  \} \\
 &-
 \textrm{Re}\{ (\mathbf{d}^{2}_{l})^H \mathbf{h}^{1}_{l}  \}
\big)
+
{\sum}_{j=0}^{K_t} \textrm{Re}\{ (d^{3}_{j})^{\ast}\mathbf{g}_{r,j} \}+c_{5,1}\nonumber\\
&\leq
{\sum}_{l=1}^{K_u}
\big(
{0.5\tau_{1,l}} \Vert\mathbf{t}^1\Vert^2_2 + (\mathbf{d}^{9}_{l})^T\mathbf{t}^1 + c_{5,4,l}\nonumber\\
&-
(-{0.5\bar{\tau}_{1,l}} \Vert\mathbf{t}^1\Vert^2_2 + (\mathbf{d}^{10}_{l})^T\mathbf{t}^1 + \bar{c}_{5,5,l})
\big)\nonumber\\
&+
{\sum}_{j=0}^{K_t} (0.5{\tau_{2,j}} \Vert\mathbf{t}^1\Vert^2_2 
+ (\mathbf{d}^{11}_{j})^T\mathbf{t}^1 + c_{5,6,j})
+c_{5,1}\nonumber\\
&={\tau_{3}} \Vert\mathbf{t}^1\Vert^2_2 + (\mathbf{d}^{12})^T\mathbf{t}^1 + c_{5,7},\nonumber
\end{align}
\end{small}%

\vspace{-0.3cm}
\noindent
where 
$c_{5,7}$
is a constant term,
$\tau_{3} \triangleq {\sum}_{l=1}^{K_u}
\big({\tau_{1,l}}/{2} + {\bar{\tau}_{1,l}}/{2} \big)+
{\sum}_{j=0}^{K_t}({\tau_{2,j}}/{2})  $
and
$\mathbf{d}^{12}\triangleq {\sum}_{l=1}^{K_u}
(\mathbf{d}^{9}_{l}-\mathbf{d}^{10}_{l})
+
{\sum}_{j=0}^{K_t} \mathbf{d}^{11}_{j}$.


%
Next, we will construct the upper-bound of the sensing SINR constraint (\ref{P11_c_1}).
Firstly,
by leveraging the arguments presented  in (\ref{MM}) again,
 a tight upper-bound of  
$ (\mathbf{h}^{1}_{l})^H
 \mathbf{D}^{2}_{l}\mathbf{h}^{1}_{l} 
 +
 \textrm{Re}\{ (\mathbf{d}^{4}_{l})^H \mathbf{h}^{1}_{l}  \}
$,
 can be formulated as
  \vspace{-0.2cm}
\begin{small}
\begin{align}
& (\mathbf{h}^{1}_{l})^H
 \mathbf{D}^{2}_{l}\mathbf{h}^{1}_{l} 
 +
 \textrm{Re}\{ (\mathbf{d}^{4}_{l})^H \mathbf{h}^{1}_{l}\} \\
&\leq
\lambda_{max}(\mathbf{D}^{2}_{l})(\Vert  \mathbf{h}^{1}_{l} \Vert^2_2+\Vert  \mathbf{h}^{1}_{l,0} \Vert^2_2)
- 2\textrm{Re}\{\lambda_{max}(\mathbf{D}^{2}_{l}) (\mathbf{h}^{1}_{l0})^H \mathbf{h}^{1}_{l}  \}\nonumber  \\
&+2\textrm{Re}\{ (\mathbf{h}^{1}_{l,0})^H  \mathbf{D}^{2}_{l} (\mathbf{h}^{1}_{l}\!-\!\mathbf{h}^{1}_{l,0}  ) \}
\!\!+\!\!(\mathbf{h}^{1}_{l,0})^H  \mathbf{D}^{2}_{l}  \mathbf{h}^{1}_{l,0}
\!\!+\!\!\textrm{Re}\{ (\mathbf{d}^{4}_{l})^H \mathbf{h}^{1}_{l} \}\nonumber\\
& = \!\textrm{Re}\{ (\mathbf{d}^{13}_{l})^H\!\mathbf{h}^{1}_{l} \} \!\!+\!\! c_{5,8,l}
\!+\!
\textrm{Re}\{ (\mathbf{d}^{4}_{l})^H\! \mathbf{h}^{1}_{l} \}
\!\!=\!\textrm{Re}\{ (\mathbf{d}^{14}_{l})^H\!\mathbf{h}^{1}_{l} \} \!\!+\!\! c_{5,8,l},\nonumber
\end{align}
\end{small}

\vspace{-0.2cm}
\noindent
where
$c_{5,8,l} $ is constant,
$
\mathbf{d}^{13}_{l} \!\!\triangleq\!\! 2( (\mathbf{D}^{2}_{l})^H\mathbf{h}^{1}_{l,0}
\!-\! \lambda_{max}(\mathbf{D}^{2}_{l})\mathbf{h}^{1}_{l,0}  )
$
and
$\mathbf{d}^{14}_{l}\! \triangleq\! \mathbf{d}^{13}_{l} \!+\! \mathbf{d}^{2}_{l} $.
Furthermore,
by using the inequality (\ref{Lemma1_2}) in Lemma \ref{Lemma_1},
a convex  upper-bound of
$\textrm{Re}\{ (\mathbf{d}^{14}_{l})^H\mathbf{h}^{1}_{l} \}$
is given as follows
 \vspace{-0.2cm}
\begin{small}
\begin{align}
& \textrm{Re}\{ (\mathbf{d}^{14}_{l})^H\mathbf{h}^{1}_{l} \}
\triangleq h^{5}_{l}(\mathbf{t}^1)\\
&= {\sum}_{i=1}^{L^{t,1}_{m,l}} \vert \mathbf{d}^{14}_{l}[i] \vert
\cos\big(  ({2\pi}/{\lambda}) (\mathbf{t}^1)^T  \mathbf{a}^{r,1}_{m,l,i} - \angle\mathbf{d}^{14}_{l}[i]  \big)\nonumber \\
&\leq
h^{5}_{l}(\mathbf{t}^1_{0})
+ \nabla^T h^{5}_{l}(\mathbf{t}^1_{0})(\mathbf{t}^1 - \mathbf{t}^1_{0})
+ {0.5\tau_{4,l}} (\mathbf{t}^1 - \mathbf{t}^1_{0})^T(\mathbf{t}^1 - \mathbf{t}^1_{0})\nonumber\\
& ={0.5\tau_{4,l}} \Vert\mathbf{t}^1\Vert^2_2 + (\mathbf{d}^{15}_{l})^T\mathbf{t}^1 + c_{5,9,l},\nonumber
\end{align}
\end{small}

\vspace{-0.3cm}
\noindent
where
$c_{5,9,l}$ is a constant term,
$\tau_{4,l} = \frac{8\pi^2}{\lambda^2} {\sum}_{i=1}^{L^{t,1}_{m,l}} \vert \mathbf{d}^{14}_{l}[i] \vert$,
$\mathbf{d}^{15}_{l} \triangleq \nabla h^{5}_{l}(\mathbf{t}^1_{0}) - \tau_{4,l} \mathbf{t}^1_{0}$,
$\nabla h^{5}_{l}(\mathbf{t}^1_{0})$
is
the gradient vector.%

Similarly,
 the  term
$\textrm{Re}\{ (d^{5}_{j})^{\ast}{g}_{r,j} \}$
 can
be convexified by leveraging (\ref{Lemma1_2}) as follows
 \vspace{-0.2cm}
\begin{small}
\begin{align}
&\textrm{Re}\{ (d^{5}_{j})^{\ast}{g}_{r,j} \}
\triangleq g^{5}_{j}(\mathbf{t}^1)\\
&= \vert d^{5}_{j}\vert \vert\mathbf{g}_{r,j}\vert 
\cos\big(  ({2\pi}/{\lambda}) (\mathbf{t}^1)^T  \mathbf{a}^{5}_{m,j} - \angle d^{5}_{j} \big)\nonumber\\
&\leq
{0.5\tau_{5,j}} \Vert\mathbf{t}^1\Vert^2_2 + (\mathbf{d}^{16}_{j})^T\mathbf{t}^1 + c_{5,10,j},\nonumber
\end{align}
\end{small}

\vspace{-0.2cm}
\noindent
with
$c_{5,10,j}$
being constants,
$\tau_{5,j} \triangleq \frac{8\pi^2}{\lambda^2}\vert d^{5}_{j}\vert \vert{g}_{r,j}\vert$,
$\mathbf{d}^{16}_{j}\triangleq \nabla  g^{5}_{j}(\mathbf{t}^1_{0}) - \tau_{5,j}\mathbf{t}^1_{0}$,
$\nabla  g^{5}_{j}(\mathbf{t}^1_{0})$
is
the gradient vector.%

And then, a convex lower bound for
$ \text{Re}\{(d^{6})^{\ast}{g}_{r,0} \}$ by applying
(\ref{Lemma1_1}) in Lemma \ref{Lemma_1} is given as follows
 \vspace{-0.2cm}
\begin{small}
\begin{align}
&\text{Re}\{(d^{6})^{\ast}{g}_{r,0}(\mathbf{t}^1) \}
\triangleq g^{6}(\mathbf{t}^1)\\
&=
\vert d^{6}  \vert \vert \mathbf{g}_{r,0} \vert   
 \cos\big(  ({2\pi}/{\lambda}) (\mathbf{t}^1)^T  \mathbf{a}^{5}_{m,0} - \angle d^{6} \big)\nonumber\\
&\geq - {0.5\tau_{6}} \Vert\mathbf{t}^1\Vert^2_2 + (\mathbf{d}^{17})^T\mathbf{t}^1_{m,n} + c_{5,11},\nonumber
\end{align}
\end{small}%

\vspace{-0.2cm}
\noindent
where
$c_{5,11}$ is constant,
$ \tau_{6} \triangleq \frac{8\pi^2}{\lambda^2} \vert d^{6}  \vert \vert \mathbf{g}_{r,0}  \vert$,
$ \mathbf{d}^{17} \triangleq \nabla g^{6}(\mathbf{t}^1) + \tau_{6} \mathbf{t}^1_{0} $,
$\nabla g^{6}(\mathbf{t}^0)$ is the
the gradient vector.

Therefore, 
a convex upper-bound of
the sensing SINR constraint (\ref{P11_c_1}) can be expressed as
\vspace{-0.1cm}
\begin{small}
\begin{align}
&{\sum}_{l=1}^{K_u}
\big(
 (\mathbf{h}^{1}_{l})^H
 \mathbf{D}^{2}_{l}\mathbf{h}^{1}_{l} 
 +
 \textrm{Re}\{ (\mathbf{d}^{4}_{l})^H \mathbf{h}^{1}_{l}  \}
\big) \\
&+{\sum}_{j=1}^{K_t}
\big(
\textrm{Re}\{
(d^{5}_{j})^{\ast}{g}_{r,j}
\}
\big)
 - \textrm{Re}\{
(d^{6})^{\ast}\mathbf{g}_{r,0}
\}+ c_{5,2}\nonumber\\
&\leq
{\sum}_{l=1}^{K_u}
\big(
{0.5\tau_{4,l}} \Vert\mathbf{t}^1\Vert^2_2 + (\mathbf{d}^{15}_{l})^T\mathbf{t}^1 + c_{5,9,l}+ c_{5,8,l}
\big)\nonumber\\
&+{\sum}_{j=1}^{K_t}({0.5\tau_{5,j}} \Vert\mathbf{t}^1\Vert^2_2 + (\mathbf{d}^{16}_{j})^T\mathbf{t}^1 + c_{5,10,j})\nonumber\\
&-(- {0.5\tau_{6}} \Vert\mathbf{t}^1\Vert^2_2 + (\mathbf{d}^{17})^T\mathbf{t}^1_{m,n} + c_{5,11})+ c_{5,2},\nonumber\\
&=
\tau_{7}\Vert\mathbf{t}^1\Vert^2_2 
+ (\mathbf{d}^{18})^T\mathbf{t}^1 + c_{5,12},\nonumber
\end{align}
\end{small}%

\vspace{-0.2cm}
\noindent
where 
$c_{5,12}$ is constant,
$\tau_{7} \triangleq
{\sum}_{l=1}^{K_u}
(
{\tau_{4,l}}/{2} 
)
+{\sum}_{j=1}^{K_t}
(
{\tau_{5,j}}/{2} 
)
 - (- {\tau_{6}}/{2} )$,
$\mathbf{d}^{18}\triangleq
{\sum}_{l=1}^{K_u}
\mathbf{d}^{15}_{l}
+{\sum}_{j=1}^{K_t}
 \mathbf{d}^{16}_{j}
- ( \mathbf{d}^{17} )$.
Inspired by the MM framework,
the constraint (\ref{P11_c_2}) can be linearized as follows
\vspace{-0.2cm}
\begin{small}
\begin{align}
\Vert \mathbf{t}^1 - \mathbf{t}^1_{m,i} \Vert_2
\geq \frac{(\mathbf{t}^1_{0} - \mathbf{t}^1_{m,i})^T}{\Vert \mathbf{t}^1_{0} - \mathbf{t}^1_{m,i} \Vert_2}
(\mathbf{t}^1 - \mathbf{t}^1_{m,i}).
\end{align}
\end{small}
\vspace{-0.2cm}
Therefore,
the optimization problem (P11) is rewritten as
\begin{small}
\begin{subequations}
\begin{align}
\textrm{(P12)}:&\mathop{\textrm{min}}
\limits_{
\mathbf{t}^1
}{\tau_{3}} \Vert\mathbf{t}^1\Vert^2_2 + (\mathbf{d}^{12})^T\mathbf{t}^1 + c_{5,7}
\\
\textrm{s.t.}\
& \tau_{7}\Vert\mathbf{t}^1\Vert^2_2
+ (\mathbf{d}^{18})^T\mathbf{t}^1
+ c_{5,12}
\leq 0,\\
& \frac{(\mathbf{t}^1_{0} - \mathbf{t}^1_{m,i})^T}{\Vert \mathbf{t}^1_{0} - \mathbf{t}^1_{m,i} \Vert_2}
(\mathbf{t}^1 - \mathbf{t}^1_{m,i}) \geq\!\! D_{t},\! \forall \!n, \!i\! \in\! \mathcal{N}_r, n \!\neq\! i,\\
& \mathbf{t}^1 \in \mathcal{C}. \label{P12_c_3}
\end{align}
\end{subequations}
\end{small}

\vspace{-0.2cm}
\noindent
Obviously,
the problem (P12) is convex 
and its optimal solution  can be obtained by CVX.

\subsection{Optimizing DL User MA's Position}

In this subsection,
we will discuss the optimization of the MA position $\mathbf{r}_{0,k}$ for the $k$-th DL user.
With given other variables,
the problem w.r.t. $\mathbf{r}_{0,k}$
is formulated as
 \vspace{-0.2cm}
\begin{small}
\begin{subequations}
\begin{align}
\textrm{(P13)}:&\mathop{\textrm{min}}
\limits_{
\mathbf{r}^0_{k} \in \mathcal{C}
}
{\sum}_{m=1}^{M_t}{\sum}_{n=1}^{M_t}
(\mathbf{h}^{0,0}_{m,k})^H  \mathbf{Q}^{1}_{k,m,n}  \mathbf{h}^{0,0}_{n,k}\label{P13_obj_1}
\\
&+ {\sum}_{m=1}^{M_t}((\mathbf{h}^{0,0}_{m,k})^H  \mathbf{Q}^{2}_{k,m}  \mathbf{h}^{0,0}_{m,k}
\!-\!\! \textrm{ Re } \{  (\mathbf{q}^1_{k,m})^H \mathbf{h}^{0,0}_{m,k} \}
)\nonumber\\
&+ {\sum}_{l=1}^{K_u} (\mathbf{h}^{3}_{k,l})^{H} \mathbf{Q}^{3}_{k,l} \mathbf{h}^{3}_{k,l}
+ c_{6,1,k},\nonumber
\end{align}
\end{subequations}
\end{small}
\vspace{-0.2cm}
\noindent
with
$c_{6,1,k}$ being constant
and
other new coefficients are given as follows
\vspace{-0.1cm}
\begin{small}
\begin{align}
& \mathbf{q}^1_{k,m}
\triangleq
2 \omega_{d,k}\beta_{d,k}^{\ast}
\mathbf{\Sigma}^{0}_{m,k}\mathbf{H}^{0}_{m,k}\mathbf{w}_{m,k},\\
& \mathbf{Q}^3_{k,l}
\triangleq
\omega_{d,k}\vert\beta_{d,k}\vert^2
q_l \mathbf{\Sigma}^{2}_{k,l}\mathbf{h}^{2}_{k,l} (\mathbf{h}^{2}_{k,l})^H (\mathbf{\Sigma}^{2}_{k,l})^H,\nonumber\\
& \mathbf{Q}^2_{k,m}
\triangleq
\omega_{d,k}\vert\beta_{d,k}\vert^2
\mathbf{\Sigma}^{0}_{m,k}\mathbf{H}^{0}_{m,k}\mathbf{W}^r_{m}\mathbf{W}^r_{m} (\mathbf{H}^{0}_{m,k})^H(\mathbf{\Sigma}^{0}_{m,k})^H,\nonumber\\
& \mathbf{Q}^1_{k,m,n}\! \triangleq\!\!\!
{\sum}_{i=1}^{K_d}
\!\omega_{d,k}\vert\beta_{d,k}\vert^2
\mathbf{\Sigma}^{0}_{m,k}\mathbf{H}^{0}_{m,k}\mathbf{w}_{m,i}
\mathbf{w}_{n,i}(\mathbf{H}^{0}_{n,k})^{\!H}\!\!(\mathbf{\Sigma}^{0}_{n,k})^H.\nonumber
\end{align}
\end{small}

Obviously,
obtaining the optimal solution to the problem (P13) is challenging due to
the non-convex nature of the objective function (\ref{P13_obj_1}).
Therefore, we still employ the MM methodology to convexify it.
Due to space constraints,
we omit the transformation process.

The update of $\mathbf{r}_{0,k}$ is achieved by solving a convex surrogate of the original problem (P13),
as follows
 \vspace{-0.2cm}
\begin{subequations}
\begin{align}
\textrm{(P14)}:&\mathop{\textrm{min}}
\limits_{
\mathbf{r}_{0,k} \in \mathcal{C}
}
{\psi_{7,k}} \Vert\mathbf{r}_{0,k}\Vert^2_2 + (\mathbf{q}^{2}_{k})^T\mathbf{r}_{0,k} + c_{6,2,k}\label{P14_obj_1}
\end{align}
\end{subequations}
The problem (P14) is convex and can be solved by CVX.
Since the units of $\mathbf{r}_{0,k}$, i.e., $x^d_{k}$ and $y^d_{k}$,
are not coupled,
the subproblem w.r.t. $x^d_{k}$
can be written as
 \vspace{-0.2cm}
\begin{subequations}
\begin{align}
\textrm{(P15)}:&\mathop{\textrm{min}}
\limits_{
x^d_{k}
}
{\psi_{7,k}} (x^d_{k})^2  + \mathbf{q}^{2}_{k}[1]x^d_{k}  \label{P15_obj_1}\\
\textrm{s.t.}\
& -{A}/{2} \leq  x^d_{k}\leq {A}/{2},\label{P15_c_1}
\end{align}
\end{subequations}
\vspace{-0.2cm}
Since $\psi_{7,k}>0$,
we can obtain the optimal solution $(x^d_{k})^{\star}$
by directly determining the position of  the axis of symmetry of the function (\ref{P15_obj_1}),
which are given as
\begin{small}
\begin{align}
(x^d_{k})^{\star}\!=\!
\left\{
\begin{aligned}
     -{A}/{2},\ &\textrm{if}\  -{\mathbf{q}^{2}_{k}[1]}/({2\psi_{7,k}}) < -{A}/{2},\\
     -{\mathbf{q}^{2}_{k}[1]}/({2\psi_{7,k}}), \ &\textrm{if}\  -{A}/{2} \leq -{\mathbf{q}^{2}_{k}[1]}/({2\psi_{7,k}}) \leq {A}/{2},\\
     {A}/{2}, \ &\textrm{if} \ -{A}/{2} < -{\mathbf{q}^{2}_{k}[1]}/({2\psi_{7,k}}).
\end{aligned}
\right.\label{R1_x}
\end{align}
\end{small}
Similarly,
the optimal solution $(y^d_{k})^{\star}$
can be directly formulated as
\begin{small}
\begin{align}
(y^d_{k})^{\star}\!=\!
\left\{
\begin{aligned}
     -{A}/{2},\ &\textrm{if}\  -{\mathbf{q}^{2}_{k}[2]}/({2\psi_{7,k}}) < -{A}/{2},\\
     -{\mathbf{q}^{2}_{k}[1]}/({2\psi_{7,k}}), \ &\textrm{if}\  -{A}/{2} \leq -{\mathbf{q}^{2}_{k}[2]}/({2\psi_{7,k}}) \leq {A}/{2},\\
     {A}/{2}, \ &\textrm{if} \ -{A}/{2} < -{\mathbf{q}^{2}_{k}[2]}/({2\psi_{7,k}}).
\end{aligned}
\right.\label{R1_y}
\end{align}
\end{small}

\subsection{Optimizing UL User MA's Position}

In this subsection,
we will investigate the update of the MA's position 
${
\mathbf{r}_{1,l}
}$ 
for the $l$-th UL user.
Firstly,
the functions $\tilde{\mathrm{R}}_{d,k}$ and $\tilde{\mathrm{R}}_{u,l}$
can be respectively rewritten as
\vspace{-0.2cm}
\begin{small}
\begin{align}
&{\sum}_{k=1}^{K_d}\tilde{\mathrm{R}}_{d,k}
\! \triangleq\!
  {\sum}_{k=1}^{K_d}\!-\!(\mathbf{h}^{2}_{k,l}(\mathbf{r}_{1,l}))^H \mathbf{Q}^{4}_{k,l} \mathbf{h}^{2}_{k,l}(\mathbf{r}_{1,l})\! +\! c_{7,1,l} ,\\
& {\sum}_{l=1}^{K_u} \tilde{\mathrm{R}}_{u,l}
\!\triangleq\!\!
{\sum}_{m=1}^{M_r}
{\sum}_{n=1}^{M_r}
(\mathbf{h}^{0,1}_{m,l}(\mathbf{r}_{1,l}))^H
\mathbf{Q}^{5}_{l,m,n}
\mathbf{h}^{0,1}_{n,l}(\mathbf{r}_{1,l})\\
& - {\sum}_{m=1}^{M_r}\textrm{Re}\{ \mathbf{q}^{3}_{l,m}\mathbf{h}^{0,1}_{m,l}(\mathbf{r}_{1,l}) \}
+c_{7,2,l},\nonumber
\end{align}
\end{small}
\noindent
with
$c_{7,1,l}$
and
$c_{7,2,l}$
being constant terms,
and
 we define
 \vspace{-0.2cm}
 \begin{small}
 \begin{align}
 & \mathbf{q}^{3}_{l,m}
 \triangleq
 2 \omega_{d,k} \beta_{d,k} \mathbf{\Sigma}^{1}_{m,l}\mathbf{H}^{1}_{m,i}\mathbf{u}_{m,l},\\
 &\mathbf{Q}^{4}_{k,l}
 \triangleq
 \omega_{d,k} \vert\beta_{d,k}\vert^2 q_l ( \mathbf{\Sigma}^{2}_{k,l} )^H \mathbf{h}^{3}_{k,l} 
 (\mathbf{h}^{3}_{k,l})^H \mathbf{\Sigma}^{2}_{k,l},\nonumber\\
  & {\mathbf{Q}}^{5}_{i,m,n}
 \!\triangleq\!\!
 {\sum}_{l=1}^{K_u}
 \omega_{d,k} \vert\beta_{d,k}\vert^2
 \mathbf{\Sigma}^{1}_{m,i} \mathbf{H}^{1}_{m,i}\mathbf{u}_{m,l}\mathbf{u}_{n,l}^H
 (\mathbf{H}^{1}_{n,i})^H\!(\mathbf{\Sigma}^{1}_{n,i})^H.\nonumber
 \end{align}
 \end{small}
Furthermore,
 the sensing SINR constraint (\ref{P1_c_1}) can be equivalent rewritten as
 \vspace{-0.2cm}
 \begin{small}
 \begin{align}
&\textrm{SINR}_{t} \geq \Gamma_r  \Longleftrightarrow\\
& {\sum}_{m=1}^{M_r}
{\sum}_{n=1}^{M_r}
(\mathbf{h}^{0,1}_{m,l}(\mathbf{r}_{1,l}))^H
\mathbf{Q}^{6}_{l,m,n}
\mathbf{h}^{0,1}_{n,l}(\mathbf{r}_{1,l})
+
c_{7,3,l},\nonumber
 \end{align}
 \end{small}%

\vspace{-0.2cm}
 \noindent
where $c_{7,3,l}$ is constant and
$
 \mathbf{Q}^{6}_{l,m,n}
 \triangleq
 \mathbf{\Sigma}^{1}_{m,l} \mathbf{H}^{1}_{m,l}\mathbf{u}_{m,0}\mathbf{u}_{n,0}^H
 (\mathbf{H}^{1}_{n,l})^H(\mathbf{\Sigma}^{1}_{n,l})^H$.
Therefore, the problem w.r.t. $\mathbf{r}_{1,l}$ can be rewritten as
 \vspace{-0.2cm}
\begin{subequations}
\begin{align}
\textrm{(P16)}:&\mathop{\textrm{min}}
\limits_{
\mathbf{r}_{1,l}
}
\!{\sum}_{k=1}^{K_d}
 \!
 (\mathbf{h}^{2}_{k,l})^H \mathbf{Q}^{4}_{k,l} \mathbf{h}^{2}_{k,l}
\!\! -\!\!\! {\sum}_{m=1}^{M_r}\textrm{Re}\{ \mathbf{q}^{3}_{l,m}\mathbf{h}^{0,1}_{m,l} \}
\nonumber \\
 &+
{\sum}_{m=1}^{M_r}
{\sum}_{n=1}^{M_r}
(\mathbf{h}^{0,1}_{m,l})^H
\mathbf{Q}^{5}_{l,m,n}
\mathbf{h}^{0,1}_{n,l}+\hat{c}_{7,1,l}\label{P16_obj_1}\\
\textrm{s.t.}\
& \!{\sum}_{m=1}^{M_r}
{\sum}_{n=1}^{M_r}
(\mathbf{h}^{0,1}_{m,l})^H
\mathbf{Q}^{6}_{l,m,n}
\mathbf{h}^{0,1}_{n,l}
\!+\!
c_{7,3,l}\! \leq\! 0, \label{P16_c_1}\\
& \mathbf{r}_{1,l} \in \mathcal{C},\label{P16_c_2}
\end{align}
\end{subequations}
with
$\hat{c}_{7,1,l}$
being a constant term.
Unfortunately,
this problem is still difficult to 
solve due to the non-convex objective (\ref{P16_obj_1}) and constraint (\ref{P16_c_1})
w.r.t $\mathbf{r}_{1,l}$.
To address these issues,
we will construct the convex surrogate functions of (\ref{P16_obj_1}) and (\ref{P16_c_1})
by  applying the MM method,
which the derivation details are omitted for the space constraints,
and then recast the problem (P11) into an explicit form of $\mathbf{r}_{1,l}$,
which is given as 
 \vspace{-0.2cm}
\begin{subequations}
\begin{align}
\textrm{(P17)}:&\mathop{\textrm{min}}
\limits_{
\mathbf{r}_{1,l}
}
\
\psi_{8,l} \Vert\mathbf{r}_{1,l}\Vert^2_2
+ (\mathbf{q}^{4}_{l})^T\mathbf{r}_{1,l} + c_{7,4,l}
\label{P17_obj_1}\\
\textrm{s.t.}\
& \psi_{10,l} \Vert\mathbf{r}_{1,l}\Vert^2_2
+ (\mathbf{q}^{5}_{l})^T\mathbf{r}_{1,l} + c_{7,5,l} \leq 0 , \label{P17_c_1}\\
& \mathbf{r}_{1,l} \in \mathcal{C}.\label{P17_c_2}
\end{align}
\end{subequations}

Now,
the problem (P17) is convex, which can be solved optimally by CVX.

Moreover,
the overall algorithm to solve (P1) is specified in Algorithm \ref{alg:A}.
 $\eta = \{\{\mathbf{w}_{m,k}, \mathbf{W}^r_m\},\{\mathbf{u}_{p,i}\},
\{q_l\},
\{\mathbf{t}^0_{m,n}, 
\mathbf{t}^1_{p,j},
\mathbf{r}_{0,k}, 
\mathbf{r}_{1,l}\}\}  $
denote the collection of all variables.
And
let $g(\eta )$ represent the sum-rate for both downlink and uplink users.

\begin{algorithm}[t]
\caption{Overall Algorithm to Solve $(\mathrm{P1})$}
\label{alg:A}
\begin{algorithmic}[1]
\STATE initialize $i=0$;
\STATE randomly generate feasible
$\{\mathbf{w}^{0}_{m,k}, (\mathbf{W}^r_m)^{0}\}$,
$\{\mathbf{u}_{p,j}^{0}\}$,\
$\{q_l\}$,\
$\{(\mathbf{t}^0_{m,n})^{0}, (\mathbf{t}^1_{m,n})^{0}\}$
and
$\{\mathbf{r}_{0,k}^{0}, \mathbf{r}_{1,l}^{0}\}$;
\REPEAT
\STATE update
$\{\omega_{d,k}\}$,
$\{\beta_{d,k}\}$,
$\{\omega_{u,l}\}$
and
$\{\beta_{u,l}\}$,
by (\ref{omega_dk}),  (\ref{beta_dk}), (\ref{omega_ul}) and (\ref{beta_ul}), respectively;
\STATE  update $\{\mathbf{w}^{i+1}_{m,k}, (\mathbf{W}^r_m)^{i+1}\}$ by solving (P3);
\STATE  update $\{\mathbf{u}_{p,l}^{i+1}\}, \forall l \in \mathcal{K}_u $ by equation (\ref{U_l});
\STATE  update $\{\mathbf{u}_{p,0}^{i+1}\}$ by solving (P8);
\STATE  update $\{(\mathbf{t}^0_{m,n})^{i+1}\}$ by solving ({P10});
\STATE  update $\{(\mathbf{t}^1_{m,n})^{i+1}\}$ by solving ({P12});
\STATE  update $\{\mathbf{r}_{0,k}^{i+1}\}$ by equations (\ref{R1_x}) and (\ref{R1_y});
\STATE  update $\{\mathbf{r}_{1,l}^{i+1}\}$ by solving ({P17});
\STATE $i++$;
\UNTIL{$
\bigg\vert
\frac{g(\eta^{i+1} )
-
g(\eta^{i} )}
{g(\eta^{i})}
\bigg\vert
\leq 
\epsilon,$}
\end{algorithmic}
\end{algorithm}

\subsection{Complexity}

In the following, we will discuss the complexity of our
proposed algorithms. According to the complexity analysis in \cite{ref_Complexity},
the complexity of solving SOCP problems (P3), (P10), (P12) and (P17) are
$\mathcal{O}( C_1M_t^{3.5}N_t^{6})$,
$\mathcal{O}( C_2M_tN_t^{1.5} )$,
$ \mathcal{O}(    C_3M_rN_r^{1.5}     )   $
and
$ \mathcal{O}(76C_4) $, respectively,
where $C_1$,
$C_2$,
$C_3$
and 
$C_4$
 are respectively the number
of iterations of solving problem (P3), (P10), (P12) and (P17) by the MM method.
The complexity of solving (P4) is $\mathcal{O}(K_u^{3.5})$.
Therefore, the total computational complexity of Algorithm \ref{alg:A}
is approximately given as 
$\mathcal{O}( C_5   ( C_1M_t^{3.5}N_t^{6} + K_u^{3.5} )  )$
with $C_5$ represented as the number of iterations to solve
problem (P1).%

\section{Numerical Results}

\begin{figure}[t]
	\centering
	\includegraphics[width=.30\textwidth]{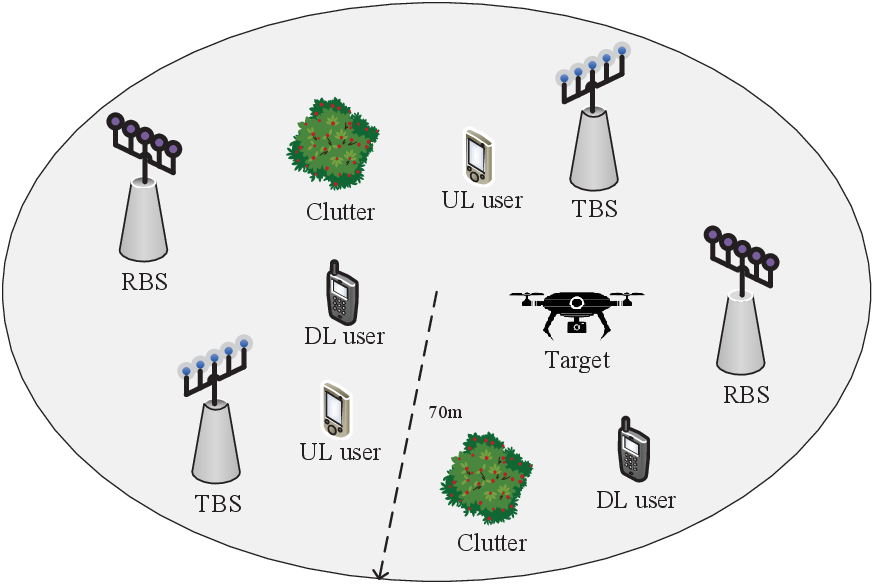}
	\caption{The simulation setup.}
	\label{fig.2}
\end{figure}

In this section,
we present numerical results to validate our proposals.
The experimental setup is illustrated in Fig. \ref{fig.2},
featuring multiple TBS and RBS units jointly attempting to detect a target amidst clutters
while simultaneously respectively serving three DL and three UL users.
In the experiment,
TBSs, RBSs, DL users, UL users, one target, and clutters
are randomly located within
a circular area
with a radius of $70$ meters (m).
We employ a geometric channel model \cite{ref_channel model}
 for the link concerning each user,
wherein the number of transmit and receive channel paths
is consistently identical, i.e.,
$L^{t,0}_{m,k} = L^{r,0}_{m,k} = L^{t,1}_{p,l} = L^{r,1}_{p,l} = L^{t,2}_{k,l} = L^{r,2}_{k,l} \triangleq L = 6,
\forall m \in \mathcal{M}_t,
\forall p \in \mathcal{M}_r,
\forall k \in \mathcal{K}_d,
\forall l \in \mathcal{K}_u
$.%
Thus, the path-response matrix for each user is diagonal,
i.e.,
the path-response matrix for between the $m$-th TBS and the $k$-th DL user
is
represented as
$\mathbf{\Sigma}^{0}_{m,k} = \textrm{diag}\{\sigma^{0}_{m,k,1}, \dots, \sigma^{0}_{m,k,L}  \}$,
where each
$\mathbf{\sigma}^{0}_{m,k,l}$ satisfying $\sigma^{0}_{m,k,l} \sim \mathcal{CN}(0, {(c^{0}_{m,k})^2}/{L}), l= 1, \dots, L $ \cite{ref_channel model},
and
$(c^{0}_{m,k})^2 = C_0d_{m,k}^{-\alpha_{loss}}$,
where $C_0$  corresponds to the path loss at the reference distance of $1$ m,
$d_{m,k}$ is the propagation distance between the $m$-th TBS and the $k$-th DL user,
$\alpha_{loss} = 2.8$ is the path-loss exponent.
The elevation and azimuth of both AoAs and AoDs 
 are assumed to be independent and identically distributed
variables following the uniform distribution over $[ -\frac{\pi}{2}, \frac{\pi}{2}]$.
The TBS-target/clutters and RBS-target/clutters are modeled as line-of-sight (LoS) channels.
The moving regions for MAs
at both the BS and users are set as  square areas of size $[-\frac{A}{2}, \frac{A}{2}] \times [-\frac{A}{2}, \frac{A}{2}]$.
In addition, $M_t = 2$ TBS equip $N_t = 4$ MAs and
$M_r = 2$ RBS equip $N_r = 4$ MAs.
The transmit power of each TBS  is set as $30$dBm.
The noise power
and the predefined target detection level of RBS are set as
$\sigma^2_{d,k} = \sigma^2_{r} = -80$dBm
 and $\Gamma_r = 3$dB,
 respectively.

\begin{figure}[t]
	\centering
	\includegraphics[scale=0.60]{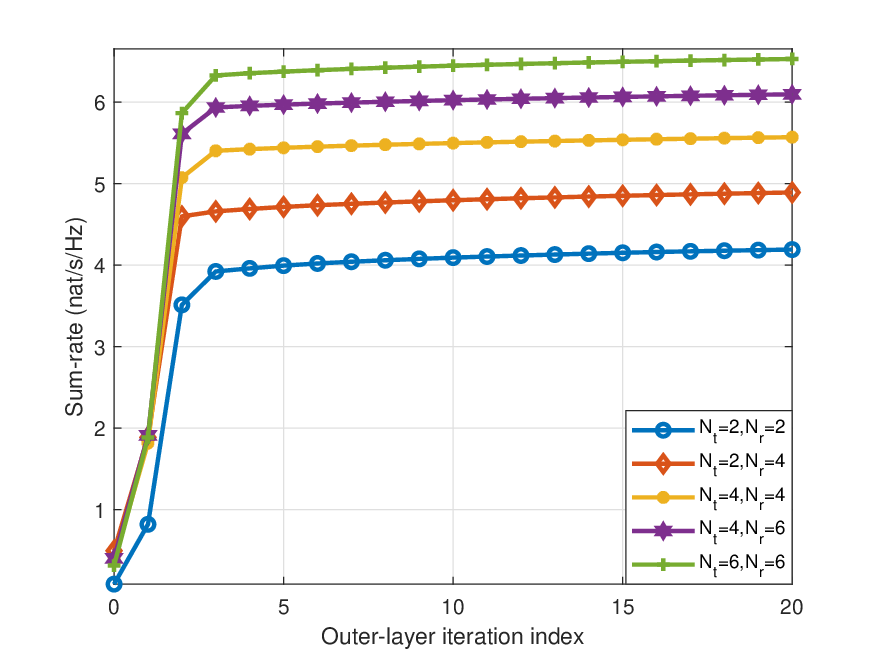}
	\caption{ Convergence of Alg. 1.}
	\label{fig.3}
\end{figure}

Fig. \ref{fig.3} examines the convergence behavior of the overall solution in Alg. \ref{alg:A}.
The obtained sum-rate iterates under varying configurations of $N_t$ and $N_r$
are illustrated in the plot.
As shown in
Fig. \ref{fig.3},
the proposed algorithm can yield monotonic improvement in sum-rate
and generally achieves significant beamforming gain within the first 10 iterations.

\begin{figure}[t]
    \centering
\includegraphics[scale=0.60]{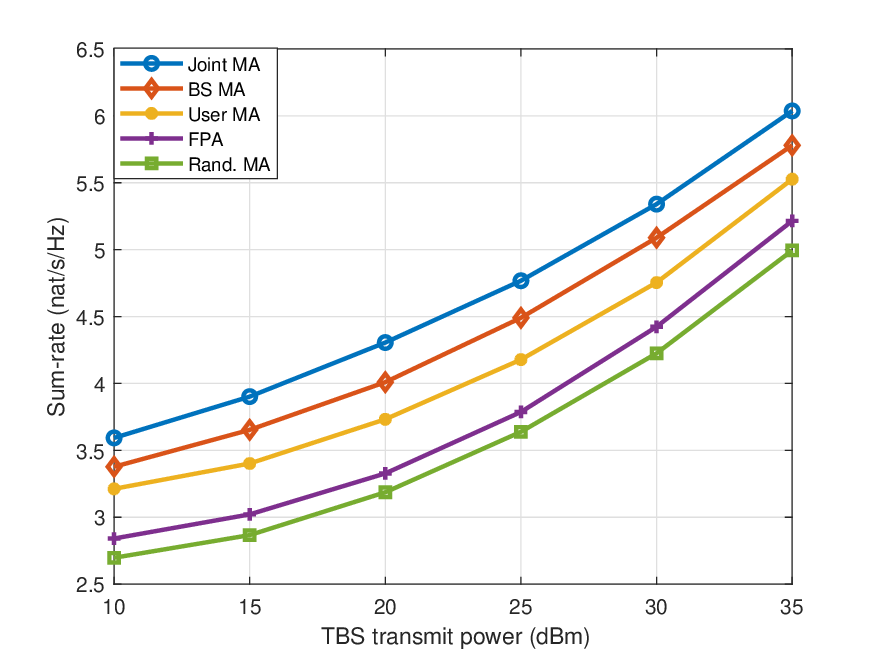}
	\caption{ The impact of TBS transmit power.}
	\label{fig.4}
\end{figure}

Alg. \ref{alg:A}  is labeled by ``Joint MA''.
For comparison in sum-rate enhancement,
we consider the following four foundational scenarios:
1) BS MA:
All BSs are equipped with the MAs,
while all users respectively employ an MA with fixed;
2) User MA:
All users respectively employ an MA,
while
the MAs of all BSs are fixed;
3) Rand. MA:
The MAs of all BSs and users have been fixed;
4)
FPA:
Both BSs and users are equipped with the FPA-based array.%

Fig. \ref{fig.4}
 compares the sum-rate for different
schemes versus the transmit power of all TBS.
The sum-rate of all schemes
increases significantly with the transmit power
and 
the three MA-based schemes, i.e., ``Joint MA", ``BS MA'', and ``User MA'' cases, 
significantly
improve the sum-rate compared to both ``FPA'' and ``Rand. MA'' cases.
For the ``Joint MA'' case, 
when the transmit power increases from 10 dBm to 35 dBm, 
the sum-rate increases by about 67\%.
In particular,
 the ``Joint MA" case shows a notable advantage over all other cases regarding sum-rate.
Specifically,
under the 20 dBm case,
 the ``Joint MA" case outperforms ``FPA'' and ``Rand. MA'' cases  by approximately 30\% and 35\%, respectively.
This is anticipated,
as the proposed ``Joint MA" scheme exploits the most spatial DoFs, 
and the other four schemes face performance loss due to less flexibility in channel reconfiguration.

\begin{figure}[t]
    \centering
\includegraphics[scale=0.60]{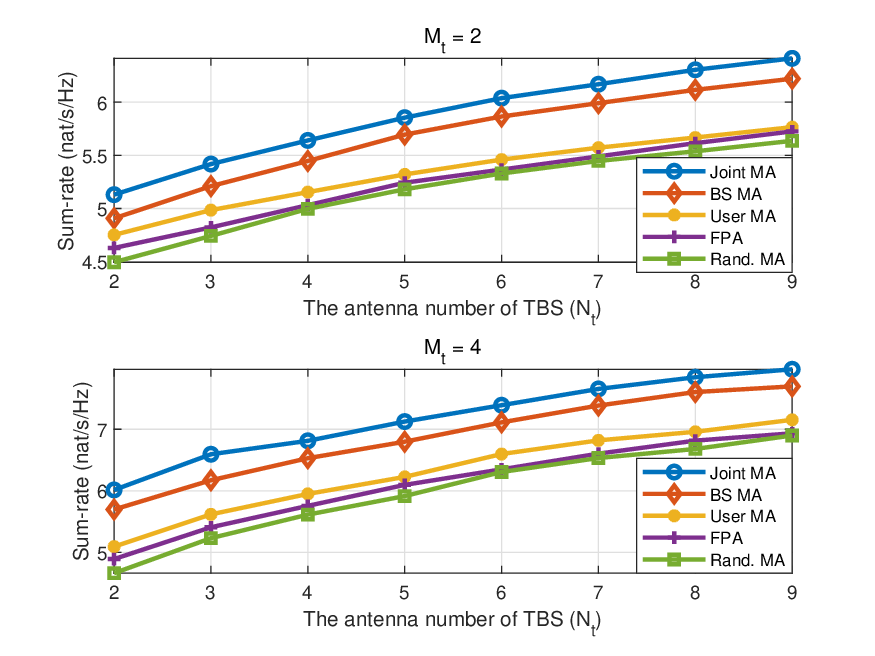}
	\caption{The impact of TBS antenna number.}
	\label{fig.5}
\end{figure}

\begin{figure}[t]
    \centering
\includegraphics[scale=0.60]{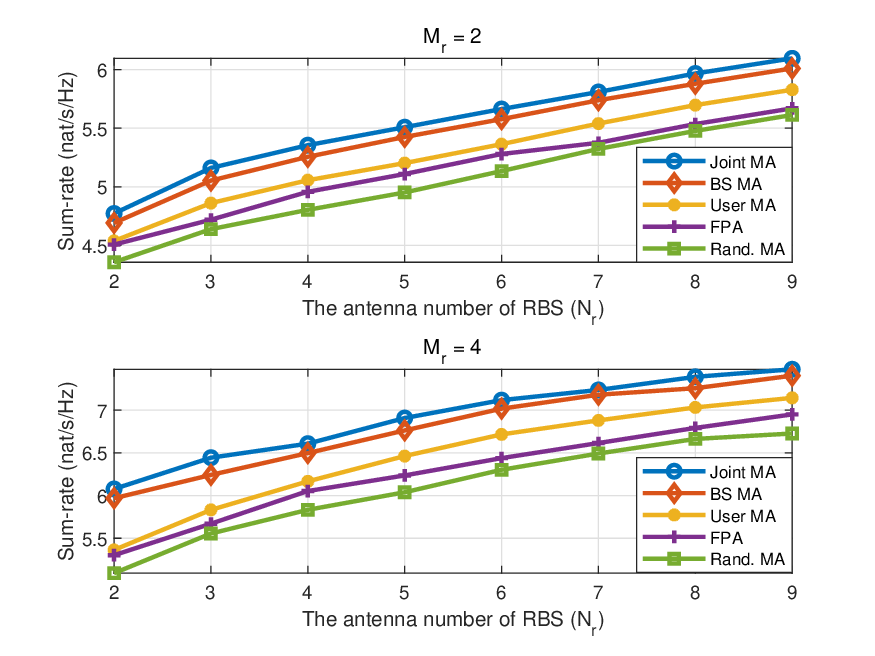}
	\caption{The impact of RBS antenna number.}
	\label{fig.6}
\end{figure}

Fig. \ref{fig.5} illustrates the impact of the number of antenna at the TBSs.
The upper and lower subplots correspond to different numbers of TBSs, respectively.
In our test,
the number of antenna 
$N_t$ varies from 2 to 9.
As reflected by Fig. \ref{fig.5},
increasing $N_t$
can improve
beamforming gain for all schemes.
Moreover,
the ``Joint MA" scheme boosts the system's sum-rate more significantly
compared to all other schemes.
For the ``Joint MA'' case, 
when the transmit antenna number increases from 2 to 9, 
the sum-rate improves by approximately 25\% for the 
``$M_t = 2$'' scenario and by about 33\% for the 
``$M_t = 4$'' scenario.

Fig. \ref{fig.6} shows the impact of the number of antenna at the RBSs.
As can be seen in both upper and lower subplots, 
which correspond to varying quantities of RBSs,
sum-rate increases when $N_r$ increases.
Compared to other benchmarks,
the deployment of MA in both BS and user can significantly boost the sum-rate.

\begin{figure}[t]
    \centering
\includegraphics[scale=0.60]{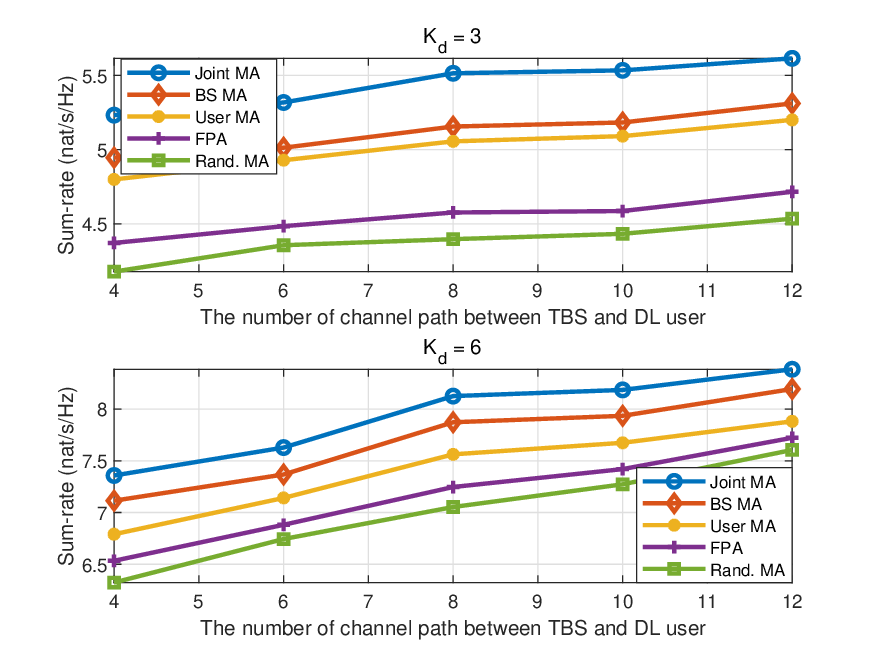}
	\caption{ The impact of the number of paths between TBS and DL user.}
	\label{fig.7}
\end{figure}

\begin{figure}[t]
    \centering
\includegraphics[scale=0.60]{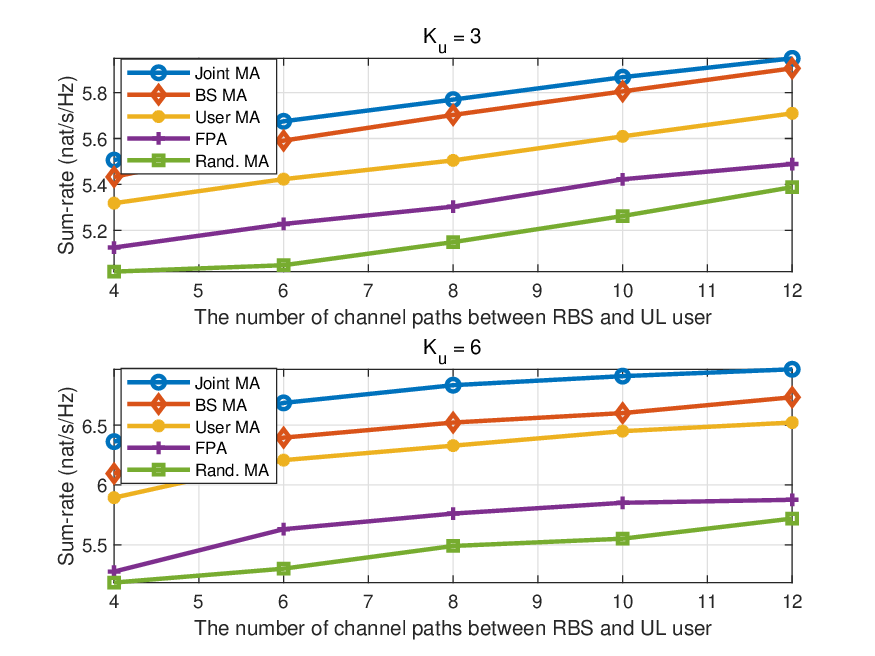}
	\caption{The impact of the number of paths between RBS and UL user.}
	\label{fig.8}
\end{figure}

Fig. \ref{fig.7}  explores the influence of the number of paths between TBSs and DL users on the sum-rate.
It can be easily observed that a larger number of
paths results in a higher achievable sum-rate across all schemes,
and our proposed algorithm demonstrates a superior performance over other schemes.
Besides, 
when $K_d = 6$,
all schemes exhibit a higher sum-rate compared to the case when $K_d = 3$.
Fig. \ref{fig.8}
compares the sum-rate for different
schemes versus the number of paths between RBSs and UL users.
It is readily apparent that an increased number of paths leads to a higher achievable sum-rate in all schemes,
with the ``Joint MA'' scheme exhibiting superior performance compared to others.
In addition,
the $K_u = 6$ case can achieve better performance than the $K_u = 3$ case.

\begin{figure}[t]
    \centering
    \includegraphics[scale=0.60]{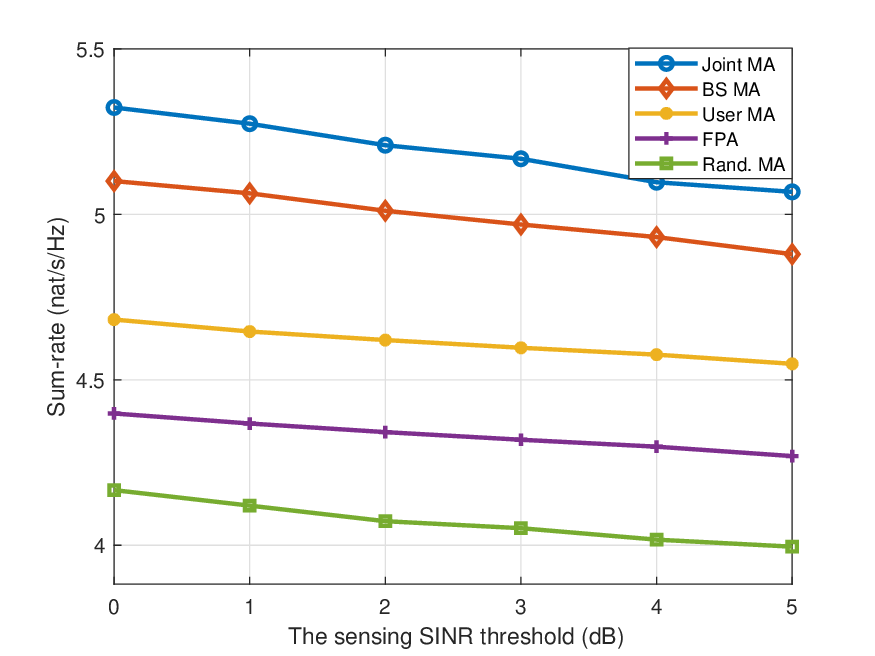}
	\caption{The impact of the sensing SINR performance.}
	\label{fig.9}
\end{figure}

The impact of varying sensing SINR requirements is illustrated in Fig. \ref{fig.9}. 
A clear trade-off between communication performance and target sensing can be observed. 
In this experiment, 
the sensing SINR ranges from $0$ dB to $5$ dB. 
As shown in the figure, 
increasing the sensing SINR results in a decrease in the sum-rate for all cases.

\section{Conclusions}

%
In this paper,
we considered a novel MA-aided networked ISAC system 
that accomplishes radar sensing as well as DL and UL communication capabilities concurrently.
Aiming to maximize the sum-rate among all DL and UL users,
we investigated a joint active beamforming and antenna position coefficients design problem.
Despite the non-convex nature of the considered problem with highly coupled optimization variables, 
we propose  an iterative algorithm to jointly optimize
BS beamforming,
UL users' power allocation,
receiving processors
and
MAs' position configuration
for enabling radar sensing and communication functionalities.
Numerical results showcase the effectiveness of our proposed solution,
highlight the advantages of deploying MA in the networked ISAC system,
and offer useful engineering insights.
Firstly,
while ensuring the sensing performance,
MA can enhance communication performance compared to the case with the FPA-based array.
Additionally, 
in comparison to benchmark schemes, 
our proposed algorithm significantly reduces the required transmit power or the number of antennas needed to 
achieve the desired level of communication performance.

\normalem



\end{document}